\UseRawInputEncoding

\documentclass[nofootinbib,pra,aps,twocolumn,floatfix,10pt,longbibliography,superscriptaddress,notitlepage]{revtex4-1}
\usepackage[english]{babel}
\usepackage{breakcites}
\usepackage{listings}

\usepackage[T1]{fontenc}
\usepackage{textcomp}
\usepackage{gensymb,braket}

\usepackage{graphicx}
\usepackage{braket}
\usepackage{color}
\usepackage{epstopdf}
\usepackage{mathtools}
\usepackage{amsmath}
\usepackage{amssymb}
\usepackage{float}
\usepackage{comment}
\usepackage{amsmath}
\usepackage{multirow}

\usepackage{amsfonts}
\usepackage{bm}
\usepackage{bbold}

\usepackage{color}

\graphicspath{{./figs/}}
\usepackage{gensymb}
\usepackage[colorinlistoftodos,prependcaption]{todonotes}
\usepackage{xargs}  
\newcommandx{\unsure}[2][1=]{\todo[linecolor=red,backgroundcolor=red!25,bordercolor=red,#1]{#2}}
\newcommandx{\change}[2][1=]{\todo[linecolor=blue,backgroundcolor=blue!25,bordercolor=blue,#1]{#2}}
\newcommandx{\info}[2][1=]{\todo[linecolor=OliveGreen,backgroundcolor=OliveGreen!25,bordercolor=OliveGreen,#1]{#2}}
\newcommandx{\improvement}[2][1=]{\todo[linecolor=Plum,backgroundcolor=Plum!25,bordercolor=Plum,#1]{#2}}
\newcommandx{\thiswillnotshow}[2][1=]{\todo[disable,#1]{#2}}
\newcommandx{\greencom}[2][1=]
{\todo[inline, color=green!40,#1]{#2}}
\newcommandx{\bluecom}[2][1=]
{\todo[inline, color=blue!40,#1]{#2}}

\usepackage[colorlinks]{hyperref}
\definecolor{winered}{rgb}{0.5,0,0}
\hypersetup
{
colorlinks=true,
linkcolor=winered,
urlcolor={winered},
filecolor={winered},
citecolor={winered},
allcolors={winered}
}

\usepackage{letltxmacro}
\LetLtxMacro{\ORIGselectlanguage}{\selectlanguage}
\makeatletter
\DeclareRobustCommand{\selectlanguage}[1]{%
  \@ifundefined{alias@\string#1}
    {\ORIGselectlanguage{#1}}
    {\begingroup\edef\x{\endgroup
       \noexpand\ORIGselectlanguage{\@nameuse{alias@#1}}}\x}%
}
\newcommand{\definelanguagealias}[2]{%
  \@namedef{alias@#1}{#2}%
}
\makeatother

\definelanguagealias{en}{english}
\definelanguagealias{EN}{english}
\definecolor{myblue}{rgb}{0.1, 0.1, 0.7}
\definecolor{myred}{rgb}{0.6, 0.1, 0.1}
\definecolor{mygreen}{rgb}{0.0, 0.45, 0.0}
\newcommand{\blue}[1]{{\color{blue}#1}}

\newcommand{\sh}[1]{{\color{olive}#1}}

\newcommand{\violet}[1]{{\color{violet}#1}}

\graphicspath{{figure/}}

\newcommand\blfootnote[1]{%
  \begingroup
  \renewcommand\thefootnote{}\footnote{#1}%
  \addtocounter{footnote}{-1}%
  \endgroup
}

\begin{document}

\title{Connecting classical and quantum 
mode theories for coupled
lossy cavity resonators using quasinormal modes} 

\author{Juanjuan Ren$^\dagger$}
\thanks{These two authors contributed equally to this work}
\affiliation{Department of Physics, Engineering Physics, and Astronomy, Queen's
University, Kingston, Ontario K7L 3N6, Canada}
\author{Sebastian Franke$^\dagger$} 
\thanks{These two authors contributed equally to this work}
\affiliation{Technische Universit\"at Berlin, Institut f\"ur Theoretische Physik,
Nichtlineare Optik und Quantenelektronik, Hardenbergstra{\ss}e 36, 10623 Berlin, Germany}
\affiliation{Department of Physics, Engineering Physics, and Astronomy, Queen's University, Kingston, Ontario K7L 3N6, Canada}
  \author{Stephen Hughes}
 \affiliation{Department of Physics, Engineering Physics, and Astronomy, Queen's University, Kingston, Ontario K7L 3N6, Canada}


\begin{abstract}
We present a quantized quasinormal approach to rigorously describe coupled lossy resonators, and quantify the quantum coupling parameters as a function of distance between the resonators. We 
also make a direct  connection between classical and quantum quasinormal modes parameters and theories, offering new and unique insights into coupled open cavity resonators. We present detailed calculations for coupled microdisk resonators and show striking interference effects that depend on the phase of the quasinormal modes, an effect that is also significant for high quality factor modes. Our results demonstrate that commonly adopted master equations for such systems are generally not applicable and we discuss  the new physics that is captured using the quantized quasinormal mode coupling parameters and show how these relate to the classical mode parameters. 
Using these new insights, we also  present several models to fix the failures of the dissipative Jaynes-Cummings type models for coupled cavity resonators. Additionally, we show how to improve the classical and quantum lossless mode  models (i.e., using normal modes) by employing a non-diagonal mode expansion based on the knowledge of the quasinormal mode eigenfrequencies,  and analytical coupled mode theory, to accurately capture the mode interference effects for 
high quality factors.
\end{abstract}
\maketitle 

\section{Introduction}\label{sec: Intro}
 \blfootnote{$^\dagger$ \text{jr180@queensu.ca, sebastian.franke@tu-berlin.de}}

A cavity mode description of nanophotonic resonators is of great significance especially when 
describing interactions with quantum emitters~\cite{kristensen_modes_2014}, and is a requirement for system level quantization in quantum nanophotonics.
The usual optical cavity mode approach to describe most optical resonators is to use normal modes (NMs), which are
only rigorously valid for closed or periodic
structures, and produce  real mode eigenfrequencies.
For open resonators, one typically  compliments such a mode picture by heuristically
adding in photon losses, usually resulting in a Lorentzian lineshape per mode of interest.

In a  quantum optics picture of resonant cavities, the NMs are first quantized with no loss (quantized harmonic oscillator modes),
and for a single NM coupled to a two-level system (TLS), this forms the basis of the quantum Rabi model~\cite{frisk_kockum_ultrastrong_2019,forn-diaz_ultrastrong_2019}; and,   following a rotating-wave approximation, the celebrated Jaynes-Cummings (JC) model is obtained.
In a bad cavity limit, the JC model~\cite{jaynes_comparison_1963},
which describes losses through
system-bath interactions at the level of Lindblad operators (dissipative JC model), 
also yields a Lorentzian lineshape response (unless more advanced bath techniques are employed),
which agrees with the classical NM models with a phenomenological decay rate.
These NM models are expected to be
accurate for high $Q$ (quality factor) systems, though overlapping 
lossy resonances in photonics 
can yield
dramatically different lineshapes, including Fano-like resonances, which is characterized with a asymmetric spectral lineshape, e.g.,  induced by interference between a broad spectral mode and a narrow mode in various hybrid optical systems~\cite{doi:10.1143/JPSJ.80.104707,barth_nanoassembled_2010,doeleman_antennacavity_2016,thakkar_sculpting_2017,limonov_fano_2017,dezfouli_molecular_2019}, as well as in some optomechanical systems~\cite{qu_fano_2013,abbas_investigation_2019,el-sayed_quasinormal-mode_2020}.


Notably, the Fano resonance 
of overlapping dissipative modes,
as shown in Refs.~\onlinecite{RosenkrantzdeLasson2015,2017PRA_hybrid}, 
are quantitatively well explained by two coupled 
quasinormal modes (QNMs), which are the formal solution
to optical cavities with open boundary conditions~\cite{lai_time-independent_1990,leung_completeness_1994,leung_time-independent_1994,leung_completeness_1996,lee_dyadic_1999,kristensen_generalized_2012,sauvan_theory_2013,kristensen_modes_2014,PhysRevA.98.043806,PhysRevX.7.021035,lalanne_light_2018,kristensen_modeling_2020}.
Unlike NMs, QNMs have 
 complex eigenfrequencies $\tilde{\omega}_{\mu}=\omega_{\mu}-i\gamma_{\mu}$, with 
 $Q_{\mu}=\tilde{\omega}_{\mu}/(2\gamma_{\mu})$, so the radiation and/or absorption losses are naturally included in QNMs, and the problem becomes
 non-Hermitian from the beginning.
Importantly,  QNMs  can also be used to accurately construct
the photon Green function (GF) with a few mode expansion, and such models  have been used with great success for describing a range of semiclassical light-matter interaction for open resonators~\cite{lee_dyadic_1999,muljarov_brillouin-wigner_2010,kristensen_generalized_2012,sauvan_theory_2013,kristensen_modes_2014,muljarov_exact_2016,lalanne_light_2018,martin-cano_chiral_2019,kristensen_modeling_2020}. Indeed, in the QNM expansion, the phase of the modes can have a significant effect, where certain modes even contribute negatively to the photonic local density of states (LDOS)~\cite{2017PRA_hybrid}.

With regards to quantizing  QNMs, 
 after some initial progress for one-dimensional structures~\cite{ho_second_1998,severini_second_2004},
 a fully three-dimensional quantization scheme was recently presented~\cite{franke_quantization_2019}.
 Moreover, 
 using a hybrid structure containing a nanoantenna and a photonic crystal cavity,
it was shown how Fano-like resonances can also be explained using both a classical and a quantized QNM approach, highlighting important 
mode interference effects that are {\it not} captured by the 
dissipative JC model~\cite{franke_quantized_2020}. Improved numerical schemes
have also been introduced for obtaining the quantum optical parameters~\cite{ren_near-field_2020}, which can readily be used for the QNM master equations.

In this paper, we 
present and apply 
both classical and quantized NM and QNM theories 
for two high $Q$  (around $10^{4}-10^{5}$) coupled lossy resonators.
We demonstrate dramatic
differences between the quantized QNMs theory and phenomenological dissipative JC models, 
and show precisely how these connect to a breakdown also of the
classical NM theories (with heuristic cavity decay rates). 
To better explain the classical QNMs of the coupled lossy microdisks system as a function of gap distance, we also
use an efficient coupled QNM theory (CQT)~\cite{ren_quasinormal_2021}, where only the QNMs of the bare resonators are required (see also Refs.~\onlinecite{vial_coupling_2016,Kristensen_coupled_modes_2017,cognee_hybridization_2020,tao_coupling_2020}), which show excellent agreement with classical full numerical dipole results for the resonators of interest, i.e., when obtained from a numerical solution of the Maxwell equations with a classical dipole source.
In addition, we study resonator gap distances
that yield QNMs close to a lossy exceptional point (EP), where two modes (both eigenfrequencies and modes)
become very close together.

In order to obtain a quantized QNMs solution, quantum overlap integrals $S_{ij}$ are required, which depend on the properties of the classical 
QNMs~\cite{franke_quantization_2019}.
Defining the QNM raising and lowering operators 
as $\tilde\alpha_i$ and $\tilde\alpha_i^\dagger$,
then these factors are defined from $S_{ij} = [\tilde \alpha_i,\tilde \alpha_j^\dagger]$.
In the limit of closed lossless system, $S_{ij}$ would be the usual Kronecker delta~\cite{franke_quantization_2019,Hughes_SPS_2019},
namely $S_{ij}=\delta_{ij}$, a result that is formally obtained only in the limit that the quality factors of the modes approach infinity~\cite{franke_fluctuation-dissipation_2020} (lossless systems). 
With finite losses, then $S_{ij}$ in general contains both  nonradiative and radiative contributions.
The nonradiative part $S^{\rm nrad}_{ij}$ and the radiative part $S^{\rm rad}_{ij}$ account for the absorption loss and the radiation loss (the power flow goes out) of the system, respectively.
To allow the construction of Fock
states, one can introduce  new operators via a symmetrizing
orthonormalization transformation~\cite{franke_quantization_2019}.
Consequently, in the bad cavity limit,  the quantum Purcell factors obtained from the quantized QNMs
recover the same answer as the classical  Purcell factors (using the classical QNMs), and we show that indeed this is the case,
where we show excellent agreement 
 as a function of gap distances and in the frequency range of interest, indicating the efficiency and validity of the classical QNMs and quantized QNMs approaches.

To help demonstrate the breakdown of the
NM theories, we compare these QNM results
with the classical NM solution with  phenomenological damping as well as the dissipative JC model, 
where  $S_{ij}=\delta_{ij}$. We show how the classical NM solutions differ from the classical QNM solutions, and connect these to the differences between the phenomenological dissipative JC model and the quantized QNM solution.
Moreover, the NMs solution coincide almost perfectly with the phenomenological dissipative JC model, which indicate the close relationship between the QNMs phases and the off-diagonal coupling
that appears in the QNM master equations.
Furthermore, we show how the non-diagonal coupling elements   contribute the most around the lossy EP, demonstrating the role of quantum mechanical coupling between the quantized QNMs.
In addition, the QNM decay rates and the emitter-QNM coupling rates due to symmetrization are presented and they are seen to maximally change around the EP.
The QNM-QNM coupling rates are also directly presented, whose absolute values decrease as the resonator gap distances increase.
Last but not the least, present a more accurate NM solutions for both classical and quantum level starting from the non-diagonal form of the QNM GF and the Hamiltonian with the bare QNM quantities, and then add the coupling parameters from CQT, which helps one understand the mode interference effects.

The rest of our paper is organized as follows:
In Sec.~\ref{sec: theory}, we introduce the main theory  of the classical and quantized QNMs formalisms used in this paper. Section~\ref{sec: cQNMs} presents the classical QNMs and Green functions (expanded in terms of the QNMs), where the generalized Purcell factors from classical QNMs are also given.
In Sec.~\ref{sec: CQNMT}, the key results of an efficient 
  CQT is reviewed~\cite{ren_quasinormal_2021}.
Section~\ref{sec: cNMs} shows the classical NMs solutions.
In order to check the validity of the classical QNMs/NMs solutions, Sec.~\ref{sec: Fulldipole} shows how full numerical dipole solutions to Maxwell's equations are obtained. 
The quantum QNM formalism is presented  in Sec.~\ref{sec: quanQNMs}, which summarizes the key results from Ref.~\onlinecite{franke_quantization_2019}.
In Sec.~\ref{sec: phase_nondiag}, a  comparison between classical and quantum interference is presented, which shows the equivalence of QNMs phases and off-diagonal coupling from the point view of the Green functions, though with a completely different interpretation. 

We then utilize the above theory and present detailed numerical results for coupled microdisk resonators in Sec.~\ref{sec: numerical}.
As the basis of the CQT, Sec.~\ref{sec: singleQNM}  presents the classical QNMs solution for the bare lossy resonators.
Using the analytical CQT, the hybrid classical QNMs (i.e., the QNMs for the coupled system) are presented in Sec.~\ref{sec: hybridQNMs}, where the QNMs phases play a significant role in those unusual lineshape for Purcell factors.
In Sec.~\ref{sec: NMsvsQNMs}, the key results of this paper are presented, where the comparison between NMs and QNMs are shown in both classical and quantum picture. We find that even for the high $Q$ resonators, the NMs theory and disspative JC model miss a significant amount of
the mode coupling effects, which are equally captured by the classical and quantized QNMs, but with quite different perspectives on the underlying physics.

Section~\ref{sec: nondiagonal_contri} discussed how the contribution of the non-diagonal terms in quantized QNMs is dominant when close to the lossy EP.
In sec.~\ref{sec: ratios}, we also present the differences of the quantum parameters between the quantum QNM model and the  dissipative JC model, specifically the respective ratios of the emitter-QNM coupling rates and the QNM decay rates as well as the photon-photon coupling constants. 
Section~\ref{sec: ImproveNM} introduces a new improved NM theory on both classical and quantum levels, starting from 
 a non-diagonal representation
 of the QNM Green function and Hamiltonian, and we demonstrate its high accuracy.
Finally, Sec.~\ref{sec: conclusion}  presents closing discussions and our conclusions.

\section{Theory}
\label{sec: theory}
In this section, we introduce the main theoretical concepts and
formulas, starting with the definition of classical QNM for open cavities, the Green function,
and the classical Purcell factor. We then summarize the key aspects of CQT, 
developed elsewhere for coupled loss-gain resonators~\cite{ren_quasinormal_2021},
as well as connect to the more usual NM theories.
We subsequently present the key results from quantized QNM approach, where the coupling between the modes has a completely different interpretation, one that is manifestly related to the fundamental quantization required to construct a meaningful Fock state basis for the resonator modes. We also discuss what new physics is obtained beyond a dissipative JC model. The correspondence between these classical and quantum pictures is then clarified in the bad cavity limit, where they can be compared directly, and where they should agree.

\subsection{Classical quasinormal modes and Green functions}\label{sec: cQNMs}

For open cavity mode problems in optics, the  QNMs~\cite{lai_time-independent_1990,leung_completeness_1994,leung_time-independent_1994,leung_completeness_1996,lee_dyadic_1999,kristensen_generalized_2012,sauvan_theory_2013,kristensen_modes_2014,PhysRevX.7.021035,lalanne_light_2018,kristensen_modeling_2020} 
 are obtained from the
vector  Helmholtz equation,
\begin{equation}\label{smallf}
\boldsymbol{\nabla}\times\boldsymbol{\nabla}\times\tilde{\mathbf{f}}_{{\mu}}\left(\mathbf{r}\right)-\left(\dfrac{\tilde{\omega}_{{\mu}}}{c}\right)^{2}
\epsilon(\mathbf{r},\tilde{\omega}_{\mu})\,\tilde{\mathbf{f}}_{{\mu}}\left(\mathbf{r}\right)=0,
\end{equation}
along with open boundary conditions, specifically the Silver-M\"uller radiation condition~\cite{Kristensen2015}.
The QNMs have complex eigenfrequencies, $\tilde{\omega}_{\mu}=\omega_{\mu}-i\gamma_{\mu}$, with a
corresponding quality factor  $Q_\mu=\omega_{\mu}/(2\gamma_{\mu})$.

For numerically obtaining the normalized QNM, $\tilde{\mathbf{f}}_{\mu}$,
of arbitrarily shaped resonators, 
one simple approach is to
obtain  the scattering fields of the cavity from a dipole source in complex frequency space, which is an efficient integration-free method~\cite{bai_efficient_2013-1}.

The QNM norm can also be defined in terms of an integration over these modes, which requires a careful regularization in general~\cite{Kristensen2015}. For example, one definition uses
the Perfectly-Matched-Layers (PML) to ensure the fields are zero at the final surfaces
of the computed mode~\cite{sauvan_theory_2013,vial_quasimodal_2014,yan_rigorous_2018,lalanne_light_2018,christopoulos_perturbation_2020}.  
For  non-dispersive and non-magnetic systems, the PML norm is
\begin{align}
&\braket{\braket{\tilde{\mathbf{f}}_{\mu}(\mathbf{r})|\tilde{\mathbf{f}}_{\mu}(\mathbf{r})}}= \nonumber \\
&{\frac{1}{2}}\int_{V-V_{\rm PML}}\bigg[
\epsilon(\mathbf{r})\tilde{\mathbf{f}}_{\mu}(\mathbf{r})\cdot\tilde{\mathbf{f}}_{\mu}(\mathbf{r})
{-\frac{\mu_0}{\epsilon_0} \tilde{\mathbf{h}}_{\mu}(\mathbf{r})\cdot\tilde{\mathbf{h}}_{\mu}(\mathbf{r})
\bigg]}dV \nonumber \\
&+\frac{1}{2\epsilon_{0}}\int_{V_{\rm PML}}\bigg[\epsilon_{0}\frac{\partial(\omega\epsilon_{\rm PML}(\mathbf{r},\omega))}{\partial\omega}\Big|_{\tilde{\omega}_{\mu}}\tilde{\mathbf{f}}_{\mu}(\mathbf{r})\cdot\tilde{\mathbf{f}}_{\mu}(\mathbf{r}) \nonumber \\
&-\mu_{0}\frac{\partial(\omega\mu_{\rm PML}(\mathbf{r},\omega))}{\partial\omega}\Big|_{\tilde{\omega}_{\mu}}\tilde{\mathbf{h}}_{\mu}(\mathbf{r})\cdot\tilde{\mathbf{h}}_{\mu}(\mathbf{r})\bigg]dV=1.
\label{eq:PML}
\end{align}
Using the normalized QNMs $\tilde{\mathbf{f}}_{\mu}$ from the dipole technique~\cite{bai_efficient_2013-1},  one will find $\braket{\braket{\tilde{\mathbf{f}}_{\mu}(\mathbf{r})|\tilde{\mathbf{f}}_{\mu}(\mathbf{r})}} =1$ (within numerical precision), as  expected. 
For simplicity we will write this norm as
\begin{equation}
\braket{\braket{\tilde {\bf f}_{\mu}|\hat \epsilon_{\mu}|\tilde {\bf f}_{\mu}}}
\rightarrow 
\int_{\rm reg} d{\bf r} \epsilon_{\mu}({\bf r}) \tilde {\bf f}_{\mu}({\bf r}) \tilde {\bf f}_{\mu}({\bf r}) =1,
\end{equation}
and only the non-PML region is important for the CQT
we use below.

The Green function for the cavity system, with permittivity $\epsilon(\mathbf{r},{\omega})$, satisfies ($k_{0}=\omega/ c$)
\begin{equation}\label{eq: Green_eq}
\boldsymbol{\nabla}\times\boldsymbol{\nabla}\times{\mathbf{G}}\left(\mathbf{r},\mathbf{r}^{\prime},\omega\right)-k_{0}^{2}
\epsilon(\mathbf{r},{\omega})\,{\mathbf{G}}\left(\mathbf{r},\mathbf{r}^{\prime},\omega\right)=k_{0}^{2}\delta(\mathbf{r}-\mathbf{r}^{\prime})\mathbf{I},
\end{equation}
which is the solution to a point dipole source, and $\mathbf{I}$ is the unit tensor.
When the QNMs are appropriately normalized as shown above, one can use them to reconstruct the photon Green function near or within the resonators~\cite{leung_completeness_1994,ge_quasinormal_2014},
\begin{equation}
\mathbf{G}\left(\mathbf{r},\mathbf{r}^{\prime},\omega\right)= \sum_{\mu} A_{\mu}\left(\omega\right)\,\tilde{\mathbf{f}}_{\mu}\left({\bf r}\right)\tilde{\mathbf{f}}_{\mu}\left({\bf r}^{\prime}\right),\label{eq: GFwithSUM}
\end{equation}
where $A_{\mu}(\omega)={\omega}/{2(\tilde{\omega}_{\mu}-\omega)}$ and $\tilde{\mathbf{f}}_{\mu}$ is the frequency independent QNM.

The Green function can easily be used to obtain the
spontaneous emission (SE) rate.
As an example, when a dipole with dipole moment $\mathbf{d} =d_{0}\mathbf{n}_{\rm d}$ is placed at $\mathbf{r}_{0}$, the corresponding {\it generalized} Purcell factor is given as~\cite{Anger2006,kristensen_modes_2014} 
\begin{align}\label{QNMpurcell}
 \begin{split}
     &F_{{\rm P}}^{\rm }(\mathbf{r}_{0},\omega) =1+\frac{\Gamma(\mathbf{r}_{0},\omega)}{\Gamma_{0}(\mathbf{r}_{0},\omega)},
\end{split}
\end{align}
with a total SE rate $\Gamma_{\rm }(\mathbf{r}_{0},\omega)$, and a background SE rate $\Gamma_{0}(\mathbf{r}_{0},\omega)$ in a homogeneous background medium, where
\begin{align}
\begin{split}
\Gamma_{\rm }(\mathbf{r}_{0},\omega)&=\frac{2}{\hbar\epsilon_{0}}\mathbf{d}\cdot{\rm Im}\{\mathbf{G}(\mathbf{r}_{0},\mathbf{r}_{0},\omega)\}\cdot\mathbf{d},\\
\Gamma_{0}(\mathbf{r}_{0},\omega)&=\frac{2}{\hbar\epsilon_{0}}\mathbf{d}\cdot{\rm Im}\{\mathbf{G}_{\rm B}(\mathbf{r}_{0},\mathbf{r}_{0},\omega)\}\cdot\mathbf{d},
\end{split}
\end{align}
where $\bf G$ (Eq.~\eqref{eq: GFwithSUM})
contains the resonant modes that dominate in the frequency band of interest (total Green function), and ${\bf G}_{\rm B}$
is the homogeneous Green function for a background medium\footnote{For a 2D TM dipole, it is ${\rm Im}\{\mathbf{G}_{\rm B}(\mathbf{r}_{0},\mathbf{r}_{0},\omega)\}={\omega^2}/{4c^2}$.}. Note the additional $1$ is added in the generalized Purcell factor, $F_{\rm P}$ , which originates from the background contribution when the dipole is placed outside the resonator~\cite{ge_quasinormal_2014}.

\subsection{Coupled QNM theory and analytical expression for the hybrid modes}\label{sec: CQNMT}

Standard (temporal) coupled mode theory (CMT) typically uses NMs with phenomenological losses, and it is able to partly investigate the hybridized (coupled) structure using  only the parameters of the individual system~\cite{haus_coupled-mode_1991,fan_temporal_2003}. This is expected to be valid for
high $Q$ resonances~\cite{Kristensen_coupled_modes_2017}.
However, since the usual CMT 
is based on NM theory, 
the formalism can be  problematic,
e.g., when dealing with complex couplings
such as those near  EPs (exceptional points)~\cite{miri_exceptional_2019,chen_revealing_2020,ren_quasinormal_2021}.
Motivated by the ongoing development of optical QNMs,  several newly CMT approaches have been developed, that are based on QNMs~\cite{vial_coupling_2016,tao_coupling_2020,ren_quasinormal_2021}.

Here we briefly review the key CQT results of Ref.~\onlinecite{ren_quasinormal_2021} (developed for loss-gain resonators), which will be used in this work.
We focus on a system consisting of two individual structures, whose QNMs, eigenfrequencies, coupling coefficients, and permittivity are labelled as $\tilde{\mathbf{f}}_{1/2}$, $\tilde{\omega}_{1/2}$, $\tilde{\kappa}_{12/21}$, and $\epsilon_{1/2}$ ($\epsilon_{\rm B}$ for background medium, $\epsilon_{\rm t}$ for hybrid system).
We define permittivity difference operator as $\hat{V}_{1/2}$, which satisfy $\hat{\epsilon}_{1/2}=\hat{\epsilon}_{\rm B}+\hat{V}_{1/2}$ and $\hat{\epsilon}_{\rm t}=\hat{\epsilon}_{\rm B}+\hat{V}_{1}+\hat{V}_{2}$. 
The subscripts `1' and `2' are general here
and represent the bare modes of two couple resonators.
  Two new eigenfrequencies are given as follows 
\begin{equation}
\label{eq: rootsRWA}
\tilde \omega_{\pm}=
\frac{\tilde\omega_1+\tilde\omega_2}{2}
\pm \frac{\sqrt{4\tilde \kappa_{12}\tilde \kappa_{21} + (\tilde\omega_1-\tilde\omega_2)^2}}{2},
\end{equation}
where the coupling coefficients are
\begin{equation}\label{QNMCMT_coup}
\tilde \kappa_{ij}=
\frac{\tilde \omega_j}{2} \braket{\braket{\tilde{\bf f}_i
|\hat V_i|\tilde{\bf f}_j}},
\end{equation}
with ($i,j=1,2$) and 
\begin{equation}
\braket{\braket{\tilde{\bf f}_i
|\hat V_i|\tilde{\bf f}_j}} 
=\int_{\Omega_{i}} d{\bf r} [\epsilon_i({\bf r})-\epsilon_{\rm B}] \tilde {\bf f}_i({\bf r}) \tilde {\bf f}_j({\bf r}).
\end{equation}
Note that this expression  is derived for a non-dispersive and non-magnetic medium, and  ${\Omega_{i}}$ represents the region of the individual resonator system $i$ (1 or 2). The generalization to dispersive media is straightforward, but is not needed for our example numerical calculations below.

In the presence of finite coupling, 
two new ``hybrid'' QNMs are then obtained,
\begin{align}\label{eq: QNMs_pm}
\ket{\tilde{\bf f}^\pm}&=
\frac{\tilde\omega_\pm-\tilde\omega_2}{\sqrt{(\tilde\omega_\pm-\tilde\omega_2)^2+ \tilde \kappa_{21}^2}}
\ket{\tilde{\bf f}_1}
\nonumber \\
&
+ \frac{-\tilde\kappa_{21}}{\sqrt{(\tilde\omega_\pm-\tilde\omega_2)^2+ \tilde \kappa_{21}^2}} 
\ket{\tilde{\bf f}_2}.
\end{align}
The corresponding two-QNM  Green function for the coupled system is
given by a two mode expansion,
in shorthand notation, reads
\begin{align}
\label{eq: GreencQNM}
\hat {\bf G}^{\rm QNM}
&=\hat {\bf G}^{\rm QNM}_{+}+\hat {\bf G}^{\rm QNM}_{-}\\
&= \frac{\omega \ket{\tilde {\bf f}^+}\bra{\tilde{\bf f}^{+*}}}
{2(\tilde \omega_+ - \omega)} 
 + \frac{\omega  \ket{\tilde{\bf f}^-}\bra{\tilde{\bf f}^{-*}}}{2(\tilde \omega_- - \omega)},
\end{align}
where $\braket{{\bf r}|\hat {\bf G}|{\bf r'}}
={\bf G}({\bf r},{\bf r}')$
and $\braket{{\bf r}|\tilde {\bf f}^+}
= \tilde{\bf f}^+({\bf r}).$

The  {\it classical Purcell factor} is
immediately obtained from only the properties of the QNMs,
\begin{align}
     F_{{\rm P}}^{\rm cQNM}(\mathbf{r}_{0},\omega) =1+\frac{\mathbf{d}\cdot{\rm Im}\{\mathbf{G}^{\rm QNM}(\mathbf{r}_{0},\mathbf{r}_{0},\omega)\}\cdot\mathbf{d}}{\mathbf{d}\cdot{\rm Im}\{\mathbf{G}_{\rm B}(\mathbf{r}_{0},\mathbf{r}_{0},\omega)\}\cdot\mathbf{d}}.\label{eq: FP_cQNM}
\end{align}
Moreover, the separate contribution from the two coupled modes in Eq.~\eqref{eq: GreencQNM} is defined as
\begin{align}
     F_{{\rm P},\pm}^{\rm cQNM}(\mathbf{r}_{0},\omega) =\frac{\mathbf{d}\cdot{\rm Im}\{\mathbf{G}_{\pm}^{\rm QNM}(\mathbf{r}_{0},\mathbf{r}_{0},\omega)\}\cdot\mathbf{d}}{\mathbf{d}\cdot{\rm Im}\{\mathbf{G}_{\rm B}(\mathbf{r}_{0},\mathbf{r}_{0},\omega)\}\cdot\mathbf{d}},\label{eq: FP_cQNM_pm}
\end{align}
which satisfy $F_{{\rm P}}^{\rm cQNM}=1+F_{{\rm P},+}^{\rm cQNM}+F_{{\rm P,}-}^{\rm cQNM}$.

\subsection{Classical normal mode limit with heuristic losses}\label{sec: cNMs}
It is useful to also compare the
QNM results with the more
common Green function expanded
in terms of NMs, namely\footnote{This is derived in a rotating wave approximation, as we can safely neglect the negative pole contribution.} from~\cite{PhysRevA.70.053823,Yao:09},
\begin{equation}
{\bf G}({\bf r},{\bf r}^{\prime},\omega) \approx \sum_\mu 
\frac{\omega\, {\bf f}_{\mu}^*({\bf r}){\bf f}_{\mu}({\bf r}^{\prime})}{2(\tilde\omega_\mu-\omega)},
\end{equation}
where the QNM modes are used in the expansion,  as these are the ones that are assumed to be computed,
and also the eigenfrequency is also
assumed to be complex. This is precisely the usual NM approach to heuristically
obtain GF expansions with a few lossy cavity modes, though for our purpose we will also assume that the modes in this expansion are correctly normalized for QNMs, as the usual NM normalization is also ambiguous in general~\cite{kristensen_generalized_2012}.

Thus, we introduce
an effective model for the NM Green function expansion, 
\begin{align}
\label{eq: GreencNM}
\hat {\bf G}^{\rm NM}
&= \frac{\omega \ket{\tilde {\bf f}^+}\bra{\tilde{\bf f}^{+}}}
{2(\tilde \omega_+ - \omega)} 
 + \frac{\omega  \ket{\tilde{\bf f}^-}\bra{\tilde{\bf f}^{-}}}{2(\tilde \omega_- - \omega)},
\end{align}
and the corresponding classical Purcell factor,
in terms of a NM expansion, is
\begin{align}
     F_{{\rm P}}^{\rm cNM}(\mathbf{r}_{0},\omega) =1+\frac{\mathbf{d}\cdot{\rm Im}\{\mathbf{G}^{\rm NM}(\mathbf{r}_{0},\mathbf{r}_{0},\omega)\}\cdot\mathbf{d}}{\mathbf{d}\cdot{\rm Im}\{\mathbf{G}_{\rm B}(\mathbf{r}_{0},\mathbf{r}_{0},\omega)\}\cdot\mathbf{d}},\label{eq: FP_cNM}
\end{align}
which differs from the QNM form,
mainly through the influence  of the
QNM phase, since it is no longer important here. It is precisely the QNM phase that causes a departure from the usual Lorentzian-like
spectrum response of the modes, which also 
requires  correction terms in the quantum models, though they appear in a different way.

\subsection{Numerical calculation of the classical Purcell factor}\label{sec: Fulldipole}
\label{sec: numericalSE}

In order to compare with the above classical QNM and NM expansions, the full numerical dipole solution to the Maxwell's equation can be directly computed 
in a numerical Maxwell solver, and for this work we use
the commercial COMSOL Multiphysics software~\cite{comsol}.

The numerical Purcell factors with a full numerical dipole method is obtained from
\begin{equation}\label{eq: Purcellfulldipole}
    F_{\rm P}^{\rm num}(\mathbf{r}_{0},\omega)=\frac{\int_{ L_{\rm c}}\hat{\mathbf{n}}\cdot {\bf S}_{\rm dipole,total}(\mathbf{r},\omega)d{L_{\rm c}} }{\int_{ L_{\rm c}}\hat{\mathbf{n}}\cdot {\bf S}_{\rm dipole,background}(\mathbf{r},\omega)d{L_{\rm c}} },
\end{equation}
where ${\bf S}(\mathbf{r},\omega)$ is the Poynting vector at a small circle (with radius of $1$~nm in our simulation) $ L_{\rm c}$, which is needed in the cases with and without the resonator, representing by the subscript `total' and `background'.

The full numerical dipole solutions are usually regarded as the benchmark for the QNM expansion, as there are no approximations (other than inherent numerical gridding ones). 
However, as seen from Eq.~\eqref{eq: Purcellfulldipole}, we note that one must run the simulation for each dipole position and also for each frequency, which is time consuming and not very useful for understanding the underlying physics.
Moreover, the full numerical approach can only give the total  Green function contribution, where the underlying origin of mode interference  would be difficult or impossible  to understand.
Such shortcomings are well solved by the classical QNMs, where the response at each position and each frequency can be obtained analytically and, simultaneously, after the corresponding Green function expansion is obtained from these  QNMs.
In a diagnonal QNM representation, the mode interference is also clearly described through the QNM phases. 

\subsection{Quantized QNM formalism}\label{sec: quanQNMs}
From  a rigorous quantum optics perspective of open cavity modes,
recent theoretical approaches have 
shown how to quantize QNMs~\cite{franke_quantization_2019,franke_fluctuation-dissipation_2020,franke_quantized_2020,franke_fermis_2021}, which
yield important differences
from a multi-mode dissipative JC model,
including quantum-induced
coupling through dissipation. 
Below, we review
the key parameters and equations that are needed, so we can compare directly with the classical picture
(using both NM and QNM expansions).

For an arbitrary lossy medium,
the quantized electric-field operator $\hat{\mathbf{E}}(\mathbf{r},\omega)$ is solved from the following Helmholtz equation~\cite{dung_three-dimensional_1998,suttorp_fano_2004}
\begin{equation}
\boldsymbol{\nabla}\times\boldsymbol{\nabla}\times\hat{\mathbf{E}}(\mathbf{r},\omega) -\frac{\omega^2}{c^2}\epsilon(\mathbf{r},\omega)\hat{\mathbf{E}}(\mathbf{r},\omega)=i\omega\mu_0\hat{\mathbf{j}}_{\mathrm N}(\mathbf{r},\omega),
\label{eq: HelmholtzQuant2}
\end{equation}
where $\hat{\mathbf{j}}_{\mathrm N}(\mathbf{r},\omega)$ represents a phenomenological quantum noise density, connecting to the annihilation operator $\hat{\mathbf{b}}(\mathbf{r},\omega)$ of the vacuum electric field and lossy medium, through
$\hat{\mathbf{j}}_{\mathrm N}(\mathbf{r},\omega)=\omega\sqrt{\hbar\epsilon_0\epsilon_I(\mathbf{r},\omega)/\pi}~\hat{\mathbf{b}}(\mathbf{r},\omega)$ ($\epsilon_I$ is the imaginary part of $\epsilon$).
The solution of Eq.~\eqref{eq: HelmholtzQuant2} is given as
\begin{equation}
\hat{\mathbf{E}}(\mathbf{r},\omega)=\frac{i}{\omega\epsilon_0}\int{\mathrm d}\mathbf{r}^{\prime}\, \mathbf{G}(\mathbf{r},\mathbf{r}^{\prime},\omega)\cdot \hat{\mathbf{j}}_{\mathrm N}(\mathbf{r}^{\prime},\omega),\label{eq: SourceEField2}
\end{equation}
where $\mathbf{G}(\mathbf{r},\mathbf{r}^{\prime},\omega)$ is again the (classical) photon Green function.

Using the QNM Green function from Eq.~\eqref{eq: GFwithSUM}, the electric field operator is rewritten as
\begin{equation}
\hat{\mathbf{E}}_{\rm QNM}(\mathbf{r})=i\sum_{\mu}\sqrt{\frac{\hbar\omega_\mu}{2\epsilon_0}}\tilde{\mathbf{f}}_\mu(\mathbf{r})\tilde{\alpha}_\mu + {\rm H.a.}, 
\end{equation}
near or within the resonators 
, where the QNM operators $\tilde{\alpha}_\mu$ is defined through a linear combinations of the noise source operators $\hat{\mathbf{b}}(\mathbf{r},\omega)$~\cite{franke_quantization_2019}.
The QNMs raising and lowering operators satisfy $[\tilde{\alpha}_{\mu},\tilde{\alpha}_{\eta}]=[\tilde{\alpha}_{\mu}^\dagger,\tilde{\alpha}_{\eta}^\dagger]=0$, and $[\tilde{\alpha}_{\mu},\tilde{\alpha}_{\eta}^\dagger]=S_{\mu\eta}$, where the {\it positive definite} $S$ factors are defined as~\cite{franke_quantization_2019,franke_fluctuation-dissipation_2020,franke_quantized_2020}
\begin{equation}
    S_{\mu\eta}=S_{\mu\eta}^{\rm rad}
    + S_{\mu\eta}^{\rm nrad},
\end{equation}
which yields a QNM overlap matrix related to the dissipation of the system, including radiative and nonradiative contributions.

For our numerical study below, we find that
$S_{\mu\eta} \approx
     S_{\mu\eta}^{\rm nrad}$,
    so we only consider the nonraditive contribution, which is calculated from 
\begin{align}
\begin{split}\label{eq: Snrad_fullw}
S_{\mu\eta}^\mathrm{nrad}=&\frac{1}{2\pi\sqrt{\omega_\mu\omega_\eta}}\int_0^\infty{\mathrm d}\omega\int_{V_{\mathrm{loss}}} {\mathrm d}\mathbf{r}\\
&\frac{\omega^2\epsilon_I(\mathbf{r},\omega)}{(\tilde{\omega}_\mu-\omega)(\tilde{\omega}_\eta^*-\omega)}\tilde{\mathbf{f}}_\mu(\mathbf{r})\cdot\tilde{\mathbf{f}}_\eta^*(\mathbf{r}),
\end{split}
\end{align}
where $V_{\mathrm{loss}}$ is the lossy cavity region.
Employing an accurate pole approximation \cite{franke_fluctuation-dissipation_2020,franke_quantized_2020,ren_near-field_2020}, which simplifies the integral over frequency, then
we have
\begin{align}
\begin{split}\label{eq: Snrad_pole}
S_{\mu\eta}^\mathrm{nrad}\approx&\frac{\sqrt{\omega_\mu\omega_\eta}}{i(\tilde{\omega}_\mu-\tilde{\omega}_\eta^*)}\int_{V_{\mathrm{loss}}} {\mathrm d}\mathbf{r}\\
&\sqrt{\epsilon_I(\mathbf{r},\omega_\mu)\epsilon_I(\mathbf{r},\omega_\eta)}\tilde{\mathbf{f}}_\mu(\mathbf{r})\cdot\tilde{\mathbf{f}}_\eta^*(\mathbf{r}).
\end{split}
\end{align}
For our examples shown in Sec.~\ref{sec: numerical}, two dominating coupled modes are included, where $\mu,\eta=+,-$ and the imaginary part of the permittivity is $\epsilon_I(\mathbf{r},\omega_\mu/\omega_\eta)={\rm Im}[n_{\rm L}^2]$ (${\rm Im}[n_{\rm R}^2]$) within lossy resonator L (R).

The formal QNM quantization scheme is obtained by constructing appropriate annihilation and creation operators via a symmetrization transformation~\cite{franke_quantization_2019,franke_fluctuation-dissipation_2020,franke_quantized_2020}
\begin{align}
\begin{split}
    \hat{a}_{\mu}=\sum_{\eta}\left[\mathbf{S}^{-1/2}\right]_{\mu\eta}\tilde{\alpha}_\eta,\\
    \hat{a}_{\mu}^{\dagger}=\sum_{\eta}\left[\mathbf{S}^{-1/2}\right]_{\eta\mu}\tilde{\alpha}_\eta^{\dagger},\\
\end{split}
\end{align}
where $\mathbf{S}$ is the above defined QNM overlap matrix related to the dissipation and the operators satisfy $[\hat{a}_{\mu},\hat{a}_{\eta}^{\dagger}]=\delta_{\mu\eta}$, as also discussed in the introduction.

Subsequently, 
employing stochastic Ito/Stratonovich calculus, the above constructed QNM operator basis can be used to derive a Lindblad
QNM master equation for density operator~\cite{franke_quantization_2019, franke_quantized_2020},
\begin{equation}\label{eq: QNMmaster}
    \partial_t {\rho} = -\frac{i}{\hbar}[{H},{\rho}]+\mathcal{L}[\mathbf{a}]{\rho},
\end{equation}
where $H$ is the Hamiltonian in the TLS-QNM basis:
\begin{equation}
{H}={H}_{\rm em} + {H}_{0} + {H}_{\rm I},
\end{equation}
with 
\begin{subequations}
\begin{align}
{H}_{\rm em}&=\hbar {\sum}_{\mu\eta} {\chi}_{\mu\eta}^{(+)} \hat{a}_{\mu}^{\dagger} \hat{a}_{\eta}\label{eq: H_QNM},\\
{H}_{0}&=\hbar\omega_{0} \hat{\sigma}^{+} \hat{\sigma}^{-},\\
{H}_{\rm I}&=-i\hbar\sum_{\mu} \tilde{g}_{\mu}^{\rm s} \hat{\sigma}^{+} \hat{a}_{\mu}
+{\rm H.a.},
\end{align}
\end{subequations}
where ${H}_{\rm em}$ describes the coupling between different QNMs due to symmetrizing transformation, ${H}_{0}$ is free term of the quantum emitter with Pauli operators $\hat \sigma^+$ and $\hat \sigma^-$ and resonance frequency $\omega_0$, and ${H}_{\rm I}$ is the emitter-QNMs coupling in a rotating wave approximation. 

The coupling between QNMs is given as
\begin{equation}\label{eq: chi_plus}
{\chi}_{\mu\eta}^{(+)}=\frac{1}{2}(\chi_{\mu\eta}+\chi_{\eta\mu}^{*}),
\end{equation}
with
${\chi}_{\mu\eta}=\sum_{\nu}(\mathbf{S}^{-1/2})_{\mu\nu}\tilde{\omega}_{\nu}(\mathbf{S}^{1/2})_{\nu\mu}$.
The emitter-QNM coupling coefficient is $\tilde{g}_{\mu}^{\rm s}=\sum_{\eta}(\mathbf{S}^{1/2})_{\eta\mu}\tilde{g}_{\eta}$ with $\tilde{g}_{\mu}=\sqrt{\omega_{\mu}/(2\epsilon_{0}\hbar)}\mathbf{d}\cdot\tilde{\mathbf{f}}_{\mu}(\mathbf{r}_{0})$, where $\mathbf{d}$ is the dipole moment of the emitter and $\tilde{\mathbf{f}}_{\mu}(\mathbf{r}_{0})$ is the QNM at the emitter position $\mathbf{r}_{0}$. 
Furthermore, the Lindblad dissipator is
\begin{equation}\label{eq: LindbladDiss_QNM}
    \mathcal{L}[\mathbf{a}]\rho=\sum_{\mu,\eta}\chi_{\mu\eta}^{(-)}(2\hat{a}_{\eta}{\rho}\hat{a}_{\mu}^\dagger - {\rho}\hat{a}_{\mu}^\dagger\hat{a}_{\eta}-\hat{a}_{\mu}^\dagger\hat{a}_{\eta}{\rho}),
\end{equation}
with 
\begin{equation}\label{eq: chi_minus}
\chi_{\mu\eta}^{(-)}=\frac{i}{2}(\chi_{\mu\eta}-\chi^{*}_{\eta\mu}).
\end{equation}

In the bad cavity limit (namely, after adiabatically eliminating the electromagnetic degrees of freedom), the master equation can also include a Lindblad dissipator for the total SE rate, i.e., $\mathcal{L}[\mathbf{\sigma^{-}}]\rho=\Gamma^{\rm qQNM}(2\hat{\sigma}^{-}{\rho}\hat{\sigma}^{+} - {\rho}\hat{\sigma}^{+}\hat{\sigma}^{-}-\hat{\sigma}^{+}\hat{\sigma}^{-}{\rho})$.
Using the  $S$ parameters (dominated by nonradiative contributions), the quantum SE rate (weak coupling) is given as~\cite{franke_quantization_2019}
\begin{align}
\begin{split}\label{eq: 110}
   \Gamma^{\rm qQNM}=\sum_{\mu,\eta}\tilde{g}_{\mu}S^{\rm nrad}_{\mu\eta}\tilde{g}_{\eta}^{\ast}K_{\mu\eta},
\end{split}
\end{align}
with 
\begin{equation}
K_{\mu\eta}=\frac{\big[i(\omega_{\mu}-\omega_{\eta})+\gamma_{\mu}+\gamma_{\eta}\big]}{\big[(\Delta_{\mu}-i\gamma_{\mu})(\Delta_{\eta}+i\gamma_{\eta})\big]},
\end{equation}
where $\Delta_{\mu(\eta)}=\omega_{\mu(\eta)}-\omega_{0}$ is the detuning between the QNMs (real part of the complex eigenfrequency) and the emitter.
Thus, the {\it quantum Purcell factor} is   
\begin{equation}\label{eq: quantumpurcell}
F_{\rm P}^{\rm qQNM}=\frac{\Gamma^{\rm qQNM}}{\Gamma_{0}}+1,
\end{equation}
where again the additional $1$ is also added as in classical QNMs formulas, counting the background contribution when the dipole is located outside the resonators~\cite{ge_design_2014}.

Furthermore, we can also separate the quantum QNM decay rate into diagonal contribution and non-diagonal contributions~\cite{franke_quantization_2019}:
\begin{equation}
     \Gamma^{\rm qQNM}=\Gamma^{\rm qQNM}_{\rm diag}+\Gamma^{\rm qQNM}_{\rm ndiag}, \label{GammaQM}
 \end{equation}
where
\begin{subequations}
\begin{align}
    \Gamma_{\rm diag}^{\rm qQNM}=\sum_{\mu}S^{\rm nrad}_{\mu\mu}\frac{2\big|\tilde{g}_{\mu}\big|^{2}\gamma_{\mu}}{\Delta_{\mu\mu}^{2}+\gamma^{2}_{\mu}},\label{eq: qGamma_diag}\\
     \Gamma_{\rm ndiag}^{\rm qQNM}=\sum_{{\mu,\eta\neq\mu}}\tilde{g}_{\mu}S^{\rm nrad}_{\mu\eta}\tilde{g}_{\eta}^{\ast}K_{\mu\eta}\label{eq: qGamma_ndiag}.
\end{align}
\end{subequations}

Next, to help better understand the underlying physics and assess if the absence of non-diagonal coupling is equivalent to the absence of the 
QNM phases in classical NM theory (this will be discussed in more detail in Sec.~\ref{sec: phase_nondiag}),  we assume, for quantum NM theory, simply 
$S_{\mu\eta}=\delta_{\mu\eta}$, then the quantum result without interference is given as
\begin{align}
\begin{split}\label{eq: qNM}
   \Gamma^{\rm qNM}=\sum_{\mu}\frac{2\big|\tilde{g}_{\mu}\big|^{2}\gamma_{\mu}}{\Delta_{\mu\mu}^{2}+\gamma^{2}_{\mu}},
\end{split}
\end{align}
which is equivalent to the well known results from the dissipative JC model in the bad cavity limit, if using classical NMs parameters.
The corresponding quantum Purcell factor (using NM theory) is now
\begin{equation}\label{eq: qNMpurcell}
F_{\rm P}^{\rm qNM}=\frac{\Gamma^{\rm qNM}}{\Gamma_{0}}+1,
\end{equation}
whose solution will be 
 compared with Eq.~\eqref{eq: quantumpurcell} and Eq.~\eqref{eq: FP_cNM}.

\subsection{Comparison between classical and quantum QNM interference terms}\label{sec: phase_nondiag}
By comparing the classical and quantum Purcell factor expressions, one can  see that the deviation from a NM result can originate from different QNM terms, namely from the QNM phase (in the classical picture) or from non-diagonal  decay contributions (in the quantum picture). 
These two different representations are deeply connected through the fundamental Green function quantity: 
$\mathbf{M}^{\rm }(\mathbf{r}_{0},\mathbf{r}_{0})\equiv\int_{\mathbb{R}^3}{\rm d}{\mathbf{r}}\epsilon_I(\mathbf{r})\mathbf{G}(\mathbf{r}_{0},\mathbf{r})\cdot \mathbf{G}^*(\mathbf{r},\mathbf{r}_{0})$
($\mathbb{R}^3$ is the whole 3D real space).
which is proportional to $\langle {0}| \mathbf{\hat{E}}(\mathbf{r}_0)\mathbf{\hat{E}}^\dagger(\mathbf{r}_0) |{0}\rangle$ (cf. Eq.~\eqref{eq: SourceEField2}), which represents vacuum fluctuations;  thus, this term is related to the photon bath correlation function that appears in the quantum equations for an emitter within the Green function quantization method, in the weak light-matter coupling limit (cf. Ref.~\onlinecite{PhysRevB.92.205420}).

As shown in Ref.~\onlinecite{franke_fluctuation-dissipation_2020}, the integral expression $\mathbf{M}^{\rm }(\mathbf{r}_{0},\mathbf{r}_{0})$ can be related to two different expressions:
\begin{align}
\mathbf{M}^{\rm }(\mathbf{r}_{0},\mathbf{r}_{0})& = {\rm Im}[\mathbf{G}(\mathbf{r}_{0},\mathbf{r}_{0})],
\label{eq: GRel1}
\end{align}
and
\begin{align}
\mathbf{M}(\mathbf{r}_{0},\mathbf{r}_{0}) &= \int_V{\rm d}{\mathbf{r}}\epsilon_I(\mathbf{r})\mathbf{G}(\mathbf{r}_{0},\mathbf{r})\cdot \mathbf{G}^*(\mathbf{r},\mathbf{r}_{0})
\nonumber \\
&-\frac{c^2}{2i\omega^2}\mathbf{I}^{\rm sur}.\label{eq: GRel2}
\end{align}
The first expression, Eq.~\eqref{eq: GRel1}, is readily associated with the LDOS at the emitter position $\mathbf{r}_0$.
In the second expression,
Eq.~\eqref{eq: GRel2},
the volume $V$ contains $\mathbf{r}_{0}$ as well as the scattering structures, in which energy is absorbed, and 
\begin{align}
    \mathbf{I}^{\rm sur}& =\oint_{\mathcal{S}_V}{\mathrm d}\mathbf{s}\left\{\mathbf{G}(\mathbf{r}_{0},\mathbf{s}) \cdot \left[\mathbf{n}_\mathbf{s}\times\left(\boldsymbol{\nabla}_{\mathbf{s}}\times\mathbf{G}^*(\mathbf{s}, \mathbf{r}_{0})\right)\right]
    \nonumber \right . \\
    & \left . - {\rm H.a.}\right\},
\end{align}
is a surface 
contribution, that is related to the radiative power flow through the surface $\mathcal{S}_V$ of $V$.


We note that for the microdisk resonator structure studies below, the radiative part related to $\mathbf{I}^{\rm sur}$ can be neglected, since the nonradiative part, related to the volume integral, is dominant.


Although in a semiclassical treatment of light and matter, both relations can be equally used to obtain an expression for the quantum SE rates of a dipole-like emitter, and thus the corresponding Purcell factor, the first relation is preferable from the calculation viewpoint, as it only involves the Green function at the emitter position $\mathbf{r}_0$ without any spatial integration. In contrast, for the full quantum treatment of light and matter within the quantized QNM model, a relation similar to the second identity, Eq.~\eqref{eq: GRel2}, must be used to construct meaningful Fock states. As a consequence, the derivation of the quantum SE rate involves spatial integrals, that are similar to the RHS (right hand side) of Eq.~\eqref{eq: GRel2}.

To elaborate further, we redefine the RHS of the first relation (Eq.~\eqref{eq: GRel1}), used in semi-classical theories, as $\mathbf{M}^{\rm cQNM}(\mathbf{r}_{0},\mathbf{r}_{0})$, and the RHS of the second relation (Eq.~\eqref{eq: GRel2}), related to the quantized QNM model, as $\mathbf{M}^{\rm qQNM}(\mathbf{r}_{0},\mathbf{r}_{0})$. Next, the full QNM expanded Green function Eq.~\eqref{eq: GFwithSUM} 
is inserted on the RHS of Eq.~\eqref{eq: GRel2} to obtain the dipole-projected QNM part
 \begin{align}
 \mathbf{d}\cdot\mathbf{M}^{\rm qQNM}&(\mathbf{r}_{0},\mathbf{r}_{0})\cdot\mathbf{d}^*\nonumber\\
 =&\sum_{\mu\eta}\int_V{\rm d}{\mathbf{r}}\epsilon_I(\mathbf{r})\mathbf{d}\cdot\mathbf{G}_\mu(\mathbf{r}_{0},\mathbf{r})\cdot \mathbf{G}^*_\eta(\mathbf{r},\mathbf{r}_{0})\cdot\mathbf{d}^*\nonumber\\
=&\sum_{\mu\eta}\frac{\pi\sqrt{\omega_\mu\omega_\eta}}{2}\mathbf{d}\cdot\tilde{\mathbf{f}}_\mu(\mathbf{r}_{0}) S_{\mu\eta}^{\rm nrad}(\omega)\tilde{\mathbf{f}}_\eta^*(\mathbf{r}_{0})\cdot\mathbf{d}^*,\label{eq: NDiagSEQNM}
 \end{align}
 where $S_{\mu\eta}^{\rm nrad}(\omega)$ is implicitly defined via the symmetrization matrix $S_{\mu\eta}^{\rm nrad}=\int_0^\infty{\rm d}\omega S_{\mu\eta}^{\rm nrad}(\omega)$ from Eq.~\eqref{eq: Snrad_pole}. Note again, that for the inspected resonator system, the surface term associated to the radiative decay was neglected, since the dominant part is the non-radiative decay.
 
Furthermore, inserting the QNM expanded Green function Eq.~\eqref{eq: GFwithSUM} into the RHS of Eq.~\eqref{eq: GRel1},
and defining $\tilde{\mathbf{f}}_{\mu}(\mathbf{r})=|\tilde{\mathbf{f}}_{\mu}(\mathbf{r})|e^{i\phi_{\mu}(\mathbf{r})}$, we obtain:
\begin{align}
\mathbf{d}\cdot\mathbf{M}^{\rm cQNM}&(\mathbf{r}_{0},\mathbf{r}_{0})\cdot\mathbf{d}^*\nonumber\\
=&\sum_\mu|\mathbf{d}\cdot\tilde{\mathbf{f}}_\mu(\mathbf{r}_{0})|^2L^{\mu}(\omega)\nonumber\\
&\times\left[\cos(2\phi_\mu(\mathbf{r}_{0}))+\frac{\omega_\mu-\omega}{\gamma_\mu}\sin(2\phi_\mu(\mathbf{r}_{0}))\right],\label{eq: DiagSEQNM}
 \end{align}
with the modified Lorentzian lineshapes, 
\begin{align}
L^{\mu}(\omega)&=\frac{\omega}{2}\frac{\gamma_{\mu}}{(\omega_{\mu}-\omega)^2+\gamma_{\mu}^{2}}.
\end{align}
This form is basically a Lorentzian for high $Q$ resonators.

Equation~\eqref{eq: DiagSEQNM} represents a diagonal form of the tensor $\mathbf{M}^{\rm{cQNM}}(\mathbf{r}_{0},\mathbf{r}_{0})$ 
, where mode-coupling effects, such as the Fano interference effect, are induced by the sum of different QNM phase terms. Here, the individual parts of the sum can be negative, although the total sum is positive~\cite{2017PRA_hybrid}, as we also clearly see from the above results.

In contrast, Eq.~\eqref{eq: NDiagSEQNM} represents a non-diagonal form of $\mathbf{M}^{\rm{ qQNM}}(\mathbf{r}_{0},\mathbf{r}_{0})$ 
. Here, the mode interference effects are induced by the off-diagonal elements $\mu\neq\eta$ through the mode overlap integrals in $S_{\mu\eta}(\omega)$. 

\section{Numerical Results for coupled lossy microdisk resonators}
\label{sec: numerical}

\begin{figure}[t]
    \centering
    \includegraphics[width=0.89\columnwidth]{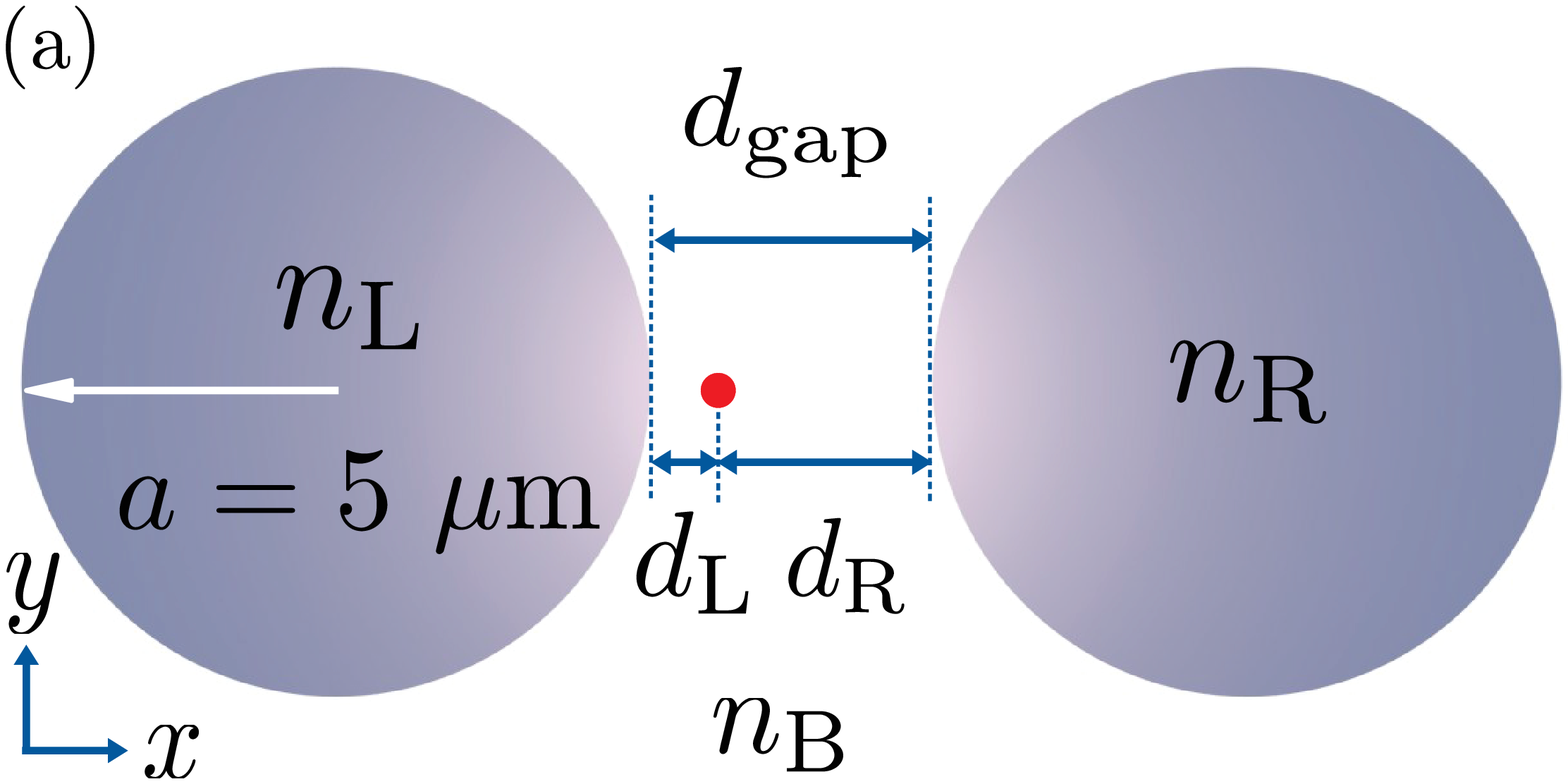}
    \includegraphics[width=0.98\columnwidth]{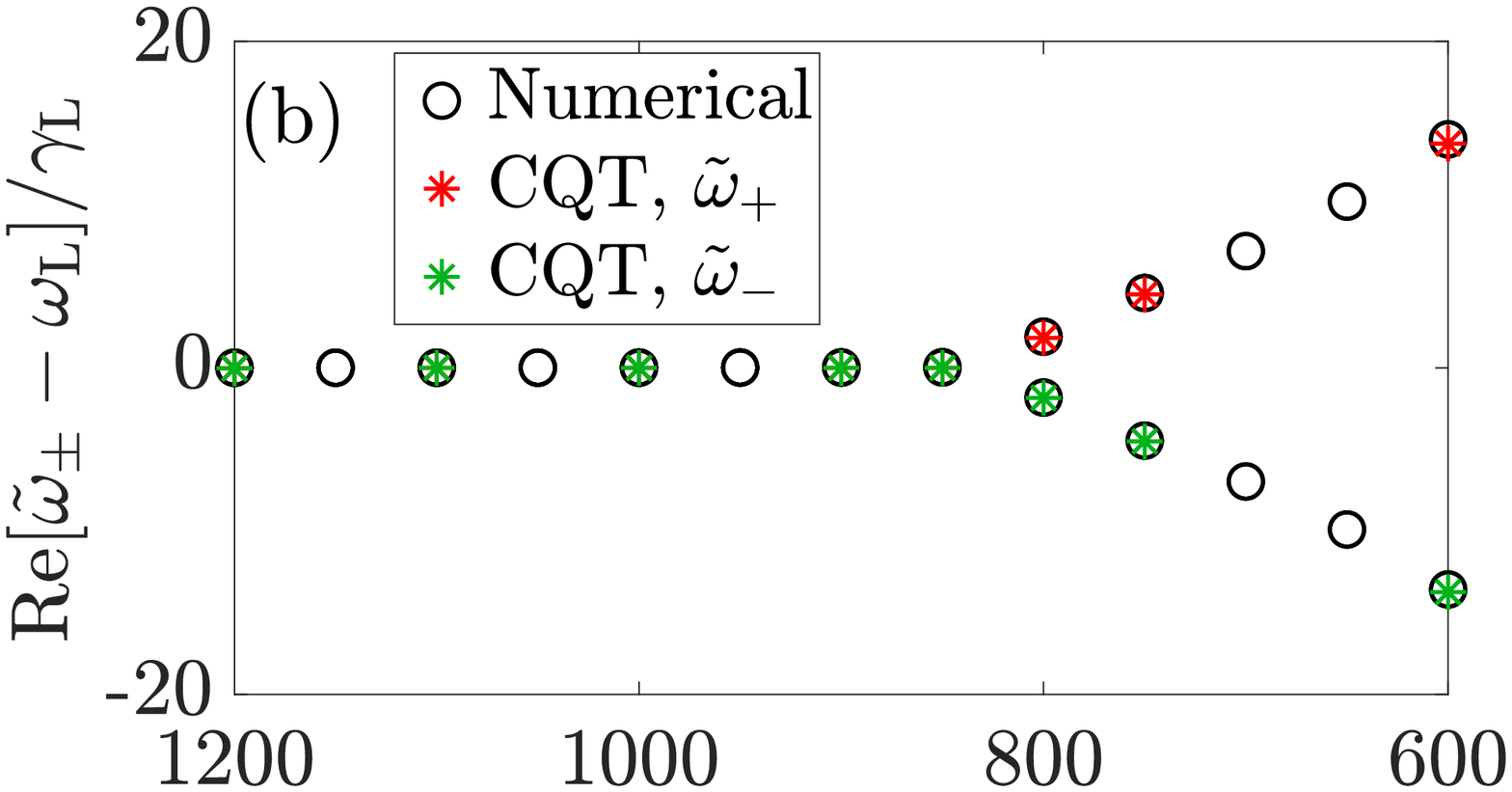}
    \includegraphics[width=0.99\columnwidth]{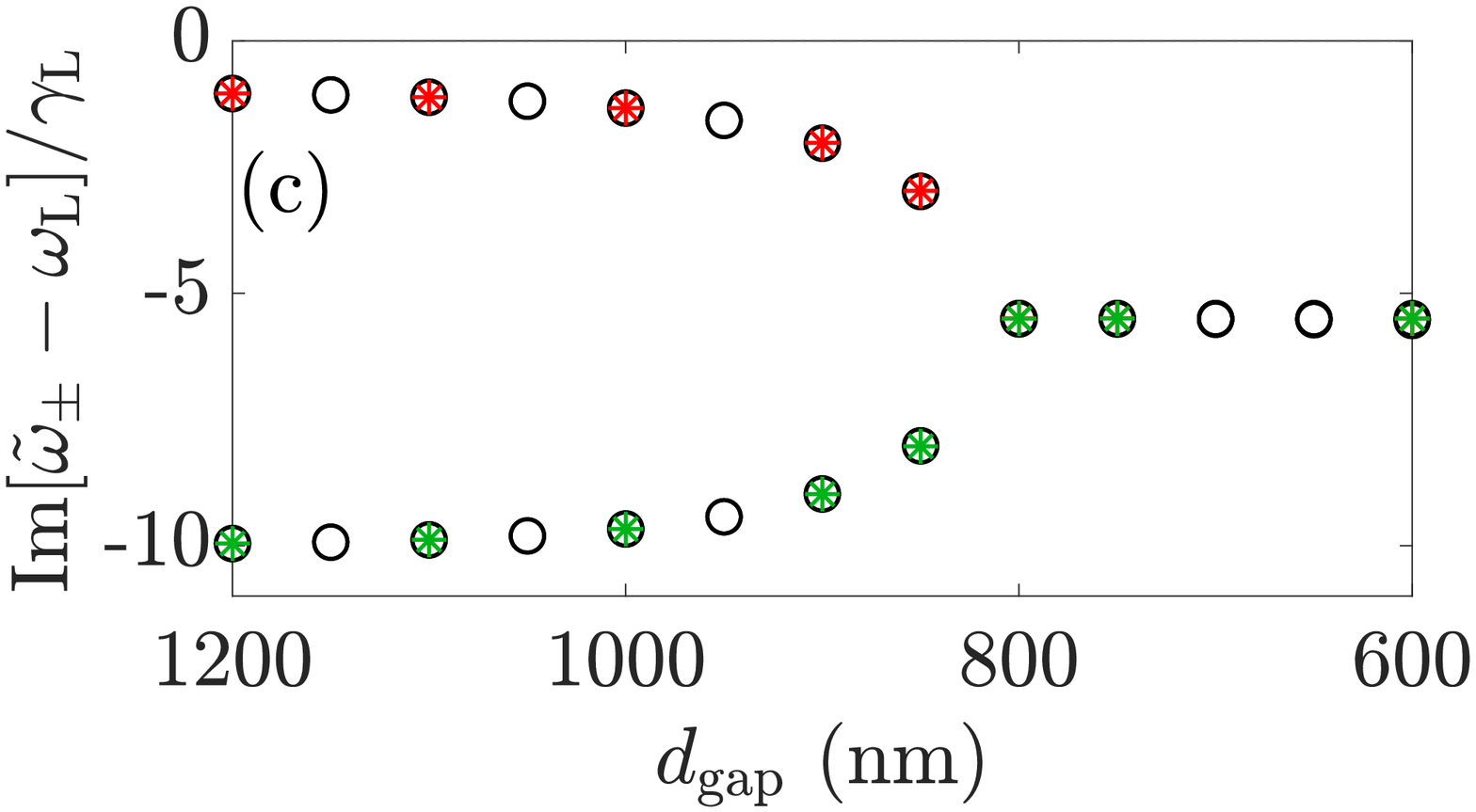}
    \centering
    \caption{
    (a) Schematic diagram of two coupled lossy microdisk resonators. The radius of both resonators is $a=5\,\mu$m. 
    The refractive indices of the lossy resonators are $n_{\rm L}=2+10^{-5}i$ and $n_{\rm R}=2+10^{-4}i$, in a background medium with  $n_{\rm B}=1$ (free space). The gap distance between the resonators is $d_{\rm gap}$, and the dipole (emitter) is placed within the gap. The distance between the dipole and the surface of resonator $\rm L$ (R) is $d_{\rm L}$ ($d_{\rm R}$), so $d_{\rm gap}=d_{\rm L}+d_{\rm R}$. The origin of the coordinate system is at the gap center. 
    (b)-(c) Complex QNM eigenfrequencies of the coupled microdisk resonators from a direct numerical eigenfrequency solver in COMSOL (numerical, black circles) and the analytical CQT 
    (red stars for $\tilde{\omega}_{+}$, green  stars for $\tilde{\omega}_{-}$), as a function of gap separation $d_{\rm gap}$. 
    Note that $\omega_{\rm L}$ and $\gamma_{\rm L}$ are for uncoupled single lossy resonator L with $n_{\rm L}=2+10^{-5}i$ ($\tilde{\omega}_{\rm L}=\omega_{\rm L}-i\gamma_{\rm L}$; while for the uncoupled single loss cavity R, with $n_{\rm R}=2+10^{-4}i$, the resonance is around at $\tilde{\omega}_{\rm R} \approx \omega_{\rm L}-i10\gamma_{\rm L}$.  
    }\label{fig: eigen}
\end{figure}

For our numerical calculations, we consider 
two lossy microdisk resonators (both with a radius of $a=5~\mu$m) that are embedded in a homogeneous medium (free space with $n_{\rm B}=1$; see Fig.~\ref{fig: eigen} (a)). 
For the lossy materials, we
 use $n_{\rm L}=2+10^{-5}i$ ($n_{\rm R}=2+10^{-4}i$) to represent the refractive index of the lossy resonator L (R) (the left (right) resonator is labeled as L (R)). 
The whispering-gallery mode (WGMs) we choose for the bare resonators have relatively high quality factors (they are around $10^{5}$ and $10^{4}$ for L and R as shown later).
When coupled together, the gap distance $d_{\rm gap}$ between the resonators  ranged from $600\sim1200$~nm, which also spans an approximate lossy EP (see Fig.~\ref{fig: eigen} (b-c)). We consider a dipole that is  located in the gap (red dot in Fig.~\ref{fig: eigen} (a)), which is $d_{\rm L}$ ($d_{\rm R}$) away from the resonator L (R) (so $d_{\rm gap}=d_{\rm L}+d_{\rm R}$).

\subsection{Quasinormal modes for a single lossy microdisk resonator}\label{sec: singleQNM}

\begin{figure*}[t]
    \centering
    \includegraphics[width=2.05\columnwidth]{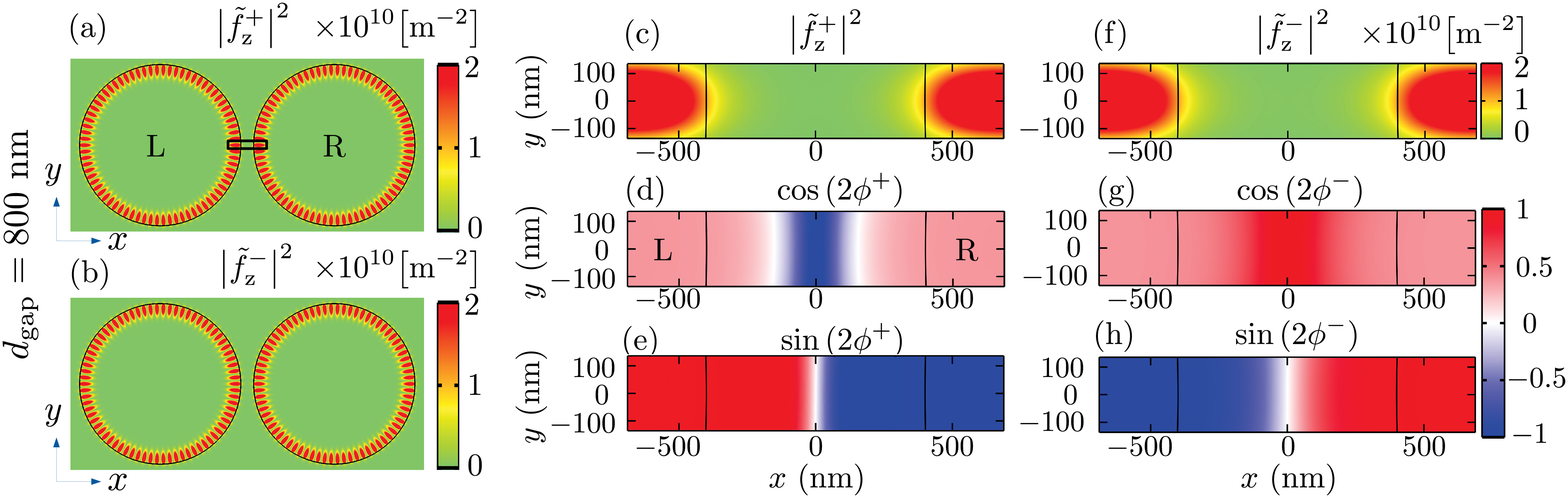}
    \includegraphics[width=2.05\columnwidth]{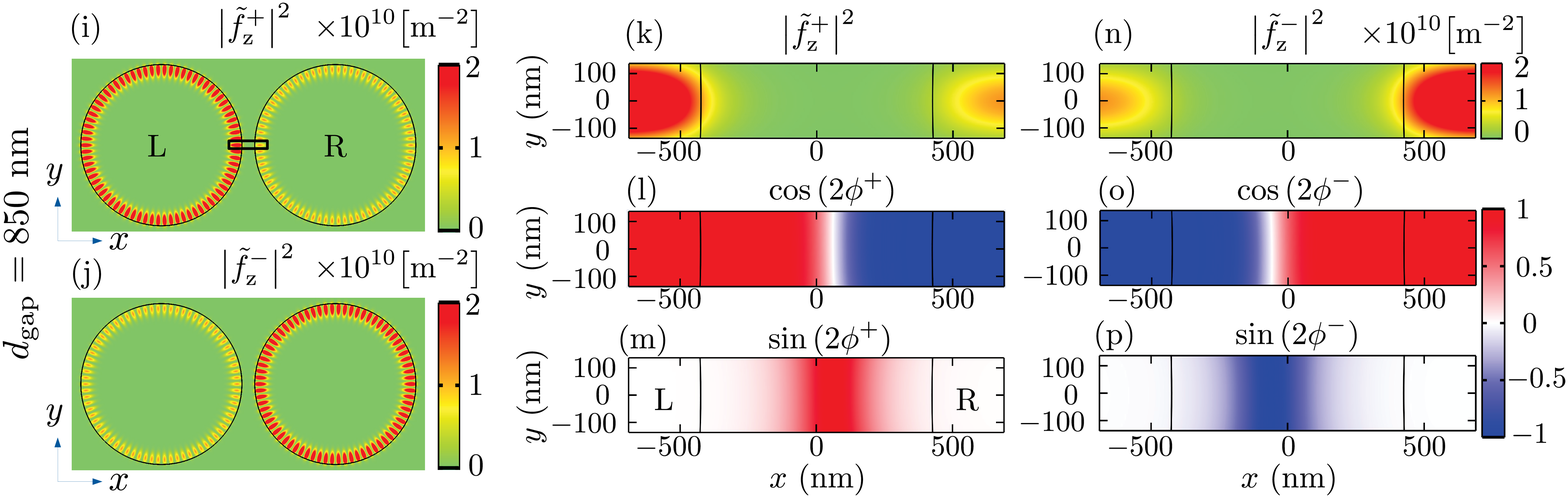}
    \caption{
    The QNMs and phase distribution for (a-h) $d_{\rm gap}=800~$nm and (i-p) $d_{\rm gap}=850~$nm (the phases are defined as $\tilde{f}^{+}_{z}(\mathbf{r})=|\tilde{f}^{+}_{z}(\mathbf{r})|e^{i\phi^{+}(\mathbf{r})}$ and $\tilde{f}^{-}_{z}(\mathbf{r})=|\tilde{f}^{-}_{z}(\mathbf{r})|e^{i\phi^{-}(\mathbf{r})}$).
    Especially, for clarity, (c-h) and (k-p) are showing the zoom-in of the small black rectangle (slightly enlarged for clarity) labeled in (a) and (f), near the regions where couple the dipole emitter. 
}\label{fig: phase_d800_d850}
\end{figure*}

As input for the CQT, the bare QNM of the single lossy microdisk are first computed, where the supporting WGMs are generally  described by two mode numbers (for the 2D cases studies here): radial mode number $q$ (the number of nodes for mode intensity along the radial direction within the resonator) and azimuthal mode number $m$ (number of wavelengths along the equator in azimuthal direction) after the polarization is determined---transverse-magnetic (TM) or transverse-electronic (TE).
It is well known that there are two degenerate WGMs for fixed mode number $q,~m$ and polarization, which propagate along opposite directions~\cite{leung_completeness_1994,mazzei_controlled_2007,teraoka_resonance_2009,cognee_cooperative_2019}:
${E}_{\rm ccw}(\mathbf{r},\phi)=E(\mathbf{r})e^{(im \phi)}$ with counter clockwise (ccw) direction and ${E}_{\rm cw}(\mathbf{r},\phi)=E(\mathbf{r})e^{(-im \phi)}$ with clockwise (cw) direction.
A linear superposition of CW and CCW modes will lead to two degenerate standing modes~\cite{mazzei_controlled_2007,teraoka_resonance_2009,cognee_cooperative_2019}.

In general, the mode with radial mode number $q=1$ and azimuthal mode number $m\gg1$ is with a strong field confinement and a high $Q$ factor~\cite{oraevsky_whispering-gallery_2002,schunk_identifying_2014}.
So in this work, we choose a TM mode (${H}_{x},{H}_{y},{E}_{z}$) with $q=1$, and $m=37$, whose resonance is in the telecommunication band.
Referring to the efficient dipole scattering approach in Ref.~\onlinecite{bai_efficient_2013-1}, we obtained the normalized QNMs for single lossy resonators. 
In the simulation, we placed an out-of-plane line current $10$~nm away from the 2D resonator (i.e., $d_{\rm L}=10~$nm for the single resonator L and $d_{\rm R}=10~$nm for the single resonator R). 
Note, similar to the work in  Ref.~\onlinecite{ren_quasinormal_2021}, only one of the two degenerate standing modes is excited using the TM dipole ($z$-polarized) located along the $x$-axis.
Moreover, the QNMs with other mode numbers are far away from the working  frequency regime, so we can safely make a single mode approximation for the single resonators.


For the single lossy resonator L, the QNM  eigenfrequency is 
$\tilde{\omega}_{\rm L}=\omega_{\rm L}-i\gamma_{\rm L}
=1.266666\times10^{15} - 6.260226\times10^{9}i$ rad/s,
with a  quality factor $Q_{\rm L}\sim10^{5}$.
For the single lossy resonator R, the eigenfrequency is
$ \tilde{\omega}_{\rm R}=\omega_{\rm R}-i\gamma_{\rm R}\approx\omega_{\rm L}-i10\gamma_{\rm L}$,  with a quality factor $Q\sim10^{4}$.

\subsection{Classical hybrid quasinormal modes for coupled lossy resonators}\label{sec: hybridQNMs}

\begin{figure*}[htb]
    \centering
    \includegraphics[width=0.92\columnwidth]{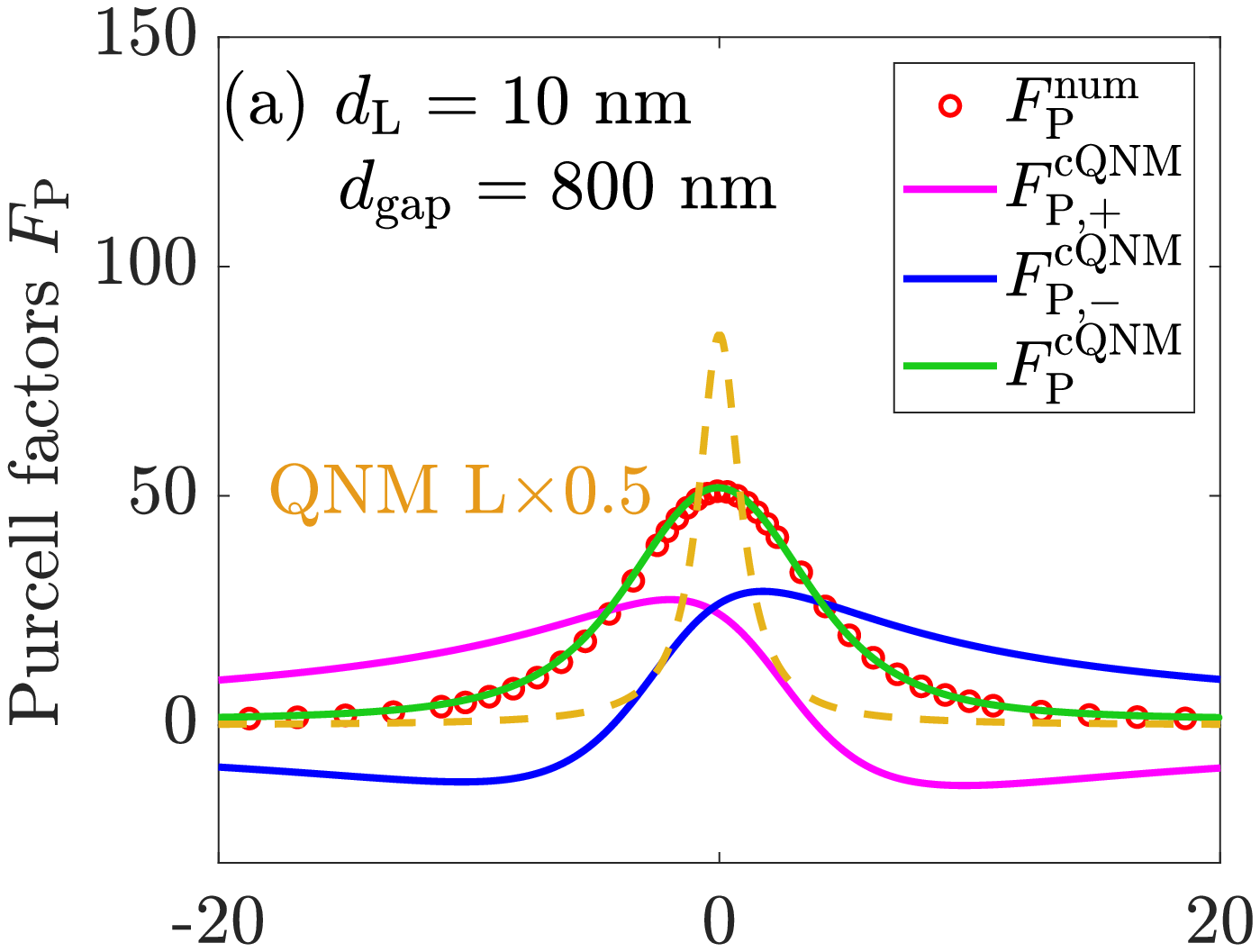}
    \includegraphics[width=0.92\columnwidth]{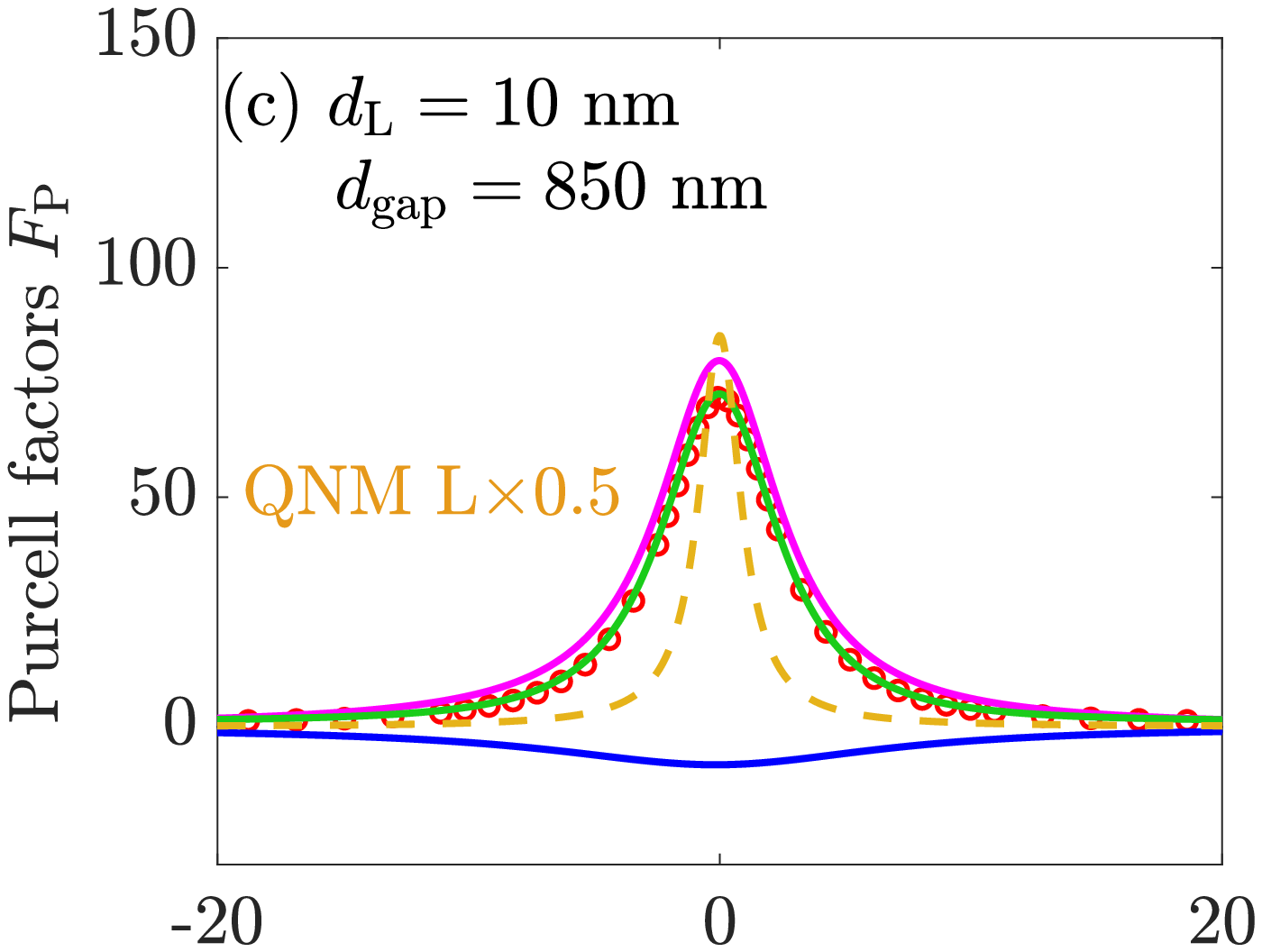}
   \includegraphics[width=0.92\columnwidth]{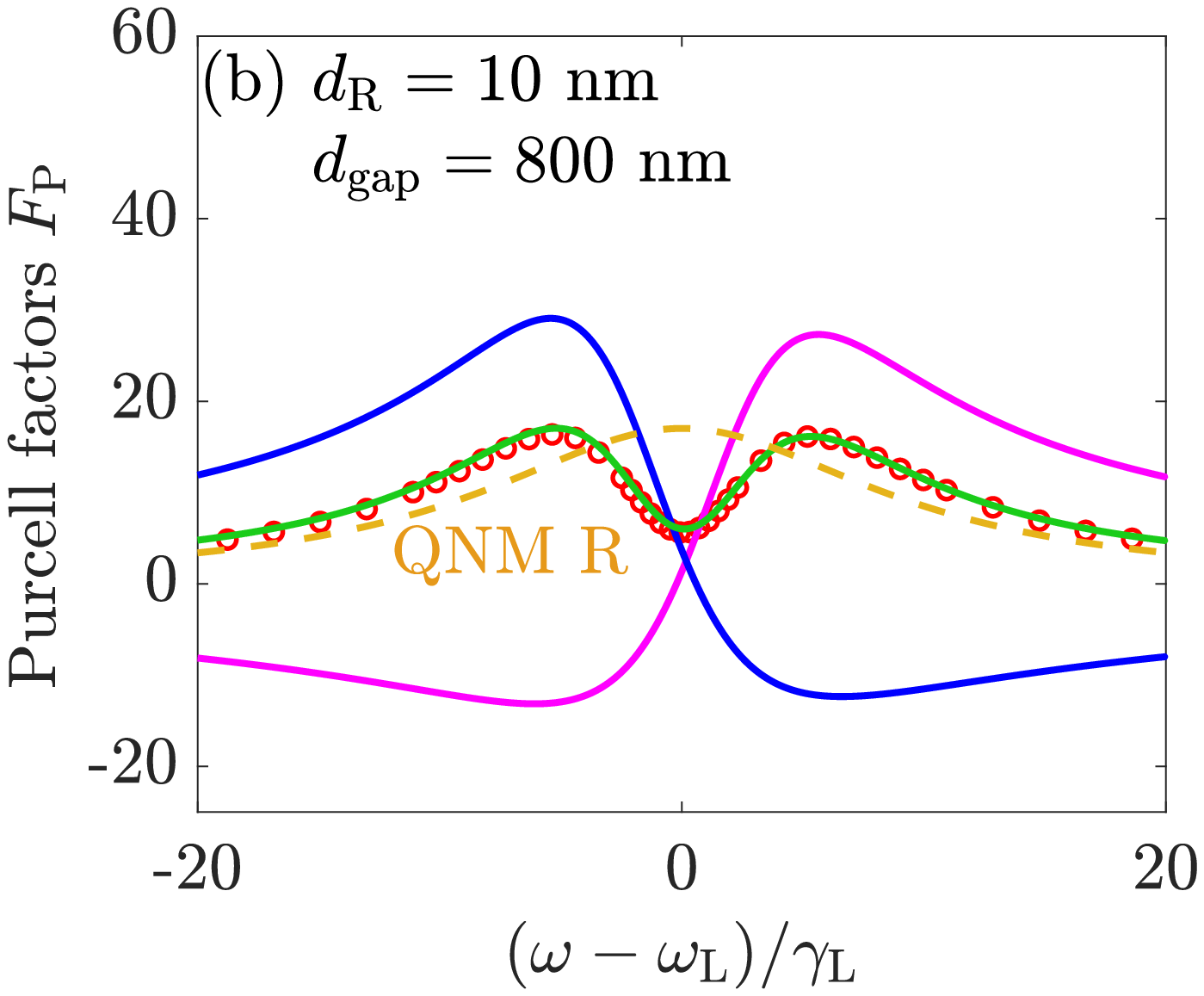}
   \includegraphics[width=0.92\columnwidth]{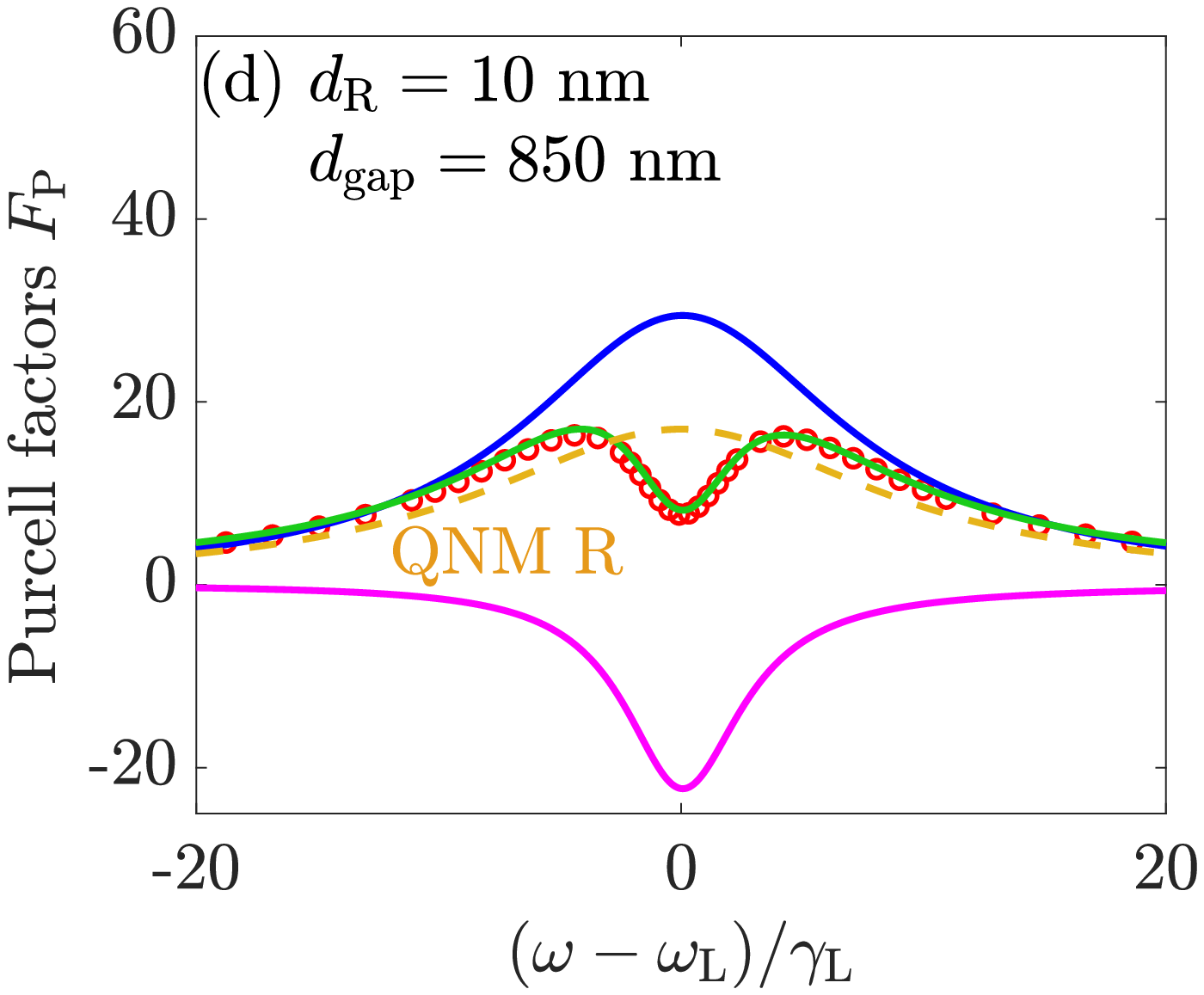}
    \caption{
    Calculated Purcell factors from the classical QNMs (total values $F_{\rm P}^{\rm cQNM}$ labelled as the solid green curves from the Eq.~\eqref{eq: FP_cQNM}; separate contribution $F_{{\rm P},\pm}^{\rm cQNM}$ labelled as the solid magenta and blue curves with Eq.~\eqref{eq: FP_cQNM_pm}) for
    two different gaps, $d_{\rm gap}=800$ nm and $850$ nm, at two different dipole positions $d_{\rm L/R}=10~$nm. 
    The dipole is $z$-polarized in all cases (same for all Purcell factor figures below). The total Purcell factors $F_{\rm P}^{\rm cQNM}$  agree extremely well with full numerical dipole solution $F_{\rm P}^{\rm num}$ (Eq.~\eqref{eq: Purcellfulldipole}, red markers).
    For comparison, the Purcell factors with single bare resonators are also shown with the dashed orange curves, where QNM L/R means the case with the only resonator L/R. 
}\label{fig: Fp_d800d850_cQNM}
\end{figure*}

In this section, we present results for the coupled QNMs that are obtained analytically by taking advantage of the CQT in Ref.~\onlinecite{ren_quasinormal_2021}, where the QNMs from single lossy cavities (shown above) are used as input.
Figures~\ref{fig: eigen}(b-c) show the eigenfrequencies $\tilde{\omega}_{\pm}$ (Eq.~\eqref{eq: rootsRWA}, red and green stars) of the analytical coupled QNMs, which agree 
extremely well with the numerical solution in COMSOL (approximate eigenfrequency solver, black circles). 
Note the same method also works well for coupled loss-gain resonators~\cite{ren_quasinormal_2021}, which is also confirmed here with loss-loss coupling, as expected. 


Next we will concentrate on two gap cases ($d_{\rm gap}=800~{\rm nm}$ and $d_{\rm gap}=850~{\rm nm}$), which are locating close to (at the both sides of) the EP (Fig.~\ref{fig: eigen}).  
First we focus on $d_{\rm gap}=800~{\rm nm}$, the corresponding spatial distribution of the coupled QNMs (Eq.~\eqref{eq: QNMs_pm}, absolute value square $|\tilde{f}^{\pm}_{z}|^2$) are shown in Figs.~\ref{fig: phase_d800_d850} (a-b), where two modes have similar intensities over both resonators.
One can also see in Figs.~\ref{fig: phase_d800_d850} (c) and (f), a   zoom-in of the intensity of the small black rectangle region schematically shown in Fig.~\ref{fig: phase_d800_d850} (a).  

We highlight that the QNMs phases always play an important role for the coupled modes.
To help show this analytically, 
 we can define $\tilde{f}^{+}_{z}(\mathbf{r})=|\tilde{f}^{+}_{z}(\mathbf{r})|e^{i\phi^{+}(\mathbf{r})}$ and $\tilde{f}^{-}_{z}(\mathbf{r})=|\tilde{f}^{-}_{z}(\mathbf{r})|e^{i\phi^{-}(\mathbf{r})}$.
The Green function is then written as
\begin{align}
&{G}_{zz}(\mathbf{r}_{0},\mathbf{r}_{0},\omega) \nonumber 
\\
&\ =A^{+}(\omega)\tilde{f}^{+}_{z}(\mathbf{r}_{0})\tilde{f}^{+}_{z}(\mathbf{r}_{0})+A^{-}(\omega)\tilde{f}^{-}_{z}(\mathbf{r}_{0})\tilde{f}^{-}_{z}(\mathbf{r}_{0})\nonumber \\
&\ =A^{+}(\omega)e^{i2\phi^{+}(\mathbf{r}_{0})}|\tilde{f}^{+}_{z}(\mathbf{r}_{0})|^2
+A^{-}(\omega)e^{i2\phi^{-}(\mathbf{r}_{0})}|\tilde{f}^{-}_{z}(\mathbf{r}_{0})|^2,
\end{align}
with $A^{+}(\omega)=\omega/[2(\tilde{\omega}_{+}-\omega)]$, and $A^{-}(\omega)=\omega/[2(\tilde{\omega}_{-}-\omega]$.  
One can see that the phases $2\phi^{+}$ and $2\phi^{-}$ contribute directly in the QNMs Green function.
Thus we also show the distribution (zoom-in of the small black box region) of $\cos{(2\phi^{+}(\mathbf{r}))}$ and $\cos{(2\phi^{-}(\mathbf{r}))}$ for $d_{\rm gap}=800$ in Figs.~\ref{fig: phase_d800_d850} (d) and (g), where the two hybrid modes have nearly the same $\cos{(2\phi)}$ when close to the resonator L and R (the value in pink region is around $0.35$ to $0.40$). In contrast, for $\sin{(2\phi^{\pm}(\mathbf{r}))}$, they have nearly opposite distribution over two resonators (nearly $1$ for red region and nearly $-1$ for blue region) as shown in Figs.~\ref{fig: phase_d800_d850} (e) and (h).

Second, we look at the case of $d_{\rm gap}=850~{\rm nm}$, the spatial distribution of the coupled QNMs (Eq.~\eqref{eq: QNMs_pm}, through the absolute value square $|\tilde{f}^{\pm}_{z}|^2$) are shown in Figs.~\ref{fig: phase_d800_d850} (i-j) (also see Figs.~\ref{fig: phase_d800_d850} (k) and (n) for zoom-in), where the two modes clearly have different intensity over both resonators.
The corresponding phases for the zoom-in region are also shown in Figs.~\ref{fig: phase_d800_d850} (l-m) and (o-p). 
In contrast to the case with $d_{\rm gap}=800~$nm, here $\cos{(2\phi^{+}(\mathbf{r}))}$ is nearly $1$ close to resonator L, while around resonator R, it is nearly $-1$. Also, now $\cos{(2\phi^{-}(\mathbf{r}))}$  has nearly the opposite distribution compared with $\cos{(2\phi^{+}(\mathbf{r}))}$.
Moreover, one will find $\sin{(2\phi^{\pm}(\mathbf{r}))}$ is nearly $0$ when close to the resonators L and R.

To elaborate further, one will find that these spatial distributions of the absolute value of the QNMs and phases are closely related with the non-Lorentizian lineshape of the Purcell factors as shown in Fig.~\ref{fig: Fp_d800d850_cQNM}(a-d),  where the Purcell factors from classical QNMs are presented (Eq.~\eqref{eq: FP_cNM}) for a dipole at $d_{\rm L}=10~$nm or $d_{\rm R}=10~$nm with gap distance $d_{\rm gap}=800/850~$nm. 
Specifically, the Purcell factors can be obtained through the imaginary part of the Green function ~\cite{el-sayed_quasinormal-mode_2020,ren_quasinormal_2021}:
\begin{align}\label{eq: purcellphase}
\begin{split}
&{\rm Im}[{G}_{zz}(\mathbf{r}_{0},\mathbf{r}_{0},\omega)]\\
=&\bigg[\cos{2\phi^{+}(\mathbf{r}_{0})}+\frac{\omega_{+}-\omega}{\gamma_{+}}\sin{2\phi^{+}(\mathbf{r}_{0})}\bigg]\Big|\tilde{f}^{+}_{z}(\mathbf{r}_{0})\Big|^2 L^{+}(\omega)\\
+&\bigg[\cos{2\phi^{-}(\mathbf{r}_{0})}+\frac{\omega_{-}-\omega}{\gamma_{-}}\sin{2\phi^{-}(\mathbf{r}_{0})}\bigg]\Big|\tilde{f}^{-}_{z}(\mathbf{r}_{0})\Big|^2 L^{-}(\omega),
\end{split}
\end{align}
with the modified Lorentzian lineshapes, 
\begin{align}
L^{+}(\omega)&=\frac{\omega}{2}\frac{\gamma_{+}}{(\omega_{+}-\omega)^2+\gamma_{+}^{2}}, \nonumber \\
L^{-}(\omega)&=\frac{\omega}{2}\frac{\gamma_{-}}{(\omega_{-}-\omega)^2+\gamma_{-}^{2}}.
\end{align}

Referring to Fig.~\ref{fig: Fp_d800d850_cQNM}(c-d) 
, one will find that the lineshape of the 
frequency-dependent Purcell factors $F_{{\rm P},\pm}^{\rm cQNM}$ (Eq.~\eqref{eq: FP_cQNM_pm}, labelled as solid magenta and blue curves) from the separate coupled modes nearly have a Lorentzian or a negative Lorentzian form, which is because their phases are $\cos{(2\phi(\mathbf{r}_{0}))}\approx 1$ or $\cos{(2\phi(\mathbf{r}_{0}))}\approx -1$ (see Figs.~\ref{fig: phase_d800_d850} (l) and (o)). As for the different linewidths and intensities, they depend on the prefactors $\big|\tilde{f}^{\pm}_{z}(\mathbf{r}_{0})\big|^2$ (shown in Fig.~\ref{fig: phase_d800_d850}(i-j), (k) and (n)) and  $ L^{\pm}(\omega)$ (which is related to the quality factor around the resonance). Finally, adding up the contributions of the two coupled QNMs, one gets the classical non-Lorentzian total Purcell factors $F_{\rm P}^{\rm cQNM}$ (Eq.~\eqref{eq: FP_cQNM}, solid green curves), which agree extremely well with full numerical dipole results ($F_{\rm P}^{\rm num}$, Eq.~\eqref{eq: Purcellfulldipole}, red markers).

In addition, one could also use the same approach to analyze the lineshapes shown in Figs.~\ref{fig: Fp_d800d850_cQNM} (a-b).
While they have a similar $\cos{(2\phi(\mathbf{r}_{0}))}$ distribution (around $0.35$ to $0.40$, see Figs.~\ref{fig: phase_d800_d850} (d) and (g))), they have nearly opposite $\sin{(2\phi(\mathbf{r}_{0}))}$ values ($0.92$ to $0.94$ or $-0.94$ to $-0.92$, see Figs.~\ref{fig: phase_d800_d850} (e) and (h)), which makes the second term in Eq.~\eqref{eq: purcellphase} also contribute to the response. This results in a highly non-Loretzian lineshape for the separate mode contributions $F_{{\rm P},\pm}^{\rm cQNM}$ (Eq.~\eqref{eq: FP_cQNM_pm}, solid magenta and blue curves) shown in Fig.~\ref{fig: Fp_d800d850_cQNM} (a-b), after considering two coupled modes that have similar intensity (Figs.~\ref{fig: phase_d800_d850} (a-b), (c) and (f)) and similar quality factors (Fig.~\ref{fig: eigen} (b-c)). Adding up the contributions from two coupled modes, one gets the total Purcell factors $F_{\rm P}^{\rm cQNM}$ (Eq.~\eqref{eq: FP_cQNM}, solid green curves), which again show excellent agreement with the full 
 numerical dipole results.

\subsection{Classical and quantum Purcell factors from a normal mode and quasinormal perspective}\label{sec: NMsvsQNMs}

\begin{figure*}[t]
    \centering
    \includegraphics[width=0.95\columnwidth]{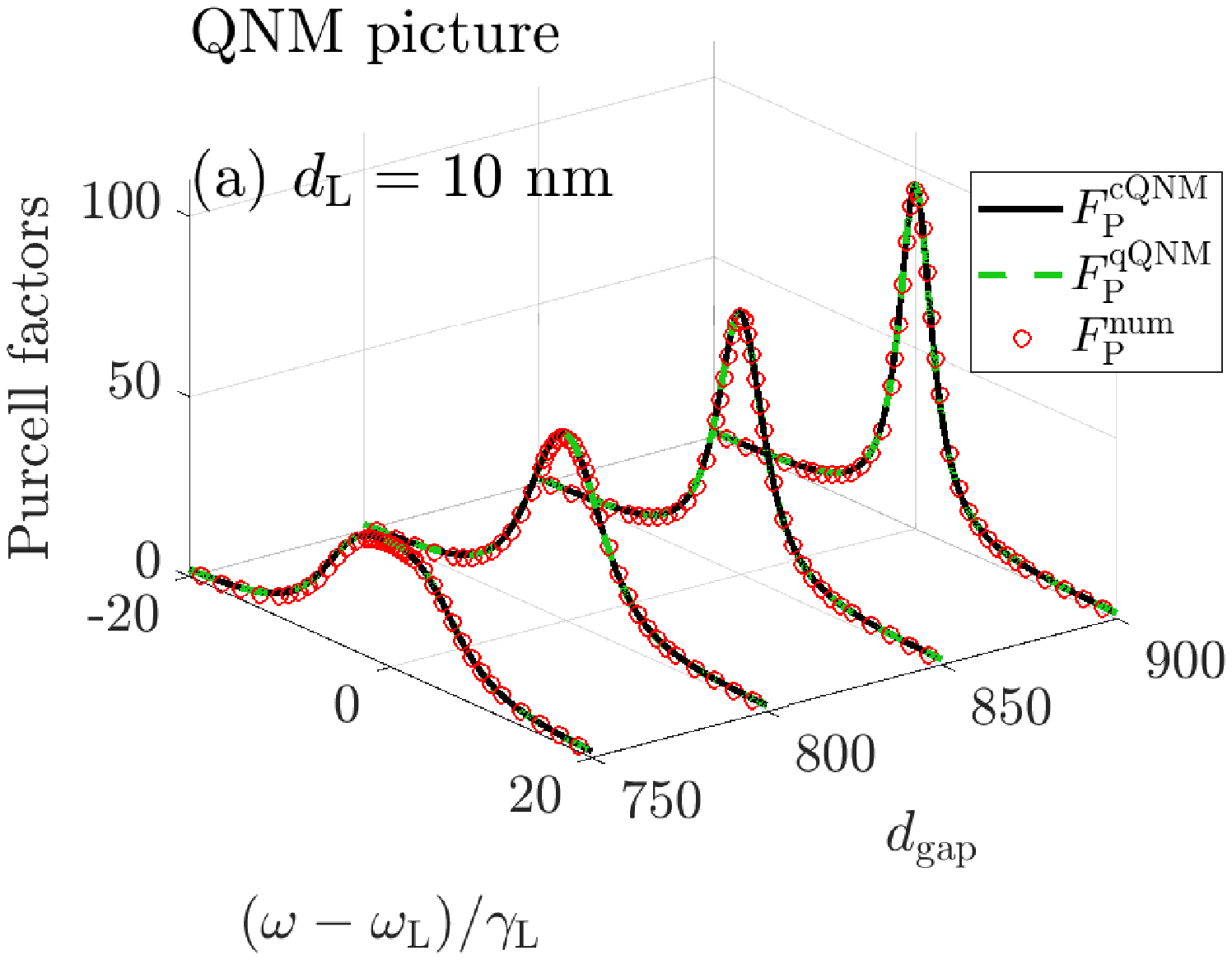}
    \includegraphics[width=0.95\columnwidth]{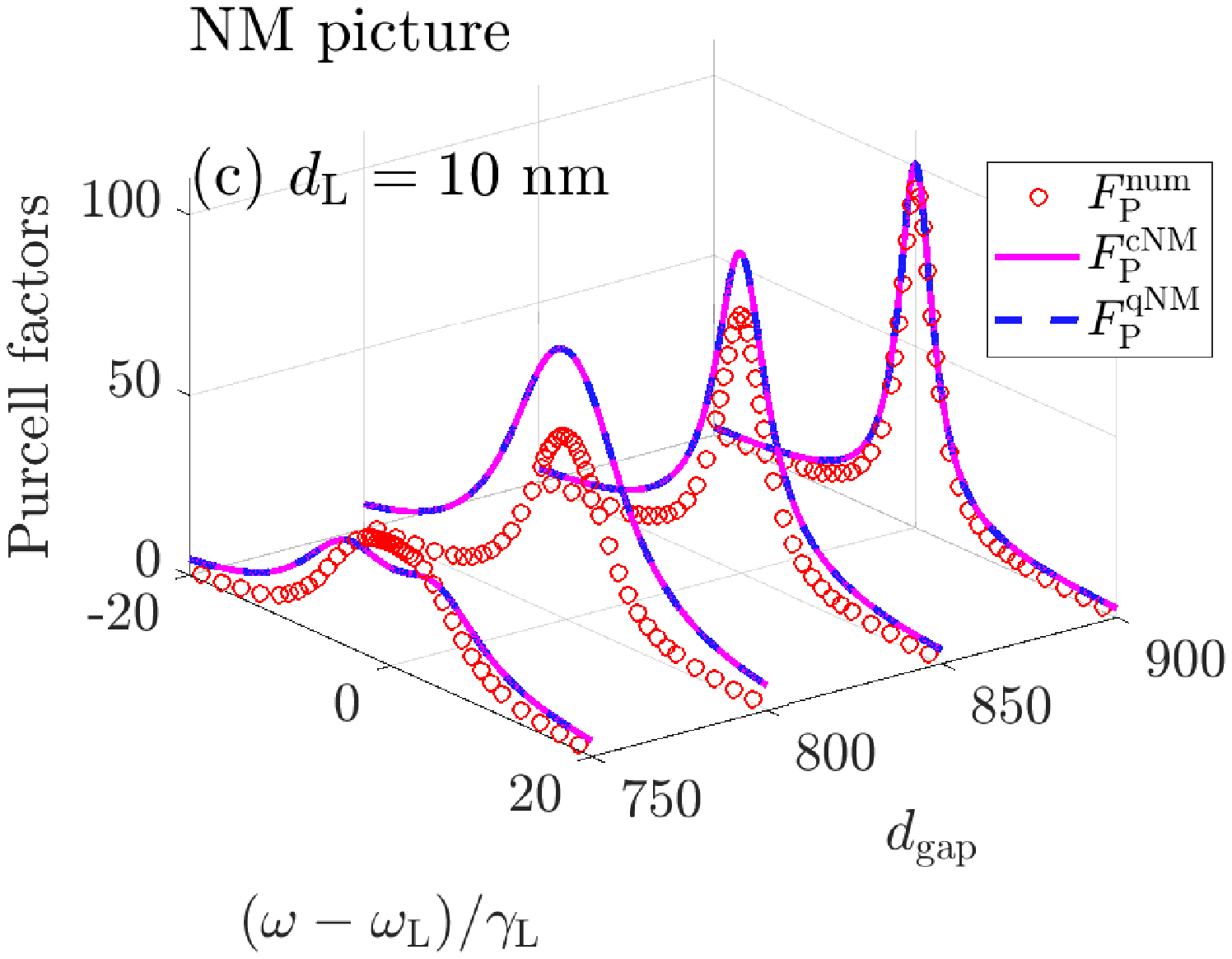}
    \includegraphics[width=0.95\columnwidth]{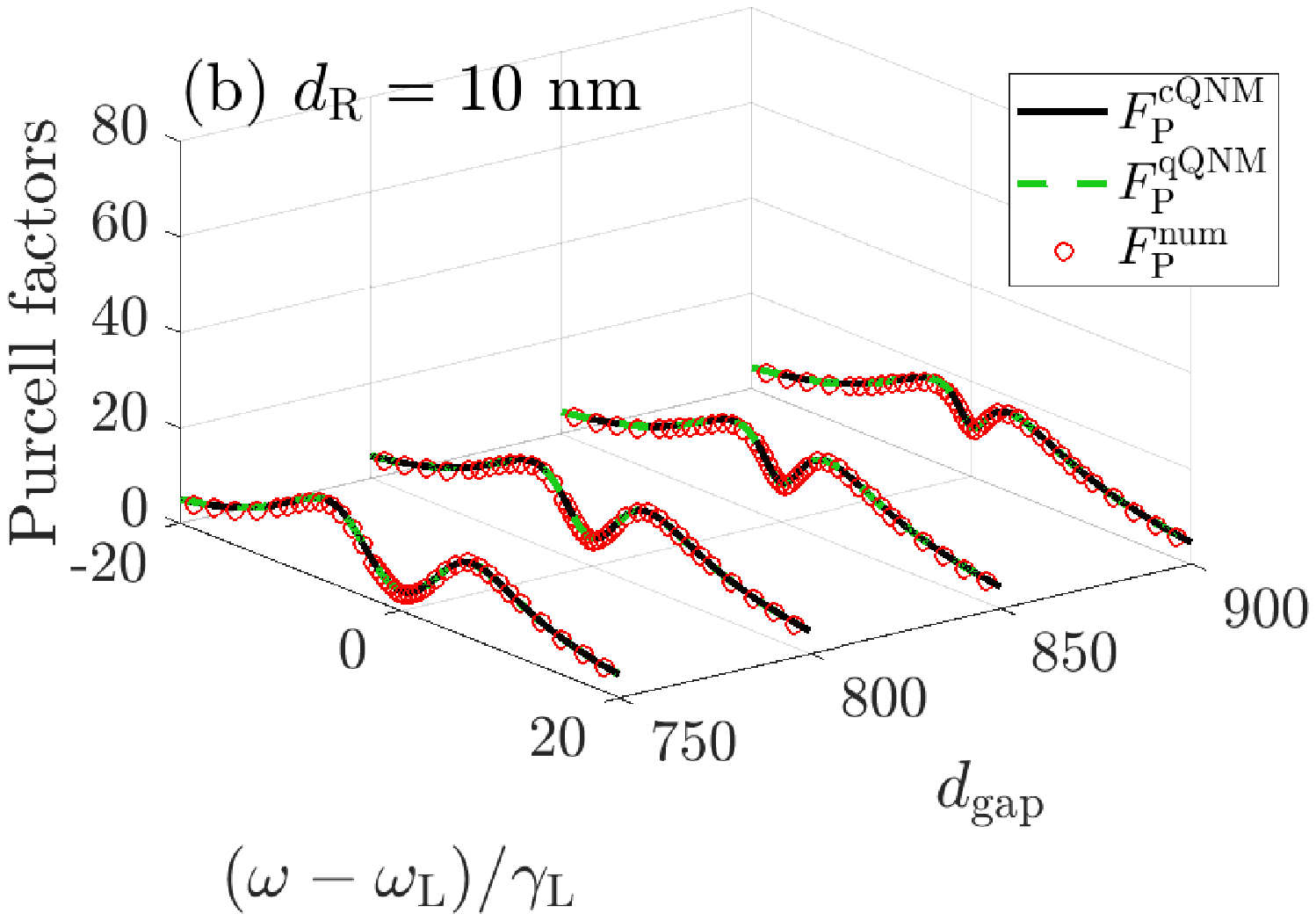}
    \includegraphics[width=0.95\columnwidth]{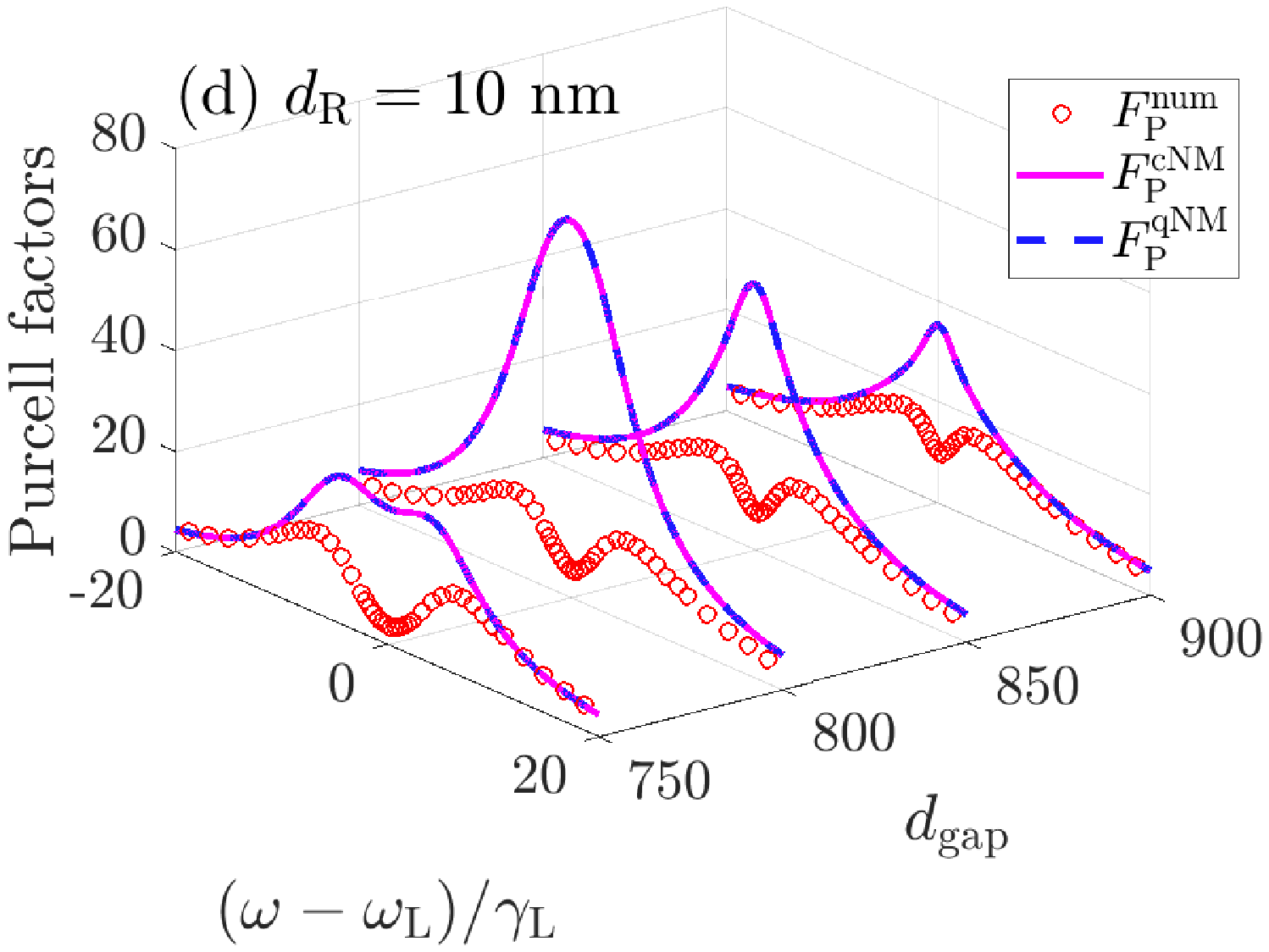}
    \caption{ (a-b) Calculated Purcell factors in a QNM picture (a,b), including results from classical QNMs ($F^{\rm cQNM}_{\rm P}$, Eq.~\eqref{eq: FP_cQNM} solid black curves) and quantum QNMs ($F^{\rm qQNM}_{\rm P}$, Eq.~\eqref{eq: quantumpurcell}, dashed green curves). (c-d) Calculated and Purcell factors in a NM picture, including results from classical NMs ($F^{\rm cNM}_{\rm P}$, Eq.~\eqref{eq: FP_cNM}, solid magenta curves) and quantum results without interference terms ($F^{\rm qNM}_{\rm P}$, Eq.~\eqref{eq: qNMpurcell}, dashed blue curves, $S_{++}=S_{--}=1$ and 
    $S_{+-}=S_{-+}=0$) for (a,c) $d_{\rm L}=10~$nm and (b,d) $d_{\rm R}=10~$ nm. Also shown are the full numerical dipole results ($F^{\rm num}_{\rm P}$, red markers).
}\label{fig: Fp_water_ind}
\end{figure*}

We next compare the classical QNM results with the quantum QNM results as well as more phenomenological NM models.
As shown in Section~\ref{sec: hybridQNMs}, taking the absolute values and corresponding phases of the two coupled classical QNMs into account, one can get the total Purcell factors
accurately from the classical QNMs ($F_{\rm P}^{\rm cQNM}$, Eq.~\eqref{eq: FP_cQNM}, Fig.~\ref{fig: Fp_d800d850_cQNM}).
For a dipole emitter at $d_{\rm L}=10~$nm, the classical QNMs solution $F_{\rm P}^{\rm cQNM}$ increases gradually and has a smaller linewidth when the gap distance increases as shown in Fig.~\ref{fig: Fp_water_ind} (a) (black solid curves), which show excellent agreements (again) with full numerical dipole results (red circles) for each gap distance, indicating the accuracy of the classical QNMs for bare resonators and the validity of the CQT. We stress that a diagonal GF expansion from a regular coupled mode theory based on NMs
would not get this level of agreement, since the QNM phase is not correctly included. However, we note, that for high-$Q$ bare resonators, there is a possibility to improve the NM model results, as shown in Section~\ref{sec: ImproveNM}.

\begin{table}[htb]
\caption {Quantum $S_{ij}$ overlap integrals (Eq.~\eqref{eq: Snrad_pole}) for the coupled microdisk resonators.
} \label{table: Sparameters} 
    \centering
    \begin{tabular}{|c|c|c|}
 \hline
\multirow{4}{7em}{$d_{\rm gap}=900~$nm} 
 & \multicolumn{2}{|c|}{$S_{++}=1.2929$} \\
 & \multicolumn{2}{|c|}{$S_{+-}=0.0001 + 0.8200i$} \\
 & \multicolumn{2}{|c|}{$S_{-+}=0.0001 - 0.8200i$} \\
 & \multicolumn{2}{|c|}{$S_{--}=1.2935$} \\
 \hline
\multirow{4}{7em}{$d_{\rm gap}=850~$nm}  
 & \multicolumn{2}{|c|}{$S_{++}=1.7768$} \\
 & \multicolumn{2}{|c|}{$S_{+-}=-0.00003 - 1.4696i$} \\
 & \multicolumn{2}{|c|}{$S_{-+}=-0.00003 + 1.4696i$} \\
 & \multicolumn{2}{|c|}{$S_{--}=1.7782$} \\
 \hline
\multirow{4}{7em}{$d_{\rm gap}=800~$nm}  
 & \multicolumn{2}{|c|}{$S_{++}=2.6439$} \\
 & \multicolumn{2}{|c|}{$S_{+-}=0.0015 + 2.4483i$} \\
 & \multicolumn{2}{|c|}{$S_{-+}=0.0015 - 2.4483i$} \\
 & \multicolumn{2}{|c|}{$S_{--}=2.6453$} \\
 \hline
\multirow{4}{7em}{$d_{\rm gap}=750~$nm}  
 & \multicolumn{2}{|c|}{$S_{++}=1.4137$} \\
 & \multicolumn{2}{|c|}{$S_{+-}=0.0003+ 0.9999i$} \\
 & \multicolumn{2}{|c|}{$S_{-+}=0.0003 - 0.9999i$} \\
 & \multicolumn{2}{|c|}{$S_{--}=1.4145$} \\
 \hline
 \end{tabular}
\end{table}

As shown in Sec.~\ref{sec: quanQNMs}, one can obtain the Purcell factors from quantized QNMs (Eq.~\eqref{eq: quantumpurcell}),
where the quantum $S$ parameters are required,
and we assume a bad cavity limit. 
For our numerical example, the radiative part contribution is negligible, and thus it is valid to only consider $S^{\rm nrad}_{\mu\eta}$ ($\mu,\eta=\pm$) as shown in Eq.~\eqref{eq: Snrad_pole}.
The detailed values of $S_{\mu\eta}$($=S^{\rm nrad}_{\mu\eta}$) for coupled lossy microdisks with a gap distance $d_{\rm gap}=750/800/850/900~$nm are shown in table~\ref{table: Sparameters}. We first recognize, that for all inspected gap cases, $S_{\mu\eta}$ has significant off-diagonals elements and even similar absolute values compared to the diagonal elements, $|S_{+-}|\lesssim S_{++},S_{--}$. Secondly, when getting closer to the lossy EP region ($d_{\rm gap}\sim 800~{\rm nm}$), $S_{\mu\eta}$ differ the most from $\delta_{\mu\eta}$, which is in line with the significant non-Lorentian behaviour of the separate classical QNM contributions at that gap distance.

Then the {\it quantum} Purcell factors $F_{\rm P}^{\rm qQNM}$ (Eq.~\eqref{eq: quantumpurcell}) are shown with the green dashed curves in Fig.~\ref{fig: Fp_water_ind} (a), which show excellent agreement with the classical QNM solution and full numerical dipole results, indicating the validity of the quantized QNMs solutions, and their ability to produce highly non-Lorentzian features, even at the two-mode level due to the off-diagonal elements of $S_{\mu\eta}$. 
We next move the  dipole emitter to $d_{\rm R}=10~$nm (closer to the right-side resonator R), and the classical QNMs solution ($F_{\rm P}^{\rm cQNM}$, black solid curves) are shown in Fig.~\ref{fig: Fp_water_ind} (b), which are smaller than those for a dipole at $d_{\rm L}=10~$nm, and show two peaks in the lineshape.
As shown in Section~\ref{sec: hybridQNMs}, these non-Lorentzian lineshape could also be successfully explained with the
classical QNM phases and intensities.
Moreover, with the quantum $S$ parameters shown in table~\ref{table: Sparameters}, and the coupled QNMs at $d_{\rm R}=10~$nm, one can get the quantized QNMs solution $F_{\rm P}^{\rm qQNM}$ (dashed green curves in Fig.~\ref{fig: Fp_water_ind} (b)), which clearly show excellent agreement with the classical QNMs solution $F_{\rm P}^{\rm cQNM}$ and the full numerical dipole result (red circles), again indicating the high accuracy of the quantized QNMs solutions, at various spatial locations. We stress that there are no 
fitting paramaters used in any of these comparisons,
which highlights the high accuracy of both classical and quantum models. To be clear, there are absolutely no fitting parameters of any kind used in these graphs.

We next focus on the results with a more standard NM picture,
 where the phases are completely ignored (they cancel out in the NM expansion).
For a dipole at $d_{\rm L}=10~$nm, the Purcell factors with classical NMs ($F_{\rm P}^{\rm cNM}$, Eq.~\eqref{eq: FP_cNM}, the phases are completely ignored) are shown in Fig.~\ref{fig: Fp_water_ind}(c) (magenta solid curves), which show obvious differences from the full numerical dipole results (red circles). This indicates the clear significance of the QNMs phases when the ``modes'' are coupled.
In addition, as discussed in Sec.~\ref{sec: phase_nondiag}, the effect of the phases of the classical QNMs is equivalent to the contribution of non-diagonal terms in the quantized QNM theory. To make this clearer,  if we neglect the non-diagonal terms, i.e., setting $S_{++}=S_{--}=1$, $S_{+-}=S_{-+}=0$, the quantum Purcell factors without such interference ($F_{\rm P}^{\rm qNM}$, Eq.~\eqref{eq: qNMpurcell}) are shown in Fig.~\ref{fig: Fp_water_ind} (c) (blue dashed curves), which again differ from the full numerical dipole results (red circles), but coincide very well with the classical NMs solutions, indicating the apparent equivalence of the absence of the classical QNMs phases and non-diagonal terms in quantized QNMs. 
Most importantly, their difference from the full dipole results show that for such high-$Q$ resonators, the NM pictures clearly fail, in both classical and quantum pictures.  

For a dipole located at $d_{\rm R}=10~$nm, the NM solutions (magenta solid curves) and quantum results without interference (blue dashed curves) are shown in Fig.~\ref{fig: Fp_water_ind} (d), where they agree well with each other but again  differ significantly to the full dipole solutions (red circles). This again indicates the significance of the classical QNMs phases and the contributions of the non-diagonal terms in the quantized QNMs models.

\subsection{Contributions of the non-diagonal terms using quantized QNMs}\label{sec: nondiagonal_contri}

\begin{figure}[htb]
    \centering
    \includegraphics[width=0.95\columnwidth]{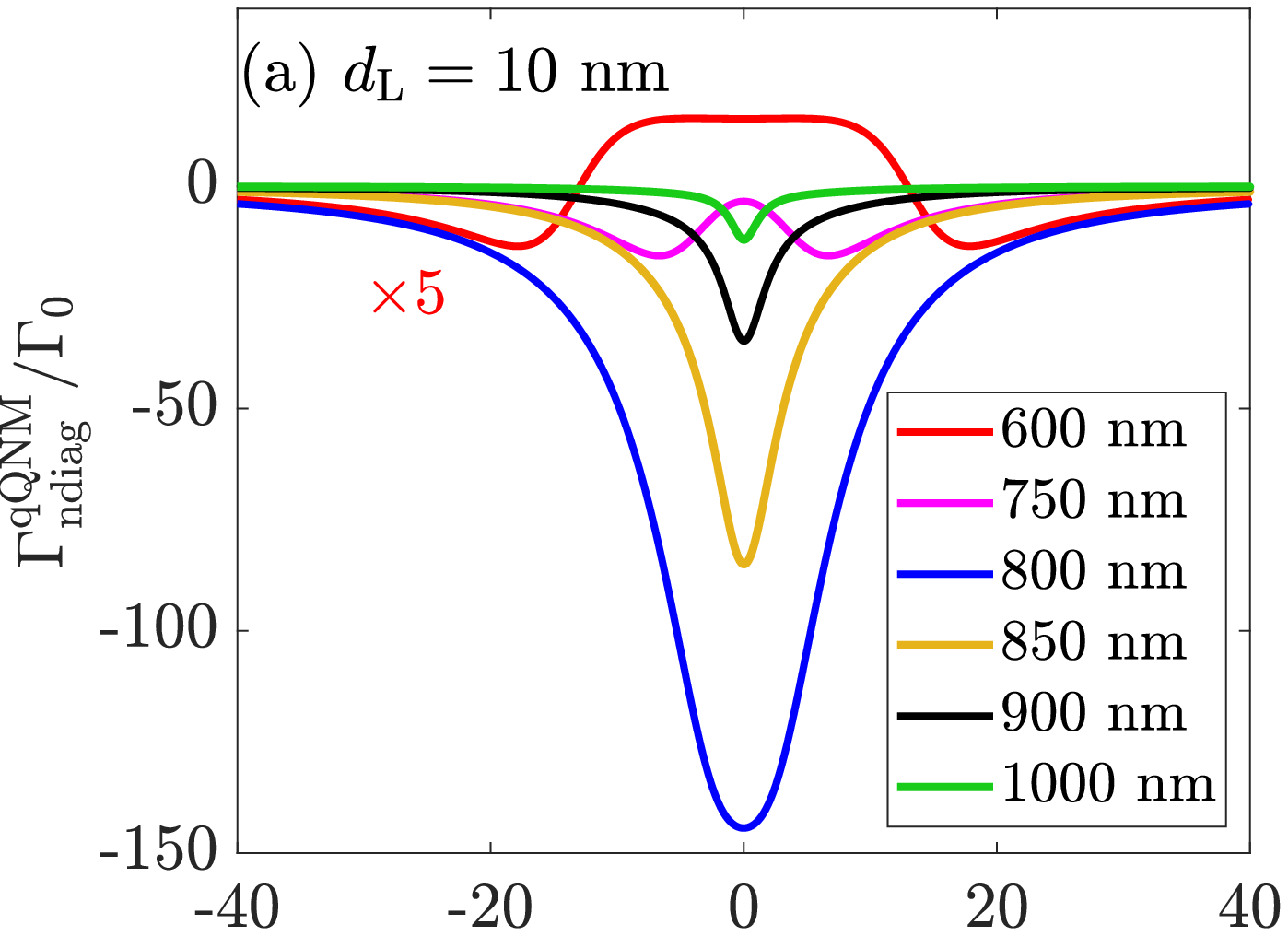}
    \includegraphics[width=0.95\columnwidth]{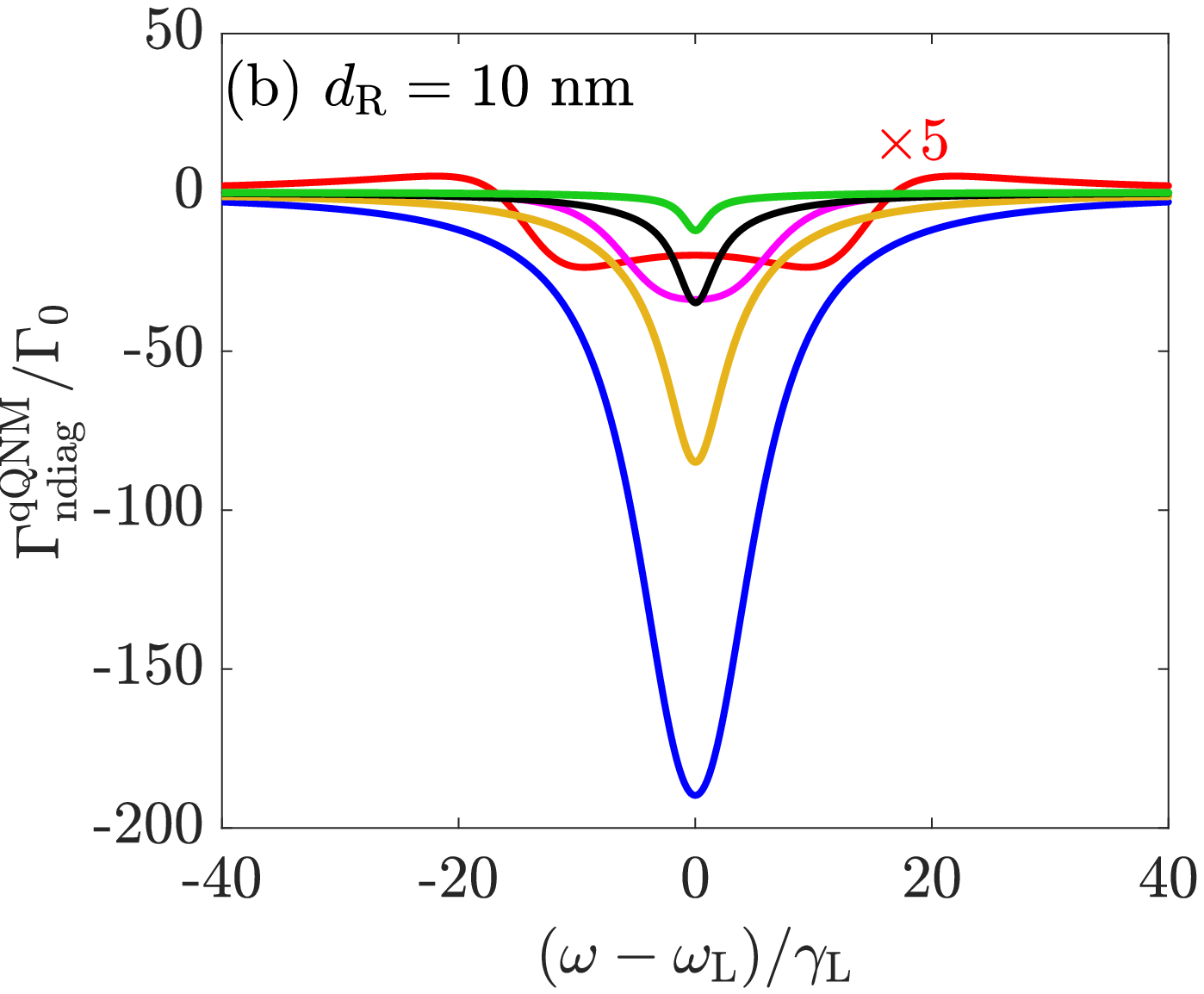}
    \caption{Contributios of the quantum Purcell factors from the non-diagonal terms (Eq.~\eqref{eq: qGamma_ndiag}, $\Gamma_{\rm ndiag}^{\rm qQNM}/\Gamma_{0}$) for (a) $d_{\rm L}=10~$nm and (b) $d_{\rm R}=10~$ nm, one can find that, with $d_{\rm gap}=800~$nm, the non-diagonal terms contribute the most. 
}\label{fig: Gamma}
\end{figure}

In the above section, we found that the contribution of the non-diagonal terms are critically important for certain gap separations. Here we investigate their detailed contribution as a function of gap distance.
For a dipole at $d_{\rm L}=10~$nm, the decay rate (Eq.~\eqref{eq: qGamma_ndiag}, $\Gamma_{0}$, is the decay rate in a background homogeneous medium) from the non-diagonal terms are shown in Fig.~\ref{fig: Gamma} (a), where except for the case with $d_{\rm gap}=600~$nm, all other contributions are negative. Taking them into account will lower the total Purcell factors as generally shown in Fig.~\ref{fig: Fp_water_ind}.
Also, one will find that the non-diagonal contribution is the most  significant for $d_{\rm gap}=800~$nm, which can be seen from Fig.~\ref{fig: Fp_water_ind} (a) and (c) (where the difference is the most).

However, note that $\Gamma^{\rm qQNM}_{\rm ndiag}/\Gamma_{0}$ (Eq.~\eqref{eq: qGamma_ndiag}) is the actual contribution from the non-diagonal terms, which is not exactly the difference between $F_{\rm P}^{\rm qQNM}$ (Eq.~\eqref{eq: quantumpurcell}) and $F_{\rm P}^{\rm qNM}$ (Eq.~\eqref{eq: qNMpurcell}) as shown in Fig.~\ref{fig: Fp_water_ind}. 
Since one can find the actual diagonal terms are $F_{\rm P}^{\rm qQNM}-\Gamma^{\rm qQNM}_{\rm ndiag}/\Gamma_{0}=\Gamma^{\rm qQNM}_{\rm diag}/\Gamma_{0}$ (Eq.~\eqref{eq: qGamma_diag}), which still use the actual quantum $S$ parameters ($S_{++}$ and $S_{--}$ shown in table~\ref{table: Sparameters}).
But for $F_{\rm P}^{\rm qNM}$ (Eq.~\eqref{eq: qNMpurcell}), one uses $S_{++}=S_{--}=1$.

For an emitter at position $d_{\rm R}=10~$nm, the results are shown in Fig.~\ref{fig: Gamma} (b), where again except for the case with $d_{\rm gap}=600~$nm, all other non-diagonal contributions are negative, and the contribution for $d_{\rm gap}=800~$nm is the most, which to some extent could be seen from Fig.~\ref{fig: Fp_water_ind}(b) and (d).

\begin{figure*}[t!]
    \centering
    \includegraphics[width=0.49\columnwidth]{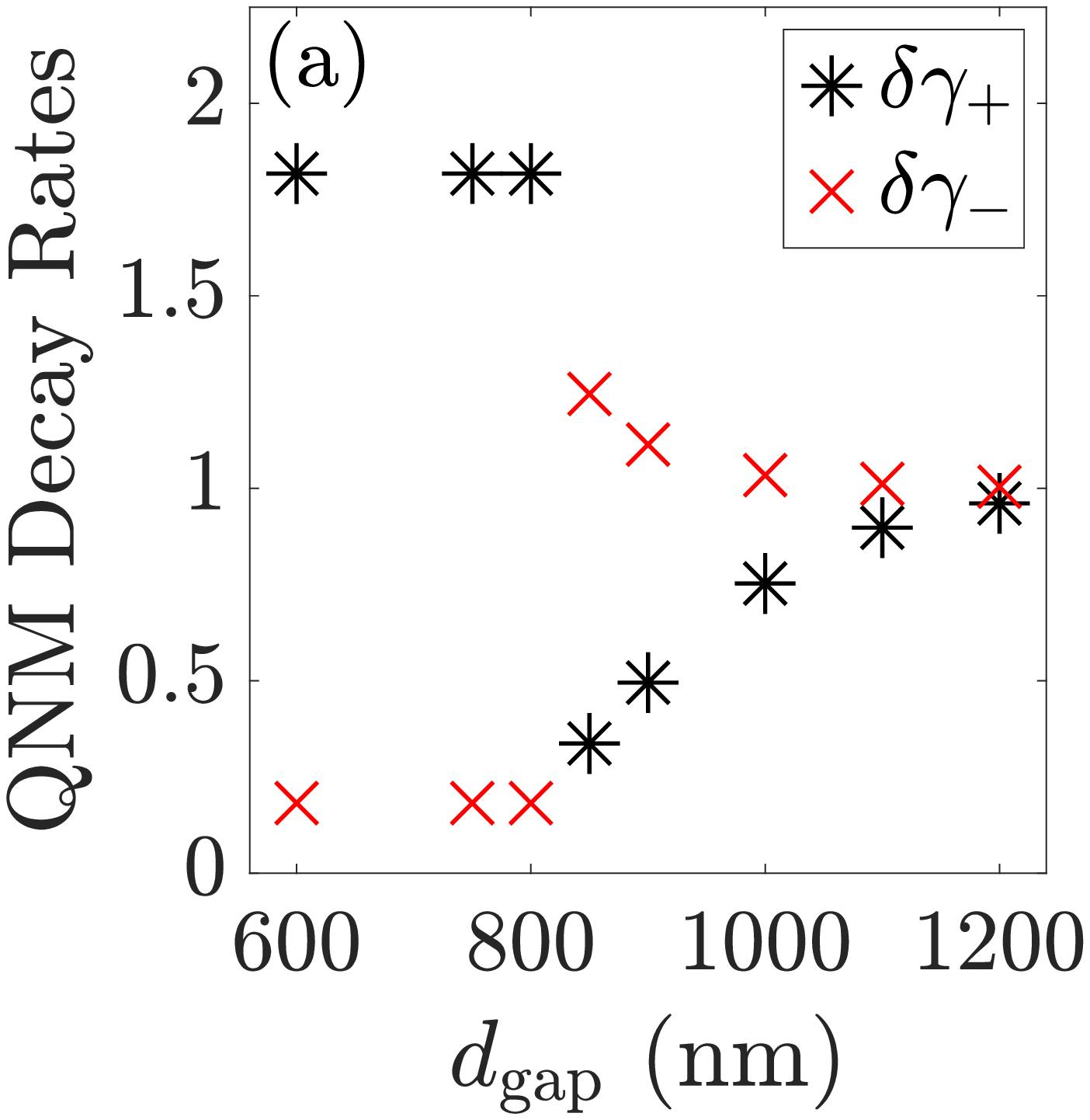}
    \includegraphics[width=0.49\columnwidth]{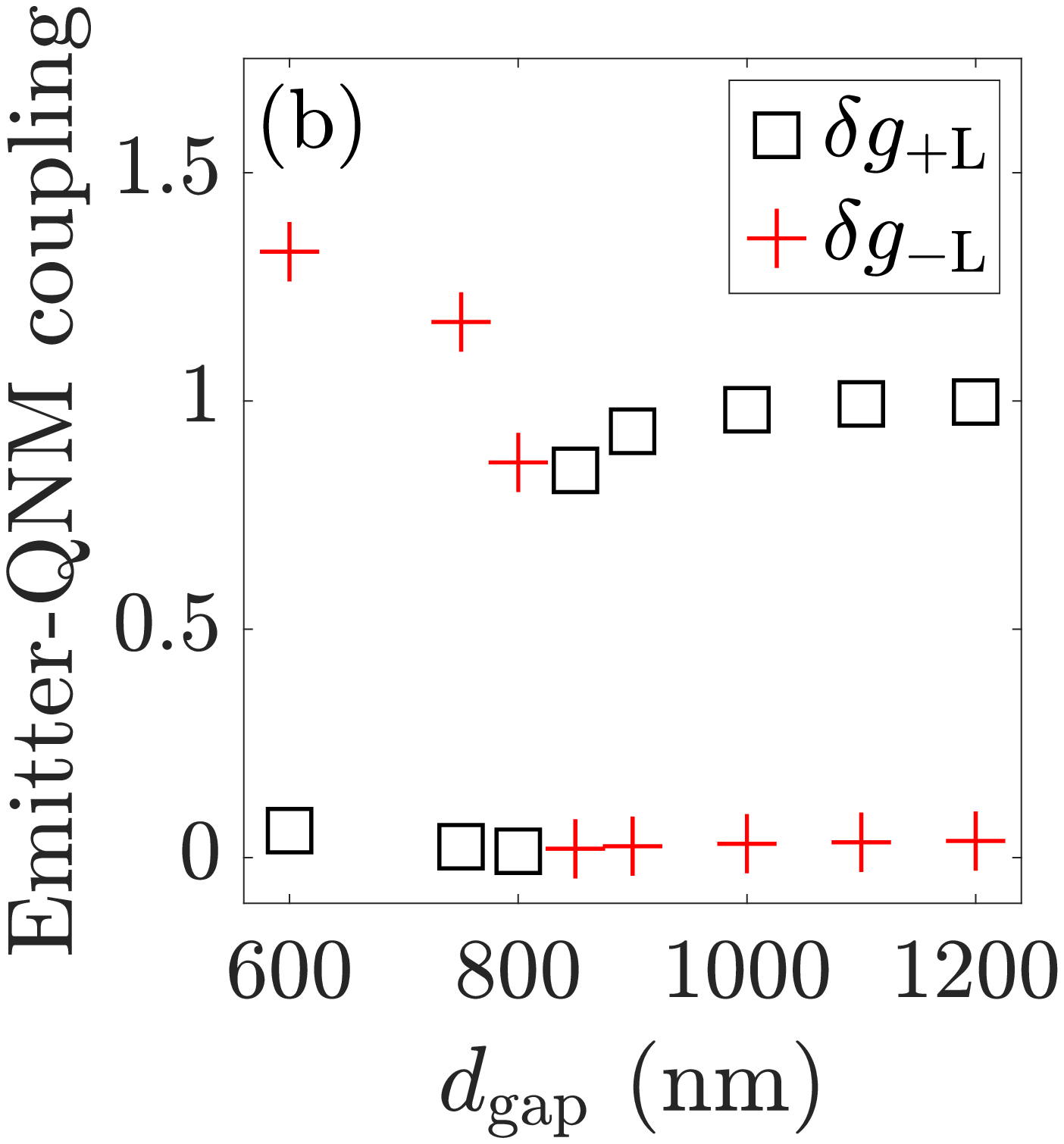}
    \includegraphics[width=0.49\columnwidth]{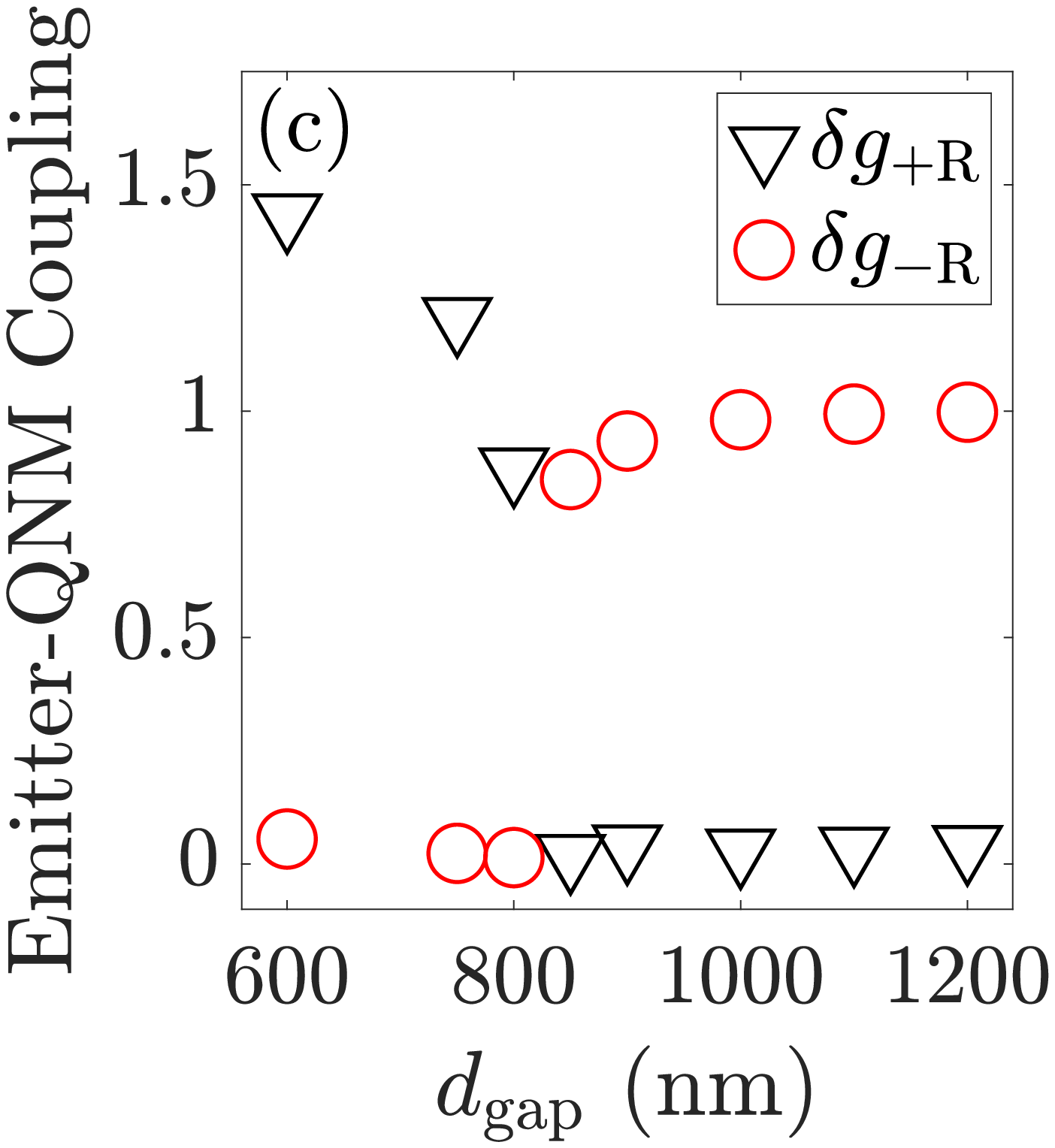}
    \includegraphics[width=0.49\columnwidth]{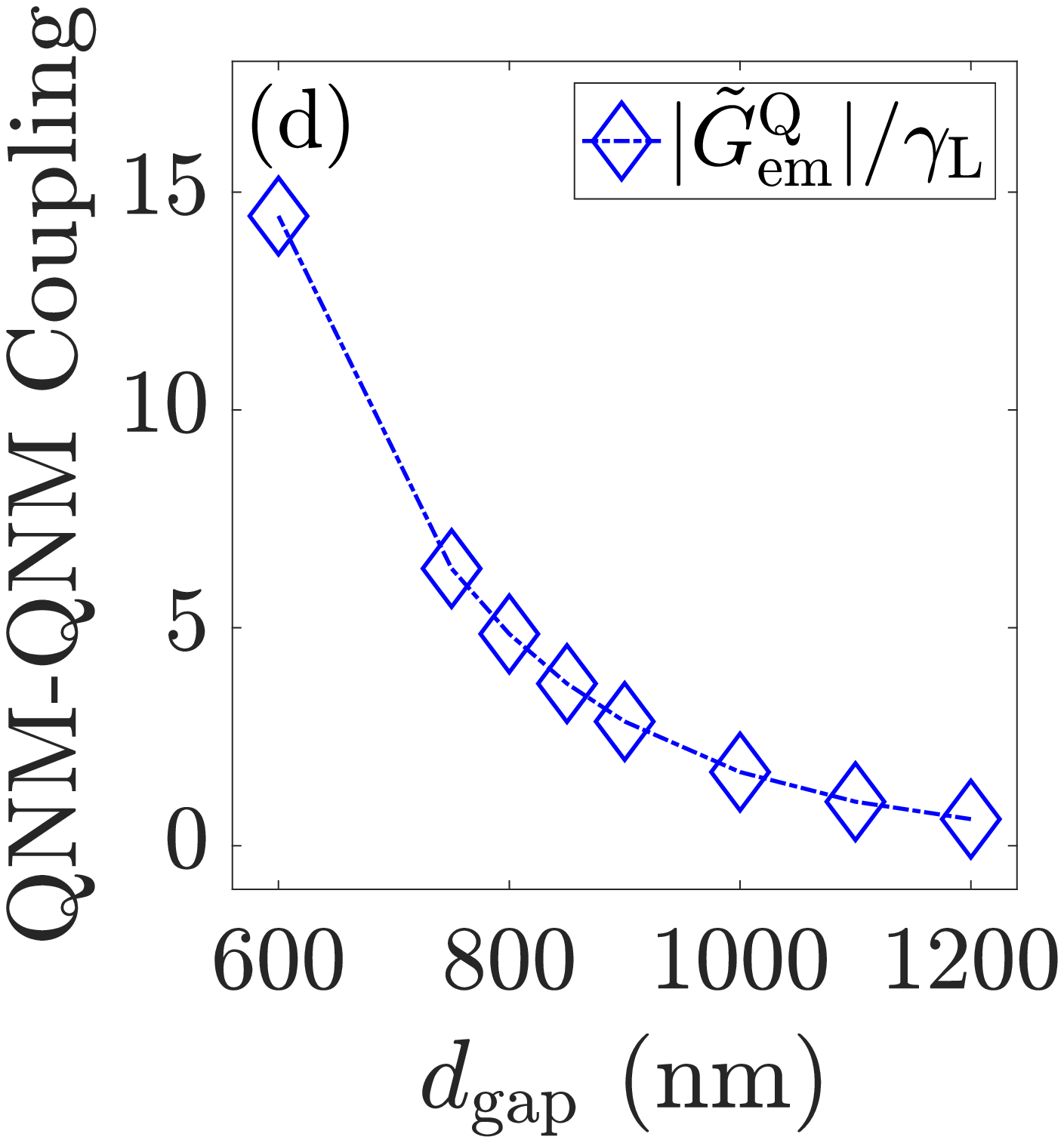}
    \caption{(a) QNM decay rates (HWHM) ratios $\delta\gamma_{\pm}$ (Eq.~\eqref{eq: Decay_ratios}), emitter-QNM coupling rate ratios $\delta g_{\pm}$ (Eq.~\eqref{eq: Coup_ratios}) for emitters at (b) $d_{\rm L}=10~$nm and (c) $d_{\rm R}=10~$nm (the subscript R/L in the figure means dipole at $d_{\rm R}/d_{\rm L}=10~$nm), and (d) QNM-QNM coupling rate $|\tilde{G}^{\rm Q}_{\rm em}|/\gamma_{\rm L}$ (normalized by the decay rate $\gamma_{\rm L}$ of the single QNM for resonator L) as a function of gap distance. 
    Note there are in general drastic changes appearing in those ratios between $d_{\rm gap}=800~{\rm nm}$ and $d_{\rm gap}=850~{\rm nm}$, as also seen from Fig.~\ref{fig: phase_d800_d850} (QNM phases show the difference). In addition, the QNM-QNM (photon-photon) coupling will decrease with the gap distance increase, as expected.
}\label{fig: AllQMParamsV3}
\end{figure*}

\subsection{QNM decay rate ratios, emitter-QNM coupling rate ratios and the coupling rates between different quantized QNMs}\label{sec: ratios}

In this section, the altering of the QNM decay rates as well as emitter-QNM coupling rates due to symmetrization and QNM-QNM coupling are formally discussed in more detail and results for the coupled resonator system are presented.

\subsubsection{Diagonalization of the QNM Lindblad dissipator}
First, since we will investigate the behavior of the full quantized model with respect to the gap distance of the lossy resonator system, here we summarize the underlying master equation of the quantized QNM model for two modes $+,-$ in a picture, where the QNM Lindblad dissipator (Eq.~\eqref{eq: LindbladDiss_QNM}) is diagonalized.
In this way, the QNM-QNM coupling induced by the symmetrization is fully captured in the Hamiltonian part, which will help to better understand the differences to a dissipative JC model. 

The underlying master equation in the transformed picture reads
\begin{equation}\label{eq: QNMmaster2}
    \partial_t{\rho} = -\frac{i}{\hbar}[H_{\rm S},{\rho}]+\sum_{\mu=\pm}\Gamma_\mu^{\rm Q}\mathcal{D}[\hat{A}_\mu]{\rho},
\end{equation}
with the system Hamiltonian 
\begin{align}\label{eq: Hamiltonian_s}
    {H}_{\rm S}=&\hbar\sum_{\mu=\pm}\Omega_\mu\hat{A}_\mu^\dagger\hat{A}_\mu + \hbar\left[ \tilde{G}^{\rm Q}_{\rm em}\hat{A}_+^\dagger \hat{A}_-+{\rm H.a.}\right]\nonumber\\
   &+\hbar\left[\sum_{\mu=\pm}\tilde{G}_\mu^{\rm Q}\hat{\sigma}^+\hat{A}_\mu + {\rm H.a.}\right]+\hbar\omega_{0}\hat{\sigma}^+\hat{\sigma}^-,
\end{align}
and the Lindblad dissipator,
\begin{equation}
    \mathcal{D}[\hat{A}]=2\hat{A}{\rho}\hat{A}^\dagger - {\rho}\hat{A}^\dagger\hat{A}-\hat{A}^\dagger\hat{A}{\rho},
\end{equation}
with $\hat{A}=\hat{A}_+,~\hat{A}_-$. Here, $A_\mu$ is the new QNM operator basis
\begin{equation}
    \hat{A}_\mu = \sum_{\eta=\pm}U^{(-)*}_{\eta\mu}\hat{a}_\eta,
\end{equation}
where $U^{(-)}_{\mu\eta}$ is the (normalized) unitary matrix that diagonalizes the Lindblad dissipator, so that 
\begin{equation}
    \sum_{\mu',\eta'=\pm}U^{(-)*}_{\mu'\mu}\chi^{(-)}_{\mu'\eta'}U^{(-)}_{\eta'\eta}=\delta_{\mu\eta}\Gamma_\mu^{\rm Q},
\end{equation}
with the new QNM decay rates $\Gamma_\mu^{\rm Q}$ as eigenvalues of $\boldsymbol{\chi}^{(-)}$ (cf. Eq.~\eqref{eq: chi_minus}). 

Furthermore, in this new picture, the photon-photon coupling matrix $\boldsymbol{\chi}^{(+)}$ (Eq.~\eqref{eq: chi_plus}) is transformed as 
\begin{equation}\label{eq: comphomatrix}
    \boldsymbol{\chi}^{(+)}\rightarrow [\mathbf{U}^{(-)}]^\dagger\cdot\boldsymbol{\chi}^{(+)}\cdot \mathbf{U}^{(-)}\equiv\begin{pmatrix}
    \Omega_+ & \tilde{G}_{\rm em}^{\rm Q}\\
    \tilde{G}_{\rm em}^{\rm Q*} &\Omega_-
    \end{pmatrix},
\end{equation}
while the TLS-QNM coupling constants $\tilde{g}_\mu^{\rm s}$ are transformed as
\begin{equation}
    \tilde{g}_\mu^{\rm s}\rightarrow \sum_{\eta, \eta'=\pm}\tilde{g}_{\eta'} [\mathbf{S}^{1/2}]_{\eta'\eta}U^{(-)}_{\eta\mu}\equiv \tilde{G}_\mu^{\rm Q}.
\end{equation}

The appearing quantum parameters can be obtained from the initial CQT quantities, and the medium/geometry parameters. 

\subsubsection{Formal interpretation of the quantum parameter ratios\label{subsub: QuantumParamsDef}}

The nominal decay rate (HWHM, linewidth) ratios are defined as
\begin{equation}\label{eq: Decay_ratios}
\delta\gamma_{\pm}=\frac{\Gamma_{\pm}^{\rm Q}}{\gamma_{\pm}},
\end{equation}
which reflects the decay rate changes due to the symmetrization and inherent dissipation. In the single QNM limit, this ratio is $1$. However, for two or more modes, it becomes only close to $1$ in the limit of very small absorption as well as initial decay, i.e. $Q_\mu\gg 1$, where $S_{\mu\eta}\rightarrow \delta_{\mu\eta}$. 

To elaborate more on this limit, for the inspected resonator structure, the non-radiative part is the dominant contribution. In the case of a finite radiative part (not strictly zero), the diagonal elements of the symmetrization matrices read $S_\mu^{\rm nrad}= \omega_\mu I^{\rm vol}_\mu \epsilon_I/(\gamma_\mu^{\rm nrad}+\gamma_\mu^{\rm rad})$ with $I^{\rm vol}_\mu=\int_V {\rm d}\mathbf{r}|\tilde{\mathbf{f}}_\mu(\mathbf{r})|^2$, while $S_\mu^{\rm rad}= \omega_\mu I^{\rm sur}_\mu/(\gamma_\mu^{\rm nrad}+\gamma_\mu^{\rm rad})$. For vanishing absorption, $\epsilon_I\rightarrow 0$, and $\gamma_\mu^{\rm nrad}\rightarrow 0$. Thus, the symmetrization factors tend to  $S_\mu^{\rm nrad}\rightarrow 0$ and $S_\mu^{\rm rad}\rightarrow \omega_\mu I^{\rm sur}_\mu/(\gamma_\mu^{\rm rad})$. For $\gamma_\mu^{\rm rad}\rightarrow 0$, the radiative contribution tends to $1$, because the surface integral $I^{\rm sur}_\mu\rightarrow 0$ with the same ``velocity'', as shown formally in Ref.~\onlinecite{franke_fluctuation-dissipation_2020}. 

However, for the artificial case of strictly vanishing radiative part (meaning a closed absorptive resonator structure), $S_\mu^{\rm rad}\rightarrow 0$, and 
\begin{equation}
    \lim_{\epsilon_I\rightarrow 0}S_\mu^{\rm nrad}=\omega_\mu I^{\rm vol}_\mu \lim_{\epsilon_I\rightarrow 0}\frac{\epsilon_I}{\gamma^{\rm nrad}_\mu[\epsilon_I]},
\end{equation}
where $\gamma^{\rm nrad}_\mu[\epsilon_I]$ is a function of the absorption. The evaluate the above limit, one needs to have access to the dependence of $\gamma^{\rm nrad}_\mu[\epsilon_I]$ on $\epsilon_I$. However, since we know, that $S_\mu\rightarrow 1$ in the limit of vanishing decay, the result must be 
\begin{equation}
    \lim_{\epsilon_I\rightarrow 0}\frac{\epsilon_I}{\gamma^{\rm nrad}_\mu[\epsilon_I]}=\epsilon_R/\omega_\mu,
\end{equation}
so that 
\begin{equation}
    S_\mu^{\rm nrad}\rightarrow \epsilon_R I^{\rm vol}_\mu =\int_V{\rm d}\mathbf{r} \epsilon_R |\mathbf{f}_\mu(\mathbf{r})|^2,
\end{equation}
which is the NM norm, and is identically $1$.

The TLS-QNM coupling rates ratios are defined as
\begin{equation}\label{eq: Coup_ratios}
\delta g_{\pm}=\frac{|\tilde{G}_{\pm}^{\rm Q}|}{|\tilde{g}_{\pm}|},
\end{equation}
which reflects the altering of the TLS-photon coupling rates due to symmetrization. In the single QNM limit, this ratio is simply the squareroot of the symmetrization factor $\sqrt{S}$. For two or more modes $\tilde{G}_{\pm}^{\rm Q}$ is a non-trivial linear combination of $\tilde{g}_{+}$ and $\tilde{g}_{-}$, which can lead to drastic changes, how the TLS couples to the quantized QNM. Again, in the limit of no dissipation, $S_{\mu\eta}\rightarrow\delta_{\mu\eta}$ (as discussed above) and $\delta g_{\pm}\rightarrow 1$.

\subsubsection{Numerical results of the ratios as function of gap distance}
As shown in above section (Sec.~\ref{subsub: QuantumParamsDef}), one could obtain the QNM decay rate ratios $\delta\gamma_{\pm}$ (Eq.~\eqref{eq: Decay_ratios}) by using the quantum decay constants $\Gamma_{\pm}^{\rm Q}$ divided by their respective CQT parameters $\gamma_{\pm}$. The results as a function of gap distance are shown in 
Fig.~\ref{fig: AllQMParamsV3}(a), where a drastic change could be found between $d_{\rm gap}=800~{\rm nm}$ and $d_{\rm gap}=850~{\rm nm}$ (close to the EP).

Similarly, the emitter-QNM coupling rate ratios $\delta g_{\pm}$ (Eq.~\eqref{eq: Coup_ratios}) are obtained by using the quantum emitter-QNM coupling constants $\tilde{G}_{\pm}^{\rm Q}$ divided by their respective CQT parameters $\tilde {g}_{\pm}$.
The coupling rate ratios $\delta g_{\pm}$ are emitter-position dependent.
Here, we still focus on two special positions $d_{\rm L/R}=10~$nm.
The corresponding results are shown in 
Fig.~\ref{fig: AllQMParamsV3}(b-c), where the addition subscript $\rm L/R$ means that the emitter is at $d_{\rm L/R}=10~$nm. 
With the gap distance increases, $\delta g_{+{\rm L}}$ ($\delta g_{-{\rm R}}$) as well as $ \delta\gamma_\pm$ approach $1$, which means the QNMs are going to decouple from each other (will reach to the bare QNM cases when the gap distance is large enough). 
This could also be seen from 
Fig.~\ref{fig: AllQMParamsV3}(d), where the direct QNM-QNM coupling $|\tilde{G}^{\rm Q}_{\rm em}|/\gamma_{\rm L}$ (non-diagonal term in complex photon matrix (Eq.~\eqref{eq: comphomatrix}), normalized by the single QNM decay rate $\gamma_{\rm L}$) are shown and it decreases with the gap distance increase as expected. 

Again, one could also find the drastic changes in coupling rate ratios $\delta g_{\pm{\rm L/R}}$, between $d_{\rm gap}=800~{\rm nm}$ and $d_{\rm gap}=850~{\rm nm}$.
Similar changes could also be found from the Fig.~\ref{fig: phase_d800_d850} in the classical picture, where the QNMs phases show apparent difference.

\subsection{Improving the normal mode theories
using a non-diagonal basis with coupled mode coefficients\\  (for high $Q$ resonators)
\label{sec: ImproveNM}}

\begin{figure*}[t]
    \centering
    \includegraphics[width=0.51\columnwidth]{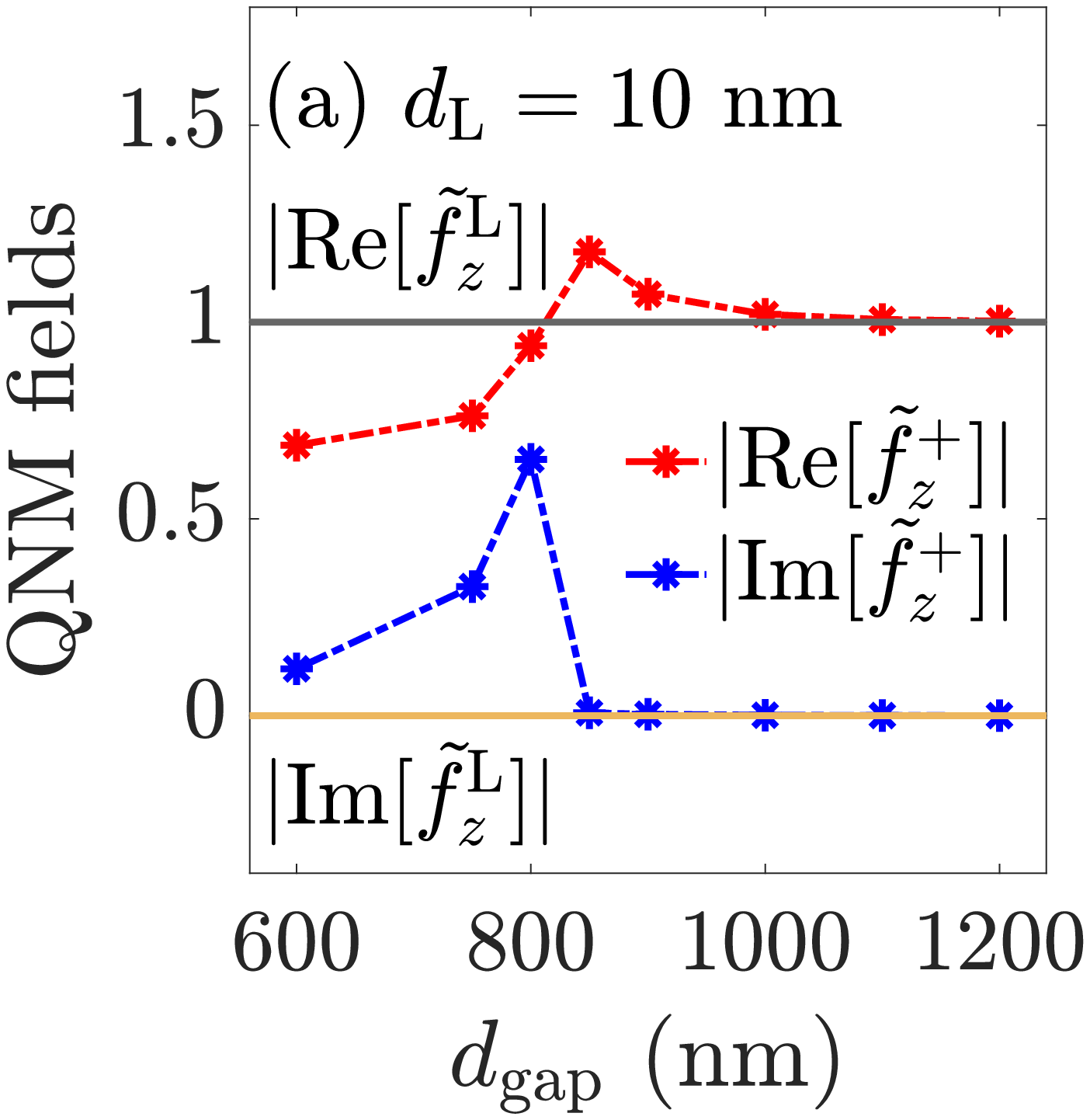}
    \includegraphics[width=0.51\columnwidth]{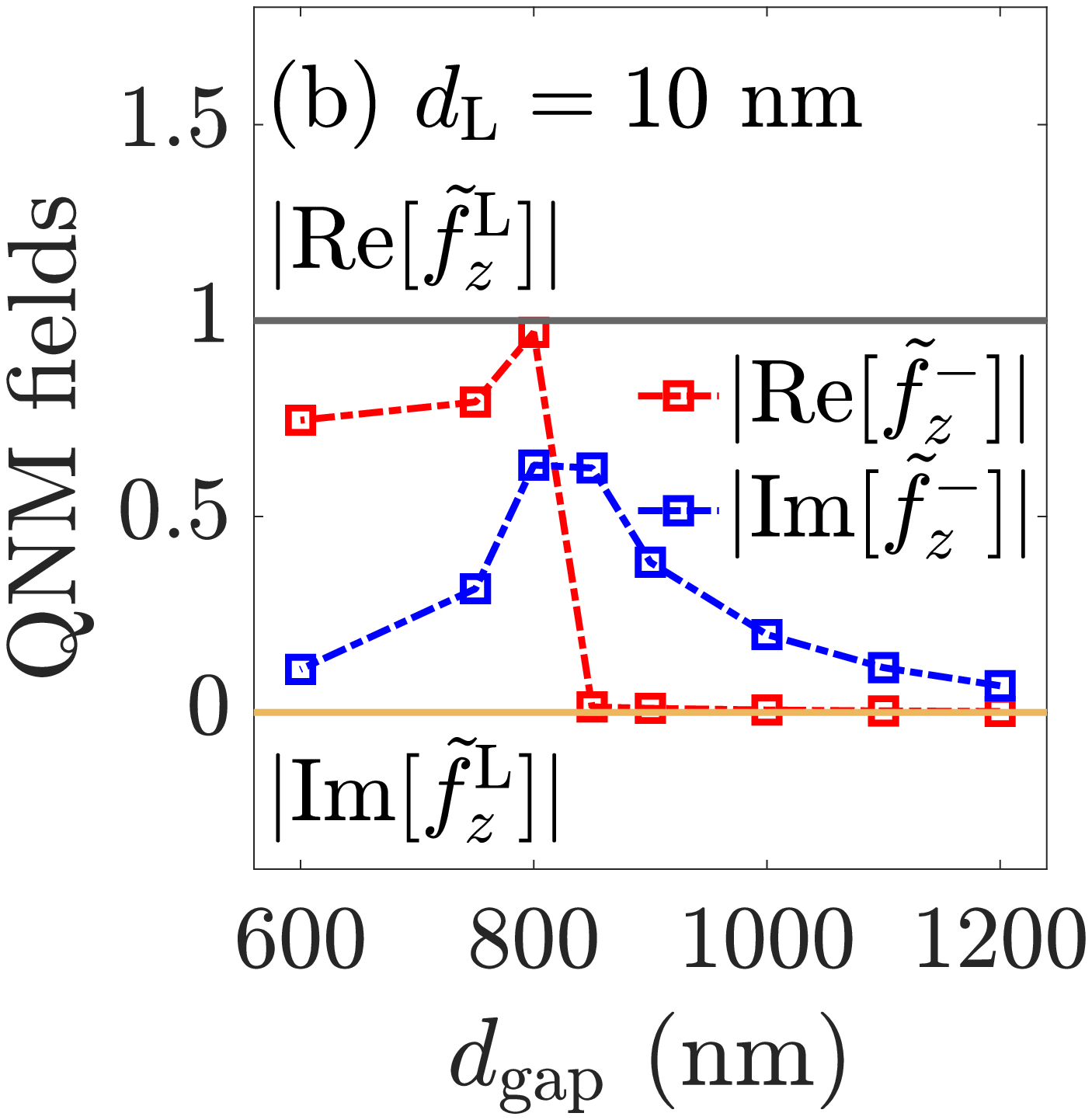}
    \includegraphics[width=0.51\columnwidth]{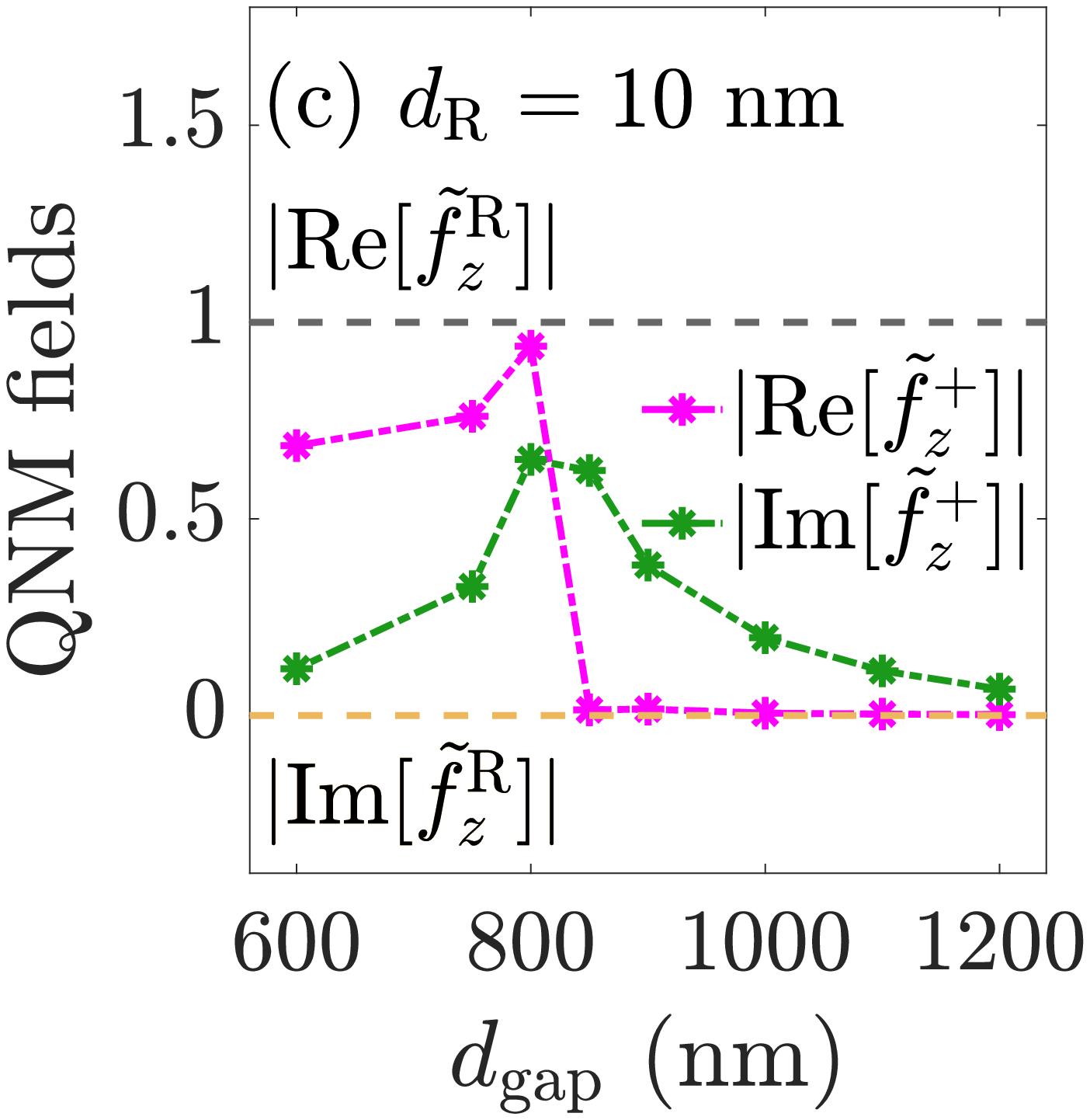}
    \includegraphics[width=0.51\columnwidth]{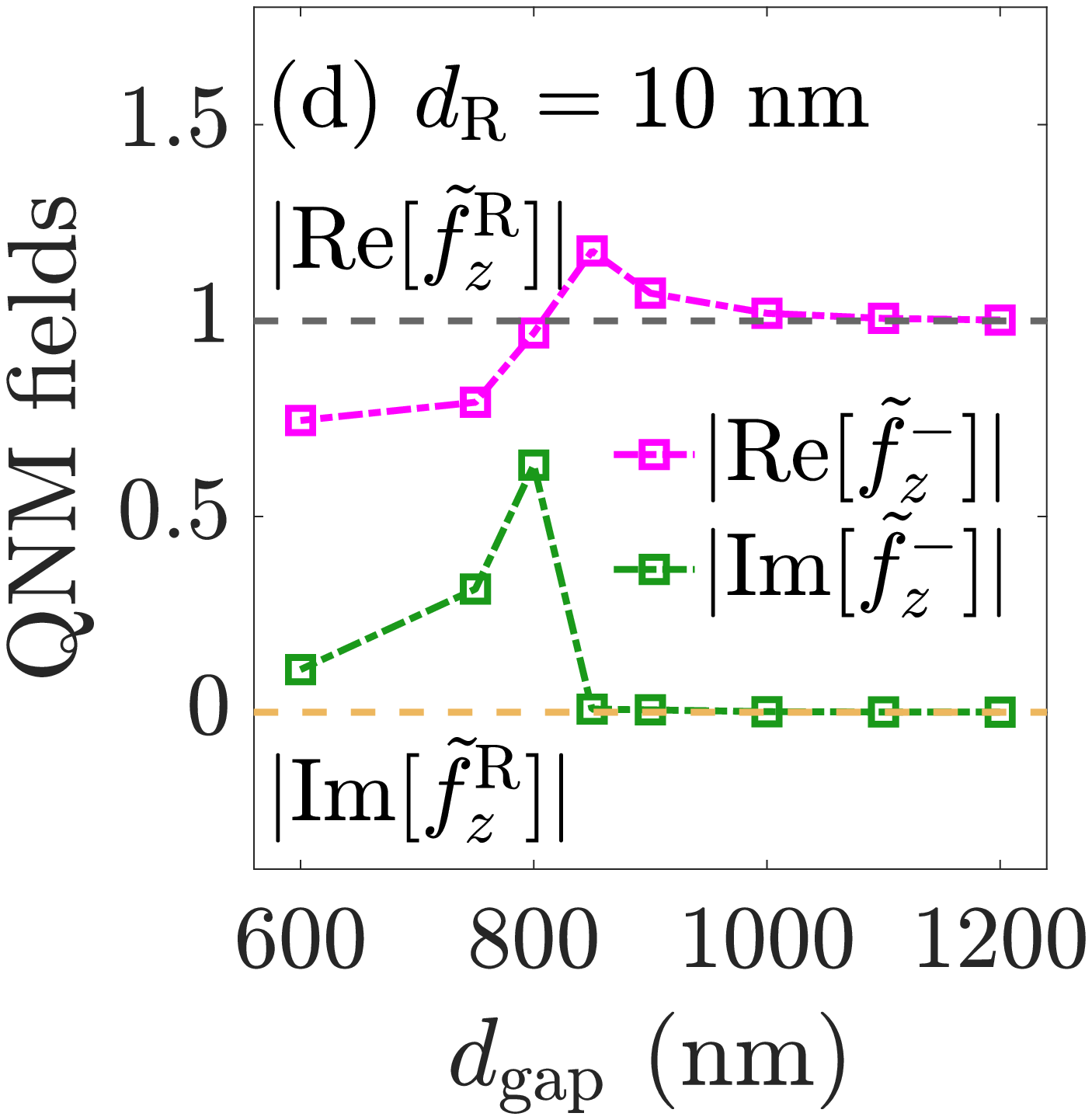}
    \caption{Normalized fields (real parts and imaginary parts) of the coupled QNMs (two modes, labeled by $\pm$),  when the dipole emitter is at (a-b) $d_{\rm L}=10~$nm and (c-d) $d_{\rm R}=10~$nm. Note all of QNMs are normalized by $|{\rm Re}[\tilde{f}^{\rm L}_{z}(d_{\rm L}=10~{\rm nm})]|$. With a small gap distances, the real part and the imaginary parts are comparable. 
    In addition, at $d_{\rm L}=10~$nm, the real part and the imaginary part of the bare QNMs $\tilde{f}_{z}^{\rm L}$ are also presented as in the solid black and orange horizontal lines in (a-b) (nearly real functions).
    As for bare QNMs $\tilde{f}_{z}^{\rm R}$ at  $d_{\rm R}=10~$nm, their real part and imaginary parts are shown as the dashed black and orange horizontal lines in (c-d), which is also nearly real functions. 
}\label{fig: QNMsReIm_bare_vs_coup}
\end{figure*}

\subsubsection{Classical perspective of an improved normal mode theory}

As explained in the previous sections, after diagonalizing to the correct
hybrid states, 
applying a NM approximation drastically fails for the hybridized expansion, since $\tilde{\mathbf{f}}^{+/-}$ have significant imaginary parts. However, in the bare basis, $\tilde{\mathbf{f}}^{\rm L/R}$ are very close to be real functions.
This is shown more clearly
in Fig.~\ref{fig: QNMsReIm_bare_vs_coup}.
%
From Eq.~\eqref{eq: QNMs_pm}, the imaginary part of $\tilde{\mathbf{f}}^{+/-}$ comes dominantly from the complex eigenfrequencies for large $Q$ (and partly from the complex $\tilde{\kappa}_{\rm LR}$), since $\tilde{\mathbf{f}}^{\rm L/R}\approx {\rm Re}[\tilde{\mathbf{f}}^{\rm L/R}]$. 

An alternative approach to expand the
diagonal-form GF in terms of the 
true hybrid modes (i.e., with coupling),
is to  start with the non-diagonal form of the QNM GF~\cite{ren_quasinormal_2021} (derived using CQT):
 \begin{align}
\hat {\bf G}
&= \frac{\frac{\omega}{2}(\tilde \omega_{\rm R}-\omega) \ket{\tilde {\bf f}^{\rm L}}\bra{\tilde{\bf f}^{{\rm L}*}}}{(\tilde \omega_{+}-\omega)(\tilde \omega_{-}-\omega)} 
+ \frac{\frac{\omega}{2}\tilde \kappa_{\rm LR} \ket{\tilde{\bf f}^{\rm L}}\bra{\tilde{\bf f}^{{\rm R}*}}}{(\tilde \omega_{+}-\omega)(\tilde \omega_{-}-\omega)}
\nonumber \\
&+\frac{\frac{\omega}{2}\tilde \kappa_{\rm RL} \ket{\tilde{\bf f}^{\rm R}}\bra{\tilde{\bf f}^{{\rm L}*}}}{(\tilde \omega_{+}-\omega)(\tilde \omega_{-}-\omega)}
+
\frac{\frac{\omega}{2}(\tilde \omega_{\rm L}-\omega) \ket{\tilde{\bf f}^{\rm R}}\bra{\tilde{\bf f}^{{\rm R}*}}}{(\tilde \omega_{+}-\omega)(\tilde \omega_{-}-\omega)}.
\end{align}
From here we can adopt a NM
approximation for the spatial modes:
\begin{equation}
\tilde{\mathbf{f}}^i\rightarrow {\rm Re}[\tilde{\mathbf{f}}^i]\equiv \mathbf{f}^i,
\end{equation}
with $i={\rm L,R}$ and
\begin{equation}
\tilde{\kappa}_{\rm LR},\tilde{\kappa}_{\rm RL}\rightarrow {\rm Re}[\tilde{\kappa}_{\rm LR}+\tilde{\kappa}_{\rm RL}]/2\equiv \kappa^{\rm NMI}.
\end{equation}
The corresponding complex eigenvalues are now: 
\begin{equation}
\tilde \omega_{\pm}^{\rm NMI}=
\frac{\tilde\omega_{\rm L}+\tilde\omega_{\rm R}}{2}
\pm \frac{\sqrt{4(\kappa^{\rm NMI})^2 + (\tilde\omega_{\rm L}-\tilde\omega_{\rm R})^2}}{2},
\end{equation}
so that 
 \begin{align}
 \label{eq: GreenwithnewcNMs}
&\hat {\bf G}^{\rm NMI}
\rightarrow \hat {\bf G}^{\rm NMI}_{\rm LL} + \hat {\bf G}^{\rm NMI}_{\rm LR} + \hat {\bf G}^{\rm NMI}_{\rm RL}+ \hat {\bf G}^{\rm NMI}_{\rm RR} \nonumber \\
&=\phantom{+}\frac{\frac{\omega}{2}(\tilde \omega_{\rm R}-\omega) \ket{{\bf f}^{\rm L}}\bra{{\bf f}^{\rm L}}}{(\tilde \omega_{+}^{\rm NMI}-\omega)(\tilde \omega_{-}^{\rm NMI}-\omega)} 
+ \frac{\frac{\omega}{2}\kappa^{\rm NMI} \ket{{\bf f}^{\rm L}}\bra{{\bf f}^{\rm R}}}{(\tilde \omega_{+}^{\rm NMI}-\omega)(\tilde \omega_{-}^{\rm NMI}-\omega)}
\nonumber \\
&
+\frac{\frac{\omega}{2}\kappa^{\rm NMI} \ket{{\bf f}^{\rm R}}\bra{{\bf f}^{\rm L}}}{(\tilde \omega_{+}^{\rm NMI}-\omega)(\tilde \omega_{-}^{\rm NMI}-\omega)}
+
\frac{\frac{\omega}{2}(\tilde \omega_{\rm L}-\omega) \ket{{\bf f}^{\rm R}}\bra{{\bf f}^{\rm R}}}{(\tilde \omega_{+}^{\rm NMI}-\omega)(\tilde \omega_{-}^{\rm NMI}-\omega)}.
\end{align}
Note we are still using aspects 
of the QNM theory, e.g., to yield the correct eigenfrequencies,
which are the ones that are typically calculated in numerical solutions with open boundary conditions. This gives  insight into 
 the use of a partial NM approximation with regards to the modes and mode expansions.

The corresponding classical Purcell factor is simply
\begin{align}
     F_{{\rm P}}^{\rm cNMI}(\mathbf{r}_{0},\omega) =1+\frac{\mathbf{d}\cdot{\rm Im}\{\mathbf{G}^{\rm NMI}(\mathbf{r}_{0},\mathbf{r}_{0},\omega)\}\cdot\mathbf{d}}{\mathbf{d}\cdot{\rm Im}\{\mathbf{G}_{\rm B}(\mathbf{r}_{0},\mathbf{r}_{0},\omega)\}\cdot\mathbf{d}}.\label{eq: FP_cNMI}
\end{align}
In addition, we can define the separate contribution from those terms in Eq.~\eqref{eq: GreenwithnewcNMs} as ($i,j={\rm L,R}$)
\begin{align}
     F_{{\rm P},ij}^{\rm cNMI}(\mathbf{r}_{0},\omega) =\frac{\mathbf{d}\cdot{\rm Im}\{\mathbf{G}_{ij}^{\rm NMI}(\mathbf{r}_{0},\mathbf{r}_{0},\omega)\}\cdot\mathbf{d}}{\mathbf{d}\cdot{\rm Im}\{\mathbf{G}_{\rm B}(\mathbf{r}_{0},\mathbf{r}_{0},\omega)\}\cdot\mathbf{d}},\label{eq: FP_cNMI_ij}
\end{align}
which satisfy $F_{{\rm P}}^{\rm cNMI}=1+F_{{\rm P,LL}}^{\rm cNMI}+F_{{\rm P,LR}}^{\rm cNMI}+F_{{\rm P,RL}}^{\rm cNMI}+F_{{\rm P,RR}}^{\rm cNMI}$.
In this way, the resulting hybridized modes now have imaginary parts, when diagonalizing the above form, because of allowing for a complex eigenfrequency. However, we stress that this
approximation likely only works for high-$Q$ resonators,  otherwise $\tilde\kappa_{\rm LR}$ and $\tilde{\mathbf{f}}^{\rm L/R}$ can have significant imaginary parts on their own, which are missing in corresponding NM theories.

\begin{figure}
    \centering
    \includegraphics[width=0.99\columnwidth]{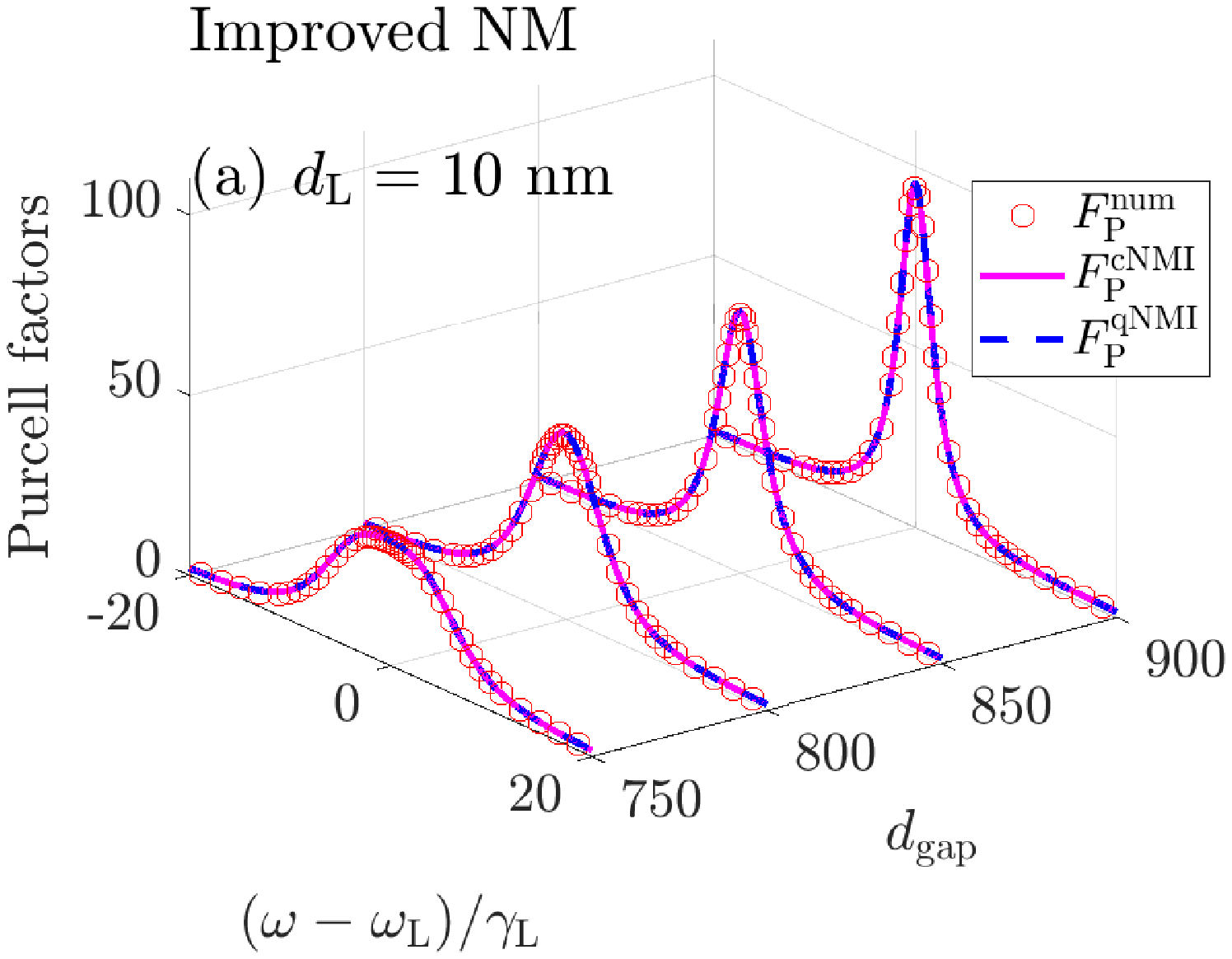}
    \includegraphics[width=0.99\columnwidth]{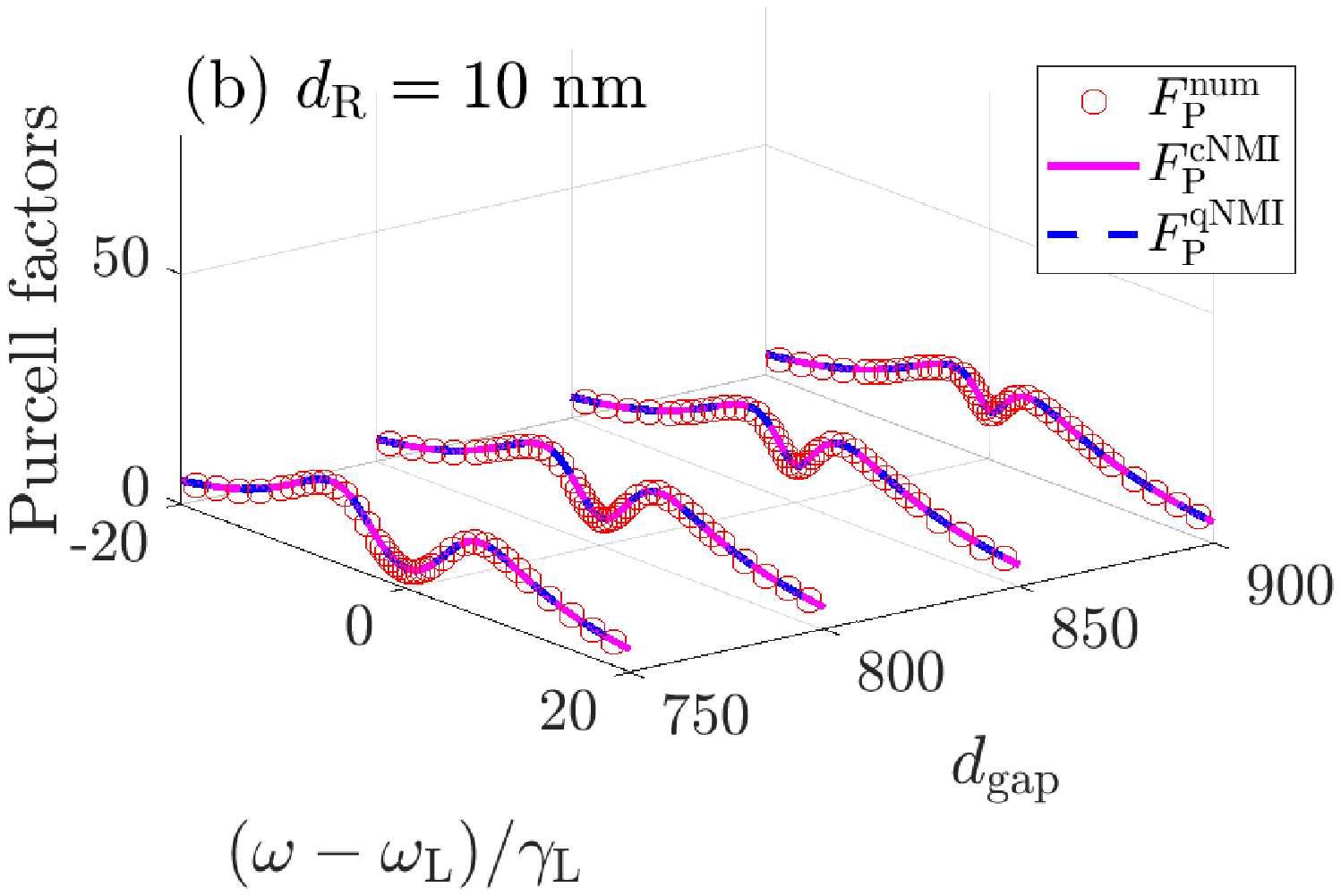}
    \caption{ 
    Computed Purcell factors from the improved
    normal mode (`NMI') solutions, using 
    classical theory ($F_{\rm P}^{\rm cNMI}$, Eq.~\eqref{eq: FP_cNMI}, solid magenta curves) and quantum theory ($F_{\rm P}^{\rm qNMI}$, Eq.~\eqref{eq: Fp_qNMI}, dashed blue curves).
    We show calculations for four gap sizes $d_{\rm gap}=750/800/850/900~$nm, when the emitter is placed at (a) $d_{\rm L}=10~$nm, and (b) $d_{\rm R}=10~$nm. IN all cases, we obtain excellent agreement with the full numerical dipole results ($F_{\rm P}^{\rm num}$, Eq.~\eqref{eq: Purcellfulldipole}, red circles).
}\label{fig: newcNMs_waterfall}
\end{figure}

For our current coupled lossy resonator examples, the Purcell factors $F_{\rm P}^{\rm cNMI}$ (Eq.~\eqref{eq: FP_cNMI}, solid mageneta curves) from the improved NMs solution (using the Green function Eq.~\eqref{eq: GreenwithnewcNMs}) are shown in Fig.~\ref{fig: newcNMs_waterfall} for $4$ gap distances $d_{\rm gap}=750/800/850/900~$nm, and dipole locations at $d_{\rm L}=10~$nm (Fig.~\ref{fig: newcNMs_waterfall} (a)) or $d_{\rm R}=10~$nm (Fig.~\ref{fig: newcNMs_waterfall} (b)). We observe an excellent agreement with the full numerical dipole results, $F_{\rm P}^{\rm num}$ (Eq.~\eqref{eq: Purcellfulldipole}, red circles).

\begin{figure}
    \centering
    \includegraphics[width=0.9\columnwidth]{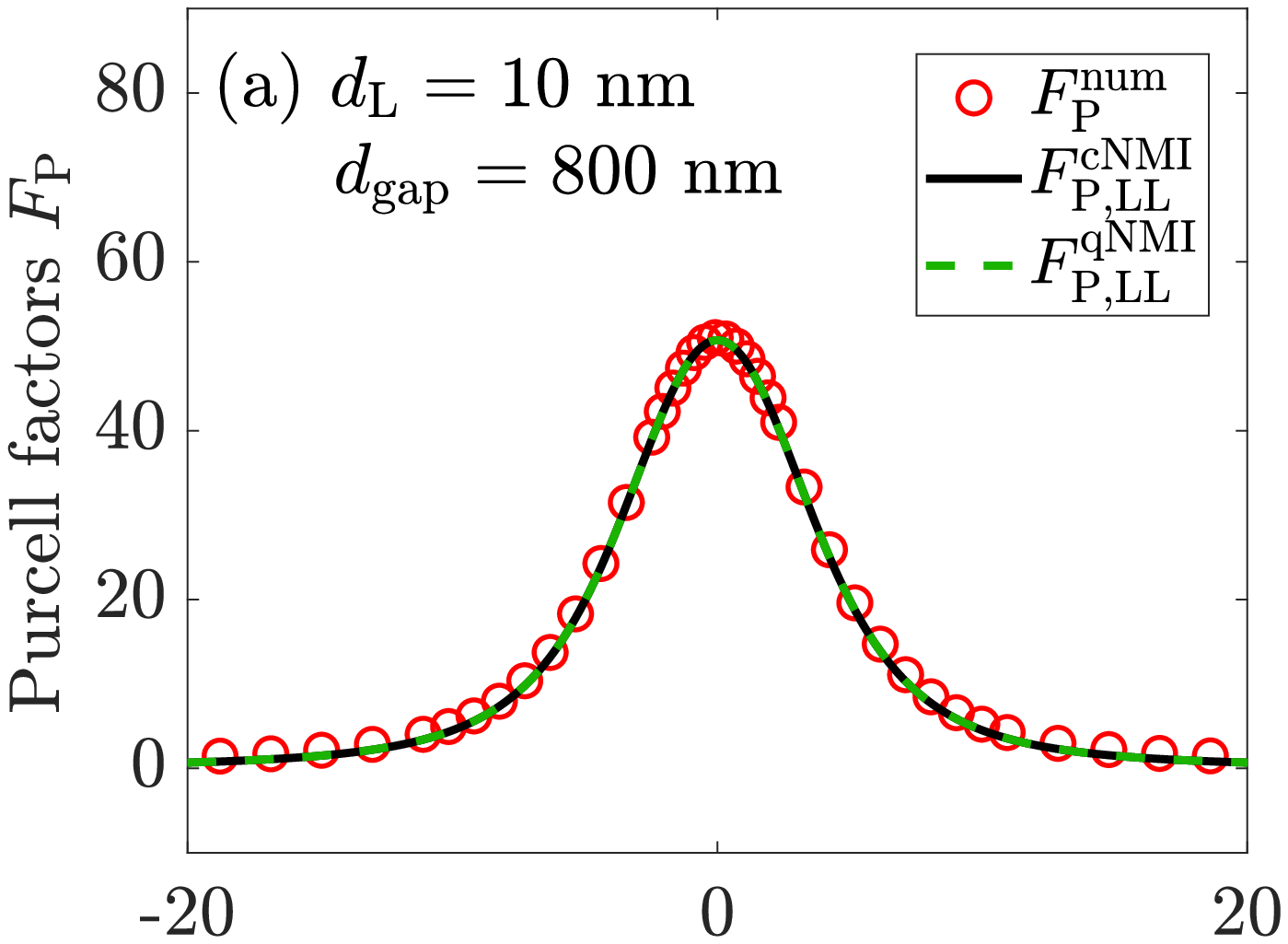}
    \includegraphics[width=0.9\columnwidth]{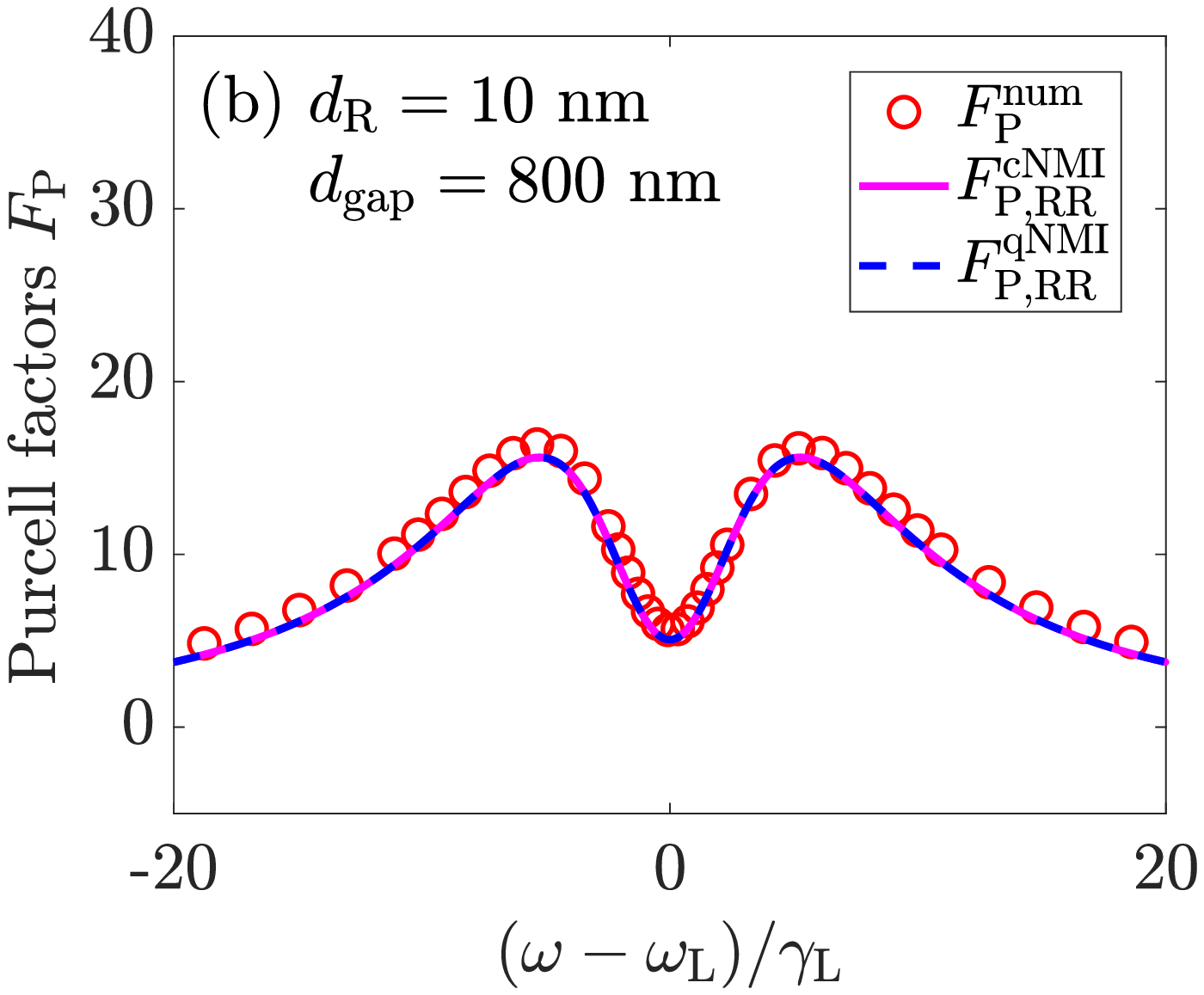}
    \caption{Dominant contributions (Eq.~\eqref{eq: FP_cNMI_ij}, classical, and Eq.~\eqref{eq: Fp_qNMI_ij}, quantum) to the Purcell factors from  improved NM theories. When the dipole emitter is located at $d_{\rm L}=10~$ nm ($d_{\rm R}=10~$nm), the LL term $F_{\rm P,LL}^{\rm cNMI/qNMI}$ (the RR term $F_{\rm P,RR}^{\rm cNMI/qNMI}$) is dominant, which agrees very wells with the full numerical dipole results $F_{\rm P}^{\rm num}$. The other terms (not shown) are negligible.
}\label{fig: newcNMs_terms}
\end{figure}

Especially, for $d_{\rm gap}=800~$nm, we show the contribution from the dominant term ($F_{{\rm P},ij}^{\rm cNMi}$, Eq.~\eqref{eq: FP_cNMI_ij}) in Fig.~\ref{fig: newcNMs_terms}, where one can clearly see that the LL term $F_{{\rm P,LL}}^{\rm cNMI}$ (the RR term $F_{{\rm P,RR}}^{\rm cNMI}$) dominates when the dipole is close to the left resonator $\rm L$ (the right resonator $\rm R$). The other terms contribute much less and 
are therefore not shown.

\subsubsection{Quantum perspective of an improved normal mode theory}

In a quantum picture, the improved classical NM approach with complex eigenfrequencies from above can be modelled through an effective (non-Hermitian) photon Hamiltonian~\cite{louisell_quantum_1961,carmichael_master_1973,zhang_transfer_2010,barlow_master_2015,wang_squeezing-enhanced_2019,PhysRevA.100.062131}
\begin{align}
    \tilde{H}^{\rm eff}_{\rm ph}=&\hbar\sum_{i=\rm L,R}\omega_{i}a^\dagger_i a_i  -i\hbar\sum_{i=\rm L,R}\gamma_{i}a^\dagger_i a_i\nonumber\\
    &-\hbar\kappa^{\rm NMI} [a^\dagger_{\rm L} a_{\rm R}+{\rm H.a.}].
\end{align}


Furthermore to model the quantum coupling to a TLS, one can assume a dipole-field Hamiltonian in the bare resonator basis, so that the total Hamiltonian would read
\begin{equation}
     \tilde{H}_{\rm S}^{\rm eff}=\tilde{H}^{\rm eff}_{\rm ph}+\hbar\omega_0\sigma^+\sigma^-+i\hbar \sum_{i=\rm L,R} g_i^{\rm NMI} [\sigma^+ a_i-{\rm H.a.}],
\end{equation}
where $g_i^{\rm NMI}=\sqrt{\omega_0/(2\hbar\epsilon_0)}\mathbf{d}\cdot \mathbf{f}^i(\mathbf{r}_{0})$ is assumed as the real-valued NM-TLS coupling constant, which uses the bare (uncoupled) resonator functions.
The above effective complex Hamiltonian can be complemented by quantum jump terms to formulate an associated Lindblad master equation of the form 
\begin{equation}\label{eq: new_masterNMs}
    \partial_t \rho = -\frac{i}{\hbar}[H_{\rm S}^{\rm eff},\rho]+\gamma_{\rm L}\mathcal{D}[a_{\rm L}]\rho +\gamma_{\rm R}\mathcal{D}[a_{\rm R}]\rho, 
\end{equation}
with an effective system Hamiltonian $H_{\rm S}^{\rm eff}$ as the Hermitian part of $\tilde {H}_{\rm S}^{\rm eff}$.

The effective dissipation in the improved quantum NM model is now captured by the imaginary parts of the bare QNM frequencies through the Lindblad dissipator terms 
while the {\it intercavity coupling constant} is equal to the NM CMT coupling. The above model is formally similar to a Jaynes-Cummings-Hubbard model~\cite{JCH_1,JCH_2} for the special case of two cavities and one quantum emitter. Note, that in the above construction, it was assumed that the photon operators for the bare cavities fulfill bosonic commutation relations, i.e., $[a_i,a_j^\dagger]=\delta_{ij}$. This is of course only true, if the bare resonators behave approximately as closed cavities. We emphasize, that besides the quality factor $Q$, the NM approximation also depends on the overlap and imaginary part of the QNM eigenfunctions, which in a coupled resonator system can be large even in a high $Q$ regime.

To access the PF from the above model, one can again apply the bad cavity limit similarly to the quantized QNM model.
In doing so, one arrives at the TLS master equation (neglecting the photonic Lamb shift)
\begin{equation}
\partial_t\rho_{\rm a}=-\frac{i}{\hbar}\left[H_{0},\rho_{\rm a}\right]+\frac{\Gamma^{\rm qNMI}}{2}\mathcal{D}[\sigma^-]\rho_{\rm a}\label{masterrhoAtom5},
\end{equation}
where $\Gamma^{\rm qNMI}$ is the NM cavity-enhanced decay rate~\cite{SI_short} 
\begin{align}\label{eq: qGammaNMI}
    \Gamma^{\rm qNMI}=2\sum_{i,j,k}g_i^{\rm NMI}g_{j}^{\rm NMI}{\rm Re}\left\{\left(\mathbf{P}^{}\right)_{ik}\frac{i}{\omega_0-\tilde\Omega_{k }^{\rm eig}}\left(\mathbf{P}^{-1}\right)_{kj}\right\},
\end{align}
which is a real valued quantity,
and the summation terms run over
 $i,j,k=({\rm L,R})$.

Similar to before, 
it is convenient to
also define the separate contributions, as
\begin{align}\label{eq: qGammaNMI_terms}
    \Gamma^{\rm qNMI}_{ij}=2\sum_{k}g_i^{\rm NMI}g_{j}^{\rm NMI}{\rm Re}\left\{\left(\mathbf{P}^{}\right)_{ik}\frac{i}{\omega_0-\tilde\Omega_{k }^{\rm eig}}\left(\mathbf{P}^{-1}\right)_{kj}\right\},
\end{align}
where $\tilde\Omega_{k }^{\rm eig}$ are the complex eigenvalues of the non-Hermitian photon-photon matrix 
\begin{equation}
    \tilde{\boldsymbol{\Omega}}=\begin{pmatrix}
    \tilde\omega_{\rm L} & \kappa^{\rm NMI}\\
    \kappa^{\rm NMI} & \tilde{\omega}_{\rm R}
    \end{pmatrix},
\end{equation}
and 
$\mathbf{P}$ is the (right) eigenmatrix of $\tilde{\boldsymbol{\Omega}}$, i.e., it diagonalizes $\tilde{\boldsymbol{\Omega}}$ so that %
\begin{equation}
    \mathbf{P}^{-1}\cdot\tilde{\boldsymbol{\Omega}}\cdot\mathbf{P}=\begin{pmatrix}
    \tilde\Omega_{\rm L}^{\rm eig} & 0\\
    0 & \tilde\Omega_{\rm R}^{\rm eig}
    \end{pmatrix}.
\end{equation}
Of course, it follows directly from the classical NM CMT problem, that the eigenvalues are $\tilde\Omega_{\rm L}^{\rm eig}=\tilde\omega_+^{\rm NMI}$ and $\tilde\Omega_{\rm R}^{\rm eig}=\tilde\omega_-^{\rm NMI}$. 

We emphasize, that in order to recapture the correct behavior of the decay of the TLS from a QNM model or a improved classical NM model, the photon interaction Hamiltonian in the quantum NM model must be constructed with a coupling constant including a negative sign, i.e., $-\hbar\kappa$. This is completely different to the usually applied phenomenological approaches, that rather involve $\hbar\kappa$ or $i\hbar\kappa$, where the Purcell factor would be erroneously predicted (see Supplemental Material~\cite{SI_short} for further details). This is not at all an obvious fix to the phenomenological approaches, since the replacement from $\tilde{\Omega}_{{\rm LR}}=\tilde{\Omega}_{{\rm RL}}=\kappa^{\rm NMI}$ to $\tilde{\Omega}_{{\rm LR}}=\tilde{\Omega}_{{\rm RL}}=-\kappa^{\rm NMI}$ or to $\tilde{\Omega}_{{\rm LR}}=i\kappa^{\rm NMI},~\tilde{\Omega}_{{\rm RL}}=-i\kappa^{\rm NMI}$ would not change the eigenfrequencies $\tilde\omega_\pm$ and the pure photon part would be invariant. The key difference is induced by the additional coupling to the quantum emitter, where signs and phases of the photon coupling constants become important, which underlines the need of a rigorous derivation of the mode parameters.

The {\it quantum Purcell factor} with this improved NM theory is   
\begin{equation}\label{eq: Fp_qNMI}
F_{\rm P}^{\rm qNMI}=\frac{\Gamma^{\rm qNMI}}{\Gamma_{0}}+1,
\end{equation}
and the modal contributions from the separate parts in Eq.~\eqref{eq: qGammaNMI_terms} are ($i,j={\rm L,R}$)
\begin{equation}\label{eq: Fp_qNMI_ij}
F_{{\rm P},ij}^{\rm qNMI}=\frac{\Gamma^{\rm qNMI}_{ij}}{\Gamma_{0}},
\end{equation}
which satisfy 
\begin{equation}
F_{\rm P}^{\rm qNMI}=1+F_{\rm P,LL}^{\rm qNMI}+F_{\rm P,LR}^{\rm qNMI}+F_{\rm P,RL}^{\rm qNMI}+F_{\rm P,RR}^{\rm qNMI}.
\end{equation}

As shown in Fig.~\ref{fig: newcNMs_waterfall}, the quantum Purcell factors $F_{\rm P}^{\rm qNMI}$ (Eq.~\eqref{eq: Fp_qNMI}, dashed blue curves) for a dipole at $d_{\rm L/R}=10~$nm agree very well with the full numerical dipole results $F_{\rm P}^{\rm num}$ (Eq.~\eqref{eq: Purcellfulldipole}, red circles) for the $4$ gap distance $d_{\rm gap}=750/800/850/900~$nm. In addition, similar to the classical results, the contribution from the dominant terms (Eq.~\eqref{eq: Fp_qNMI_ij}) are given in Fig.~\ref{fig: newcNMs_terms}, where the LL term $F_{{\rm P,LL}}^{\rm qNMI}$ is dominant when the dipole is at $d_{\rm L}=10~$nm. Similarly, the
the RR term $F_{{\rm P,RR}}^{\rm qNMI}$) is dominant when the dipole is at $d_{\rm R}=10~$nm.

Although our numerical simulations clearly
show excellent agreement again will full numerical
dipole simulations, we add the same word of caution: similar to the general classical NM model, this quantum model would also break down for smaller $Q$, since we are still applying a partial NM approximation to the fields and the coupled mode coefficients. For example, the improved NMs can qualitatively break down for low $Q$,
e.g., using 
coupled photonic crystal cavities in the region of Fano-like interference~\cite{RosenkrantzdeLasson2015}, where we have checked this breakdown explicitly. 


\section{Conclusions}\label{sec: conclusion}

We have presented both classical and quantized QNM theories for coupled high-$Q$ lossy microdisk resonators, showing how both approaches yield the Purcell factor limit in different ways and how these models connect to each other.

Starting with the bare classical QNMs of single microdisk resonators (as input), the hybrid classical QNMs were also obtained analytically using an efficient CQM approach, which agree extremely well with full numerical dipole results as a function of gap distance in the frequency of interest.
Furthermore, a significant difference between the classical QNMs solution and classical NMs solutions are presented, indicating the significance of the QNMs phase terms and interference effects.
This occurs despite the fact that the resonances have a large quality factor (around $10^{4}-10^{5}$).

In a {\it modal} quantum picture, we show how the {\it incorrect} classical NM theory recovers the same result as the commonly used  dissipative JC model, whose equivalence is also discussed rigorously from the viewpoint of photon Green functions.
More importantly, after obtaining the required quantum $S$ parameters (formally quantum overlap integrals, $S_{ij}$), the more correct quantized QNM approach includes essential 
off-diagonal coupling terms, which captured the underling mode interference and exactly recovers  the Purcell factors from classical QNMs and full numerical dipole solution (within numerical precision).
Moreover, the contribution from these non-diagonal terms is maximized when close to the EP, where changes to the QNM decay rate ratios and the coupling rate ratios for separate QNMs are also significant.  
These results show the power and importance of using QNMs in the construction of mode theories of coupled resonators, in both classical and quantum optics pictures.
Even for high $Q$ resonators, common NM theories and dissipative JC model can drastically break down, which is essential to capture the correct 
underlying physics and to guide 
and explain emerging experiments works. 
Clearly such effects are also important for applications in optical sensing and lasing. 

Another useful application is for developing more accurate ways of improving quantum approaches at the system level for these complex resonators, offering a quantitatively accurate theory with just input from the classical QNM parameters. We have presented a first example of such an improvement, where the QNM complex eigenfrequencies are used instead of the real NM eigenfrequencies, to the commonly used lossless CMT, complemented by a Jaynes-Cummings model with an intercavity coupling in the bare resonator regime. Indeed, this improved NM model fully recovers the interference effects in the Purcell factor calculations for our example coupled resonator structure with no fitting parameters. However, likely this level of accuracy is related to using  high-$Q$ bare resonator systems, which underlines the importance of the fully quantized QNM model. Future work will look at lower $Q$ systems including coupled metallic systems and low $Q$ dielectric cavity systems, using the three-dimensional QNMs.\\

\section*{Acknowledgements}
We  acknowledge funding from Queen's University,
the Canadian Foundation for Innovation, 
the Natural Sciences and Engineering Research Council of Canada, and CMC Microsystems for the provision of COMSOL Multiphysics.
We also acknowledge support from 
the Alexander von Humboldt Foundation through a Humboldt Research Award.
We thank Andreas Knorr
for discussions and  support.

\bibliography{refs}

\end{document}